\documentclass[a4paper,11pt]{article}
\pdfoutput=1 % if your are submitting a pdflatex (i.e. if you have
\usepackage{jcappub} % for details on the use of the package, please
\usepackage[T1]{fontenc} % if needed
\usepackage{hyperref}

\usepackage{xfrac}
\usepackage{graphicx}
\usepackage{rotating}
\usepackage{physics}
\usepackage{multirow}
\usepackage[normalem]{ulem}
\usepackage{amssymb,amsmath}
\usepackage[dvipsnames]{xcolor}
\usepackage{tikz}
\usetikzlibrary{shapes,arrows}
\usepackage{pdflscape} 
\usepackage{array}
\newcolumntype{L}{>{$}l<{$}}
\newcolumntype{C}{>{$}c<{$}}

\newcommand{\join}{\cup}

\newcommand{\rd}{r_\mathrm{d}}

\newcommand{\Mpcoverh}{\left[ h^{-1}\mathrm{Mpc}\right]}

\newcommand{\hoverMpc}{\left[ h\mathrm{Mpc}^{-1} \right]}

\title{\boldmath Model-agnostic interpretation of 10 billion years of cosmic evolution traced by BOSS and eBOSS data}

\author[1,2]{Samuel Brieden}
\author[1]{H\'ector Gil-Marín}
\author[1,3]{Licia Verde}

\affiliation[1]{ICC, University of Barcelona, IEEC-UB, Mart\'i i Franqu\`es, 1, E-08028 Barcelona, Spain}
\affiliation[2]{Dept. de F\'isica Qu\`antica i Astrof\'isica, Universitat de Barcelona, Mart\'i i Franqu\`es 1, E-
08028 Barcelona, Spain}
\affiliation[3]{ICREA, Pg. Llu\'is Companys 23, Barcelona, E-08010, Spain}

\emailAdd{sbrieden@icc.ub.edu}
\emailAdd{hectorgil@icc.ub.edu}
\emailAdd{liciaverde@icc.ub.edu }

\abstract{We present the first model-agnostic analysis of the complete set of Sloan Digital Sky Survey III (BOSS) and -IV (eBOSS) catalogues of luminous red galaxy and quasar clustering in the redshift range $0.2\leq z \leq 2.2$ (10 billion years of cosmic evolution), which consistently includes the baryon acoustic oscillations (BAO), redshift space distortions (RSD) and the shape of the transfer function signatures, from pre- and post-reconstructed catalogues in Fourier space. This approach complements the standard analyses techniques which only focus on the BAO and RSD signatures, and the full-modeling approaches which assume a specific underlying cosmology model to perform the analysis. These model-independent results can then easily be interpreted in the context of the cosmological model of choice. In particular, when combined with $z>2.1$ Ly-$\alpha$ BAO measurements, the clustering BAO, RSD and {\it Shape} parameters can be interpreted within a flat-$\Lambda$CDM model yielding $h=0.6816\pm0.0067$, $\Omega_{\rm m}=0.3001\pm0.0057$ and $10^{9}\times A_s= 2.43\pm0.20$ (or $\sigma_8=0.858\pm0.036$) with a Big Bang Nucleosynthesis prior on the baryon density. 
Without any external dataset, the BOSS and eBOSS data alone imply $\Omega_{\rm m}=0.2971\pm 0.0061$ and $10^{9}\times A_s=2.39^{+0.24}_{-0.43}$ (or $\sigma_8=0.857\pm0.040$).
For models beyond $\Lambda$CDM, eBOSS data alone (in combination with Planck) constrain the sum of neutrino mass to be $\Sigma m_\nu< 0.40$ eV with a BBN prior ($\Sigma m_\nu <0.082$ eV) at 95\% CL, the curvature energy density to $\Omega_\mathrm{k} = -0.022_{-0.038}^{+0.032}$ ($\Omega_\mathrm{k} = 0.0015\pm 0.0016$) and the dark energy equation of state parameter to $w=-0.998_{-0.073}^{+0.085}$ ($w=-1.093_{-0.044}^{+0.048}$) at 68\% CL without a BBN prior.
These results are the product of a substantial improvement of the state-of-the-art methodologies and represent the most precise model-agnostic cosmological constrains using spectroscopic large-scale data alone.}

\begin{document}
\maketitle
\flushbottom

\section{Introduction}
\label{sec:intro}

Observations of the Cosmic Microwave Background (CMB, e.g., \cite{2013ApJS..208...19H, Aghanim:2018eyx}) have been pivotal in establishing the $\Lambda$CDM model as the standard model for cosmology.
The interpretation of CMB observations is very sensitive to the (linear, and, within the standard cosmological model, simple and well understood) physics of the early Universe. 
However, one of the main puzzles of modern cosmology, the cosmic acceleration, is a late-time ($z\lesssim1$) phenomenon, hence cosmological constraints from the late-time Universe observations are of crucial importance to study dark energy. 
Complementary to Supernovae observations, which first provided evidence for cosmic acceleration, the large-scale structure (LSS) of the Universe, provides a unique window into the evolution of the late-time Universe. 

The development of massive spectroscopic surveys of galaxies and quasars over wide areas of the sky over the past two decades (e.g., \cite{WiggleZ,SDSS,2dF})
has propelled the study of clustering of LSS into the realm of precision cosmology.
The clustering of galaxies and other dark matter tracers (such as quasars or the Ly-$\alpha$ forest) provides precise measurements of the cosmic expansion history with baryon acoustic oscillations (BAO) and measurements of the rate of structure growth with redshift space distortions (RSD). 

Perturbations in the photon-baryon fluid of the early Universe leave an imprint in the late-time clustering of cosmic structure as a feature (the BAO, \cite{EH_TransferFunction}) observable in the LSS power spectrum and first detected by \cite{SDSS,2dF}. The BAO feature offers a standard ruler whose length can be calibrated by early-time physics, but also, when observed in the late-time clustering, can be used to determine the expansion history of the Universe via the Alcock-Paczynski (AP) effect \cite{1979Natur.281..358A}. The tracer's power spectrum yields two scaling parameters
-- $\alpha_\parallel, \alpha_\perp$-- respectively along and across the line-of-sight (LOS) direction. 
The information extracted is purely geometrical and model-independent, it is only mildly affected by non-linear physics, making the BAO one of the most robust probes of the late-time Universe. 
To reduce the potential bias on the BAO feature induced by non-linearities and to boost the BAO signal-to-noise, it is customary to apply the reconstruction technique \cite{Eis2007,burden_reconstruction_2015}. Reconstruction effectively generates an additional catalogue and thus additional `post-recon' power spectra. These are highly correlated to the `pre-recon' ones --hence their covariances must be carefully taken into account-- but do add significant information and are used only for the BAO part of the analysis. 

Furthermore, gravitationally-induced peculiar velocities give rise to deviations from the Hubble flow which imprint RSD on the three-dimensional map produced by redshift surveys. Pioneered by \cite{kaiser_clustering_1987}, RSD enclose information about the combination of the amplitude of velocity fluctuations with the dark matter amplitude perturbations. As such, they trace the growth history of cosmic structures, offering thus important insights into the nature of gravity. 

Most analyses of state-of-the-art surveys, e.g., \cite{alam_clustering_2017, eboss_collaboration_dr16}, adopt what we refer to as the `classic' approach in order to extract cosmological information from the tracers' clustering. With the help of a template of the power spectrum, the clustering data are compressed into few (three per redshift interval considered corresponding to two scaling parameters and one growth rate parameter) physical observables, or compressed variables, which are only sensitive to late-time physics. The resulting constraints on these compressed variables can then be re-interpreted {\it a posteriori} as constraints on (cosmological) parameters for a given cosmological model (or family of models).

This `classic' approach is conceptually different from the way, for example, the CMB power spectrum is analyzed, and from the analysis of LSS data pre-BAO era (see for e.g., \cite{Efstathiou:2000fk,2dFGRS:2001csf,SDSS:2003tbn}), which we refer to as `full modeling' (or FM). 
After selecting a cosmological model {\it ab initio}, the measured power spectrum is compared directly to the model's prediction, and the model's parameters are then constrained by standard statistical inference. The procedure is repeated for every model under consideration. 

If clustering is analyzed without external datasets or priors, the application of the FM approach to state-of-the-art redshift surveys (e.g., \cite{DAmico:2019fhj,Ivanov:2019pdj, philcoxetal22}) produces much tighter constraints on cosmological parameters than the classic approach. In a joint CMB+LSS analysis however the two perform very similarly.

In other words, the compression employed by the classic approach, disentangles the late-time physics from the early-time one, isolates the part of the cosmological signal least affected by systematics and makes the resulting constraints as model independent as possible (e.g., \cite{eisenstein_robustness_2007, seo_baryonic_2005} and refs. therein). But has the drawback that the compression is not lossless.
Full modeling approaches are model-dependent, and computationally more demanding both in terms of analysis and of modeling of the signal, but --compared to the `classic' approach-- extract additional information mostly from the broadband shape of the power spectrum. 

A simple one-parameter extension of the `classic' approach, ShapeFit, was proposed by \cite{ShapeFit}.
Rather than being a perturbative approach, Shapefit is a compression technique that can be applied to any perturbative model. It improves the information content offered by fixed-template approaches to the point that it becomes competitive with full-modelling approaches.

The ShapeFit phenomenological parameter $m$ is related to the shape of the power spectrum on very large scales and to the shape of the matter transfer function; it was designed to capture a series of early-time processes that affect the broadband power spectrum shape in the linear regime.
ShapeFit provides constraints on a suite of model-independent variables, capturing from the data a series of physical effects and processes without relying specifically on any cosmological model. The extraction of these model-independent quantities for the biggest  galaxy redshift survey to date represents the main result of this paper. Additionally these model-independent results can be transformed into, or interpreted as, constraints on parameters for specific cosmological models.

The application of ShapeFit to the large volume, high resolution, PT challenge \cite{2020PhRvD.102l3541N} simulations suite \cite{ShapeFitPT} demonstrates that this approach is effectively unbiased even for a survey volume 10 times larger than that probed by future surveys. Ref.~\cite{BriedenPRL21} presents the ShapeFit analysis of the Sloan Digital Sky Survey-III BOSS data and demonstrates that it matches the constraining power of FM approaches performed to date on the same data. 

Here we consider the full BOSS and eBOSS observations campaigns~\cite{eboss_collaboration_dr16} representing the 
final use of the Apache Point Observatory 2.5m Sloan Telescope for galaxy redshift surveys designed to measure cosmological parameters using BAO and RSD techniques. Four generations of Sloan Digital Sky Survey culminated with the eBOSS data release, which probes $\sim 10$ billion years of cosmic evolution through more than 2 million spectra.
We apply the ShapeFit analysis on these data and present the resulting constraints on the physical parameters.
We argue that ShapeFit extracts virtually all the robust and model-independent cosmological information carried by LSS clustering. 
The constraints on the physical parameters are then interpreted in light of a suite of popular cosmological models including the standard $\Lambda$CDM and its common one-parameter extensions. 

It is worth highlighting that in producing the ShapeFit constraints on the physical parameters, no assumption is made about the underlying cosmological model. A Friedmann-Lemaitre-Robinson-Walker metric is assumed and thus statistical homogeneity and isotropy, although General Relativity (GR) is not assumed on large scales. However, Newtonian dynamics (hence GR) is assumed at small, mildly non-linear, scales when reconstruction is applied to boost the BAO signal. Within ShapeFit, no explicit scale-dependence of the growth rate is considered, hence the measured growth rate should be considered as effective, suitably weighted across the relevant scales. No assumption is made about early-time physics, the nature of dark energy, of dark matter or spatial curvature. However, unless otherwise stated, the interpretation of the constraints on the shape parameter $m$ assumes a power-law primordial power spectrum with fixed spectral slope; moreover a Big Bang nucleosynthesis (BBN) prior is adopted when converting the compressed variable constraints into cosmological parameters. Unless otherwise stated, galaxy bias is assumed to be local in Lagrangian space. This set of assumptions only affects the shape parameter constraints and not the other compressed variables. 

The rest of the paper is organized as follows. In section~\ref{sec:shapefit} the theory and methodology are described. This section is mostly a review of material covered elsewhere in the literature, but its presentation is tuned to the current application.
The data set used is presented in section~\ref{sec:data} along with the simulated mock surveys which are employed to estimate the relevant covariance matrices. Model-independent results, and the main results of this paper, on the physical variables are presented in section~\ref{sec:results} and their re-interpretation under the $\Lambda$CDM and a suite of extensions to this model are reported in section~\ref{sec:cosmo}, as a demonstration of how our main results
can be interpreted under the light of the most popular models in the literature.
Section~\ref{sec:sys} reports a suite of systematic checks performed on synthetic catalogues and estimates the overall systematic error budget.  Finally we present the main conclusions of this work in section~\ref{sec:conclusions}. The appendices quantify the impact on the final results of several assumptions ranging from the nature of tracer's bias, fiber collisions and prior choices.

\section{Methodology, Theory and Data compression techniques} \label{sec:shapefit}

As mentioned in the introduction, there is more than a single way to perform a cosmological analysis to spectroscopic galaxy survey data.
In this section we review and summarize approaches that are already in the literature, in particular \cite{ross_information_2015,Percival:2008sh,ShapeFit,Eis2007,burden_reconstruction_2015} and references therein. This section however, also serves to highlight differences, similarities and connections among them. 
In this work, we focus on the fixed template approach, where the measured galaxy power spectra are compressed into physical variables $\mathbf{\Theta}_\mathrm{phys}$ at each redshift bin, which in turn can be interpreted in light of a cosmological model and its parameters $\mathbf{\Omega}$. Note that, within the compression step, the power spectra are fitted in a model-agnostic way, without imposing any of the $\Lambda$CDM-type of relations among (physical) parameters. In this way, (cosmological) model's assumptions are introduced only at the very late stages of the analyses. This has several advantages, for example, there is no need to re-do the fit if the cosmology paradigm changes, or when a novel model or class of models needs to be tested. This is one of the main reasons for adopting this philosophy of interpreting the spectroscopic data, rather than direct fits (or full modeling fits). 

All model-agnostic fixed-template compression techniques rely on two fundamental steps. First, a fiducial cosmology $\mathbf{\Omega}^\mathrm{fid}$ is needed to generate a reference coordinate and unit system. The coordinate system depends on the distance-redshift relation and the unit system depends on the fiducial linear matter power spectrum template $P_\mathrm{lin}(k,\mathbf{\Omega}^\mathrm{fid})$ as function of wavenumber $k$. 
Second, the fiducial template is transformed as to be compared with the observable galaxy power spectrum multipoles in redshift space, $P_\mathrm{model}^{(\ell)}(k,\mathbf{\Theta_\mathrm{model}}|\mathbf{\Omega}^\mathrm{fid})$ given a certain model or compression type. Here the `|' sign indicates that the dependence of the fiducial cosmology is implicit rather than explicit. As we explain in more detail in section~\ref{sec:data-prepostcomb}, the final constraints on physical parameters do not depend on the template's choice.\footnote{Previous studies have checked that there is a residual dependence which is very sub-dominant with respect to BOSS and eBOSS statistical errors, even for cases where this reference template is many standard deviations away from best-fit CMB anisotropy cosmologies. For detailed studies on how the arbitrary choice of the fixed template can impact the cosmological results, we refer the reader to Appendix B of \cite{ShapeFit}.} The symbol $\mathbf{\Theta_\mathrm{model}}$ corresponds to a set of physical and nuisance parameters $\mathbf{\Theta}_\mathrm{model} = \mathbf{\Theta}_\mathrm{phys}^\mathrm{model} \join \mathbf{\Theta}_\mathrm{nuis}^\mathrm{model}$ that is used to \textit{i)} probe all late-time dynamics effects (geometry and/or growth) in the most generic, model-independent way, and {\it ii)} once the physical parameters are constrained at each redshift bin, use them to test cosmological models. This compression step is described for three different cases in section \ref{sec:theory_Pell}. 

In particular, we review the `classic' BAO and RSD analyses in addition to the recently introduced ShapeFit compression.

\subsection{Fiducial cosmology} \label{sec:theory_fidcosmo}

As anticipated above, the purpose of adopting a fiducial cosmology $\mathbf{\Omega_\mathrm{fid}}$ is twofold. 

First, it is needed to generate a \textit{coordinate system}. The galaxy positions provided by BOSS and eBOSS are measured in terms of angles and redshifts. These coordinates are transformed to distances based on a distance-redshift relation determined by the fiducial cosmology, in particular (within $\Lambda$CDM) by the matter density today $\Omega_\mathrm{m}$ and the Hubble expansion rate today (or `little $h$') $H_0=100\,h\,\mathrm{km}/\mathrm{s}/\mathrm{Mpc}$. This coordinate transformation is essential to extract the full three-dimensional clustering statistics from galaxy catalogs, as different cosmological models affect the distances along and across the LOS differently.

Second, the fiducial cosmology is needed to generate a \textit{unit system} for the distances (akin to interpreting the hatching of a ruler). This is provided by the fiducial matter power spectrum template $P_\mathrm{lin}(k,\mathbf{\Omega}^\mathrm{fid})$, whose shape is predominantly (but not solely) determined by the sound horizon at radiation drag epoch, $r_\mathrm{d}$, the so-called `standard ruler'. In the template it manifests itself via the location of the wiggles on one hand (measured by the BAO analysis), and as a characteristic suppression scale on the other hand (measured by ShapeFit). The latter effect is somewhat degenerate (at the scales of interest for galaxy clustering) with the overall power spectrum slope determined by the scale of equality between matter and radiation, $k_\mathrm{eq}$, and the primordial tilt, $n_s$. But the power of ShapeFit is to measure the slope in a model-independent way.  

In principle, one could adopt different fiducial cosmologies for the coordinate and the unit system. But, as it is customary, here we use the same fiducial cosmology for both tasks. For simplicity, we use the same fiducial cosmology $\mathbf{\Omega}_\mathrm{fid}$ employed in the official BOSS and eBOSS analyses, with parameter values listed in table~\ref{tab:cosmo}. 

Throughout this work we denote by `fid' the quantities evaluated at that cosmology. The quantities without this notation denote the true underlying values of the sample we fit (either mock or actual data). 

\subsection{Modeling the power spectrum multipoles} \label{sec:theory_Pell}

In general, because the power spectrum of the observed galaxy map is constructed adopting a fiducial coordinate \textit{and} unit system, the modeled power spectrum multipoles need to be rescaled to that system in order to be compared to the data. These coordinate and unit conversions (called late-time and early-time rescaling respectively in \cite{ShapeFit}) are almost perfectly degenerate, which is why they are often represented by the following scaling parameters,  

\begin{equation} \label{eq:thery_alpha_scaling}
    \alpha_\parallel(z) \equiv \frac{D_H(z)/r_{\rm d}}{[D_H(z)/r_{\rm d}]^{\rm fid}},\quad \alpha_\perp(z) \equiv \frac{D_M(z)/r_{\rm d}}{[D_M(z)/r_{\rm d}]^{\rm fid}}~,
\end{equation}
where $D_H(z) \equiv c/H(z)$ and $D_M(z) \equiv \int_0^{z} c/H(z^\prime) dz^\prime$ are the distances along and across the LOS respectively, with Hubble expansion rate $H(z)$. These scaling-parameters are used to transform the power spectrum multipoles into the correct observable coordinates in units of the standard ruler and they are allowed to vary freely.

Note that the scaling parameters as defined in eq.~\eqref{eq:thery_alpha_scaling} depend on the arbitrary choice of the template, but once they are converted to the physical distances in units of the BAO scale, $D_H/r_{\rm d}$ and $D_M/r_{\rm d}$, this dependence vanishes. Hence, we use both notations interchangeably, in particular we use $\left\lbrace \alpha_\parallel,\alpha_\perp \right\rbrace$ when referring to the template fits and $\left\lbrace D_H/r_{\rm d},D_M/r_{\rm d} \right\rbrace$ when referring to their cosmological interpretation.

The modeled power spectrum multipoles for a given reference template based on the cosmology $\mathbf{\Omega}_\mathrm{fid}$ are usually written as,
\begin{equation} \label{eq:theory_Pell_general}
    P_\mathrm{model}^{(\ell)} (k) =  \frac{(2\ell+1)}{2\alpha_\perp^2 \alpha_\parallel} \int_{-1}^1 \! P_\mathrm{model} (\widetilde{k}(k,\mu), \widetilde{\mu}(\mu), \mathbf{\Omega}_\mathrm{fid}, \mathbf{\Theta}_\mathrm{model}) \mathcal{L}_\ell(\mu) \, d\mu + g^{(\ell)}(\mathbf{X}_\mathrm{model}) ~,
\end{equation}
where $\mathbf{\Theta}_\mathrm{model}$ includes physical and nuisance parameters of the compression method of choice, $g^{(\ell)}(\mathbf{X}_\mathrm{model})$ represents an arbitrary function accounting for the broadband signal which depends on multipole $\ell$ and extra free nuisance parameters, $\mathbf{X}_\mathrm{model}$, the coordinates $(k, \mu)$ are the wavevector in units $[\mathrm{Mpc}^{-1}h]$ and the cosine of the separation angle, $\mathcal{L}_\ell$ is the Legendre polynomial of order $\ell$ and the rescaled coordinates $(\widetilde{k}, \widetilde{\mu})$ are defined as,

\begin{equation}
    \widetilde{k} = \frac{k}{\alpha_\perp} \left[ 1 + \mu^2 \left( \frac{\alpha_\perp^2}{\alpha_\parallel^2} -1 \right)\right]^{1/2}, \quad \widetilde{\mu} = \mu \frac{\alpha_\perp}{\alpha_\parallel} \left[ 1 + \mu^2 \left( \frac{\alpha_\perp^2}{\alpha_\parallel^2} -1 \right)\right]^{-1/2}. 
\end{equation}

The exact model implementation of $P_\mathrm{model}$ and the corresponding parameter-sets $\mathbf{\Theta}_\mathrm{model}$ and $\mathbf{X}_\mathrm{model}$ depend on the type of compression used to analyze the data. Different choices are summarized below.

\subsubsection{BAO compression} \label{sec:theory_BAO}
For the BAO analysis, only the oscillatory feature within the power spectrum, at wavenumbers determined by the sound horizon at radiation drag $r_\mathrm{d}$, is of interest. Therefore it is customary to separate the fiducial linear power spectrum template into a no-wiggle ($P_\mathrm{nw}^\mathrm{fid} = P_\mathrm{lin}^\mathrm{fid}-P_\mathrm{wig}^\mathrm{fid}$) and a wiggle ($P_\mathrm{wig}^\mathrm{fid}$) part, such that the scaling only affects the oscillatory part $\mathcal{O}_\mathrm{lin}^\mathrm{fid}=P_\mathrm{lin}^\mathrm{fid}/P_\mathrm{nw}^\mathrm{fid}$, while the no-wiggle broadband shape is marginalized over. In practice, this is achieved by setting the following model power spectrum $P_\mathrm{model}$ into eq.~\eqref{eq:theory_Pell_general} \cite{Beutler:2016ixs,Gil-Marin:2020bct}:
\begin{equation}
    P_\mathrm{BAO}(k,\mu) = B^2 (1+\beta \mu^2 R)^2 P_\mathrm{nw}^\mathrm{fid}(k) \left[ 1 + \left( \mathcal{O}_\mathrm{lin}^\mathrm{fid}(k) - 1 \right) e^{-\frac{1}{2} \left( \mu^2 k^2 \Sigma_\parallel^2 +   (1-\mu^2) k^2 \Sigma_\perp^2 \right)} \right] , 
\end{equation}
where $B$ represents a global amplitude parameter, $\beta$ (defined as $\beta=f/b_1$)\footnote{Within the BAO analysis we do not use this parameter to measure the growth rate $f$, but rather marginalize over it.} incorporates linear (Kaiser) redshift space distortions, the damping terms ($\Sigma_\parallel,\Sigma_\perp$) include the anisotropic, non-linear damping of the BAO-amplitude and $R$ is either the smoothing scale used in reconstruction (see section \ref{sec:theory_recon}), or set to zero in case the BAO fit is performed on pre-reconstruction measurements.

In addition, the broadband power spectrum is marginalized over by adding to each power spectrum multipole the following polynomial expansion of order $N=5$ for BOSS LRGs and $N=3$ for eBOSS LRGs:
\begin{equation} \label{eq:theory_broadband_marginalization}
    g^{(\ell)}(\mathbf{X}_\mathrm{BAO}) = \sum_{i=1}^{N} A_i^{(\ell)}k^{2-i}~.
\end{equation}
Hence, our BAO-model is fully described by 2 physical parameters $\mathbf{\Theta}_\mathrm{phys}^\mathrm{BAO} = \left\lbrace \alpha_\parallel,\alpha_\perp \right\rbrace$ and 23 (15) nuisance parameters for BOSS (eBOSS) LRGs $\mathbf{\Theta}_\mathrm{nuis}^\mathrm{BAO} = \left\lbrace \beta, B_\mathrm{N/S}, A_{i,N/S}^{(\ell)} \right\rbrace$ per redshift bin, where subscripts `N' and `S' stand for the north and south galactic caps. The damping terms ($\Sigma_\parallel,\Sigma_\perp$) are not varied freely but are  set to fiducial values estimated from the mocks of each sample (see section~\ref{sec:mocks}). This is the standard procedure of the official BOSS and eBOSS papers which  we adopt  here  for a  transparent comparison.  This approach has been extensively  validated by exploring the impact of relaxing these fixed values, or varying them with Gaussian priors (see table 12 of \cite{Bautistaetal21} as an example), where no significant shift on the BAO position or its error has been found. 

\subsubsection{RSD compression} \label{sec:theory_RSDcomp}
In addition to the BAO analysis, where only the `horizontal' information (coming from the wiggle position as a function of the angle to the LOS) is considered, the RSD analysis aims to gain cosmological insight also from the `vertical' information (coming from the relative broadband amplitude as a function of the angle to the LOS).

Therefore, the first ingredient of the RSD compression are the scaling parameters $\left\lbrace \alpha_\parallel,\alpha_\perp \right\rbrace$ defined in eq.~\eqref{eq:thery_alpha_scaling}
which capture the AP-effect and are also sensitive to the absolute position of the BAO at drag epoch. 
Although strictly speaking the AP effect affects all scales (and not only the BAO scale) it has been shown (see appendix D of \cite{ShapeFit}) that the BAO signal greatly dominates over the rest of the scales, and therefore it is common practice in the literature to treat the BAO scale as fully degenerate with the scale dilation parameters (see also sections 2 and 3 of \cite{ShapeFit} for a further discussion on this topic). 

The redshift space distortion effects on the other hand are sensitive to the following combination of parameters: the growth of structure $f$ times the amplitude of matter fluctuations at the scale of $8\,\mathrm{Mpc}\,{h}^{-1}$,
\begin{equation}
    f(z)\times \sigma_8(z)\equiv f\sigma_8(z)=  \Omega_{\rm m}^\gamma(z) \times \left[ \int_0^\infty dq\, q^2 P_{\rm lin}(q;z) W_{\rm TH}(qR_8)\right]^{1/2}
\end{equation}
where $\gamma=6/11$ for General Relativity, $W_{\rm TH}$ is the top-hat function, which in this case smooths the fluctuations of the matter field in a scale of $R_8\equiv 8\,{\rm Mpc}\,{h}^{-1}$ .

In practice, this is implemented within the fixed template fits as follows: the amplitude of matter fluctuations $\sigma_8$ is fixed by the template, which provides a `standard amplitude' in a similar fashion to the standard ruler $r_\mathrm{d}$. The free parameter of the RSD compression is the growth rate $f$, which enters the galaxy power spectrum $P_g$ in redshift space following the TNS \cite{Taruya:2010mx} model,
\begin{equation} \label{eq:theory_RSD}
\begin{aligned}
P_\mathrm{RSD} (k,\mu) = \left(1+\left[ k\mu\sigma_P \right]^2/2 \right)^{-2} &\left[ \,P_{g,\delta\delta}(k|\mathbf{\Omega}_\mathrm{fid},\mathbf{b}) + 2 f \mu^2 P_{g,\delta\theta}(k|\mathbf{\Omega}_\mathrm{fid},\mathbf{b}) + f^2 \mu^4 P_{g,\theta\theta}(k|\mathbf{\Omega}_\mathrm{fid}) \right. \\
&\left. ~+~b_1^3 A^\mathrm{TNS}(k,\mu,f/b_1)  + b_1^4 B^\mathrm{TNS}(k,\mu,f/b_1) \,\right] ~,
\end{aligned}
\end{equation}
where the density (`$\delta\delta$'), velocity (`$\theta\theta$') and cross (`$\delta\theta$') contributions to the galaxy power spectrum are obtained by applying two-loop Re-summed Perturbation Theory (2LRPT) to the fiducial power spectrum template as described in \citep{Gil-Marin:2020bct}. The power spectrum terms also depend on a set of bias parameters $\mathbf{b} = \left\lbrace b_1, b_2, b_{s2}, b_{3\mathrm{nl}} \right\rbrace$ \cite{Beutler:2013yhm}, where we assume the non-local bias parameters to follow the local Lagrangian prediction \citep{Baldauf_2012,Saito:2014qha,kwanetal:2012} of the co-evolution model $b_{s2}=-4/7(b_1-1)$ and $b_\mathrm{3nl}=32/315(b_1-1)$. These choices are justified by the findings on N-body mocks with a HOD consistent with BOSS LRGs \cite{PTchallenge:data,ShapeFit}. We study the impact of relaxing these assumptions in Appendix~\ref{app:nonlocal_bias}. 
The functions $A^\mathrm{TNS},B^\mathrm{TNS}$ are provided by \cite{Taruya:2010mx} and the Fingers of God effect (FoG, highly nonlinear RSD along the LOS on small scales, \cite{10.1093/mnras/156.1.1P}) are modeled via the Lorentzian damping term in front of eq.~\eqref{eq:theory_RSD} with free parameter $\sigma_P$. 
The model of eq.~\eqref{eq:theory_RSD} has a limited scale-range of validity due to certain effects such as shell-crossing, extra velocity terms caused by non-linear redshift space distortions, the complexity of galaxy formation encoded by only few galaxy bias parameters, or the effective Fingers-of-God pre-factor term. Because of these, we truncate our model at some effective scale we refer as $k_{\rm max}$, which filters out all the effects present in the data, but not described by our model. We calibrate the value of $k_{\rm max}$ using accurate N-body simulations whose galaxies have been populated according to state-of-the-art techniques and validated by the official BOSS collaboration to describe the data we work with. In section \ref{sec:sys_budget} we quantify the systematic error-bars of our model with the truncation scale used on the data sample, finding these are negligible compared to the statistical errors of the BOSS/eBOSS data set when we set $k_{\rm max}=0.15\,h{\rm Mpc}^{-1}$ for the $z<1$ and $k_{\rm max}=0.30\,h{\rm Mpc}^{-1}$ for the $z>1$ dark matter tracers.

Finally, we also take into account deviations from Poisson shot noise, $P_\mathrm{Poisson}$, in the monopole by setting into eq.~\eqref{eq:theory_Pell_general} 
\begin{equation}
    g^{(\ell=0)}(\mathbf{X}_\mathrm{RSD}) = P_\mathrm{Poisson} \left[ \frac{A_\mathrm{noise}}{\alpha_\parallel\alpha_\perp^2} -1 \right]~,
\end{equation}
where the $P_\mathrm{Poisson}$ values are provided by BOSS and eBOSS and $A_\mathrm{noise}$ is a free parameter for each redshift bin. Hence, the free parameters of the RSD compression consist of 3 physical parameters $\mathbf{\Theta}_\mathrm{phys}^\mathrm{RSD} = \left\lbrace \alpha_\parallel,\alpha_\perp, f\sigma_8 \right\rbrace$ and 8 nuisance parameters $\mathbf{\Theta}_\mathrm{nuis}^\mathrm{RSD} = \left\lbrace b_1^\mathrm{N/S}, b_2^\mathrm{N/S}, \sigma_P^\mathrm{N/S}, A_\mathrm{noise}^\mathrm{N/S} \right\rbrace$ per redshift bin.

\subsubsection{ShapeFit compression} \label{sec:theory_shapefit}
The ShapeFit method is a simple, yet powerful, extension of the BAO+RSD compression, that has been developed and validated in \cite{ShapeFit}, applied to BOSS DR12 data in \cite{BriedenPRL21} and successfully verified on high-volume N-body mocks \cite{ShapeFitPT} in the context of the blind PT challenge \cite{2020PhRvD.102l3541N}. Below we briefly introduce the extra parameter of ShapeFit and the relevant cosmological interpretation.

First of all, it is important to stress that in the fixed template method what is fixed is actually not the  amplitude $\sigma_8$ at a fixed scale $R_8$, but the amplitude $\sigma_{s8} =\sigma(R_{s8})$ at the scale 
\begin{equation} \label{eq:theory_rs8}
R_{s8} \equiv s \cdot 8\,{\rm Mpc}{h}^{-1}, \qquad s=\frac{r_\mathrm{d}}{r_\mathrm{d}^\mathrm{fid}}~, 
\end{equation}
because all scales within the fixed template method can only be expressed in units of the standard ruler $r_\mathrm{d}$. Further explanation is provided in section 3.2 of \cite{ShapeFit} and eq. (3.6) therein. 

Then, in addition to the RSD compression parameters from section \ref{sec:theory_RSDcomp}, we include the shape parameter $m$ proposed in \cite{ShapeFit} (same parameterization and parameter values as eqs. (3.7) and (3.8) therein),  which aims to capture information from the no-wiggle linear matter transfer function $T_\mathrm{nw}(k)$. Indeed, the shape of the transfer function for any model is predominantly determined by its slope in the transition region between the very large scales (where $T_{\rm nw}(k)$ is constant) and small scales (where it behaves like a power law). In particular, the scale-dependent slope reaches its maximum at the pivot scale $k_p = \pi/r_\mathrm{d} \sim 0.03\,h{\rm Mpc}^{-1}$, related to the standard ruler.
The reason for scaling $k_p$ with $r_\mathrm{d}$ instead of with the equality scale $k_\mathrm{eq}$, which also determines the broadband shape of the linear power spectrum, is twofold. First, the presence of baryon perturbations induces an abrupt suppression at a scale related to $r_\mathrm{d}$. Second, within our fixed-template approach we have to choose a single scale to set our \textit{unit system}. We therefore rely on the assumption that the baryon density and hence the sound horizon are nonzero. However, this assumption could be in principle relaxed, if a different choice of \textit{unit system} is preferred. 
In practice the pivot scale $k_p$ as defined above is in a sweet spot: it is on large, linear scales where the scale dependence of the  measured galaxy power spectrum is expected to be a faithful tracer of the transfer function shape and non-linearities are unimportant, yet it is a scale that is small enough to be sampled reasonably well by state-of-the-art and upcoming surveys. 

Consequently, the measurement of $m$ can be interpreted within any model of choice as
\begin{equation} \label{eq:theory_m}
%    m(z)= a \ln\left( \frac{P_{\rm lin}(k,z)}{P_{\rm lin, fid}(k,z)} \right)/ \tanh\left[a\ln\left(\frac{k}{k_p}\right) \right]
    m = \frac{d}{dk} \left( \ln \left[ \frac{ \left( \frac{\rd^{\rm fid}}{\rd} \right)^3 \cdot T_\mathrm{nw}^2\left( \left(\frac{\rd^{\rm fid}}{\rd}\right) \cdot k\right)}{{T_\mathrm{nw}^\mathrm{fid}}^2\left(k\right)} \right] \right) \Bigg|_{k=k_p}.
\end{equation}
Note that, within the $\Lambda$CDM model, $m$ does not depend on redshift, but a suite of physical processes might in principle introduce a (real or effective) redshift dependence \footnote{Effects of systematics or physics beyond the $\Lambda$CDM can leave signatures on $m$ see \cite{BriedenPRL21}}.  Hence, when we fit the BOSS and eBOSS data, $m$ is recovered as a function of redshift.
Only in the later stage, under the interpretation of $m$ within a specific model, redshift-independence is imposed.

Finally, our baseline ShapeFit parameter set for each redshift bin contains 4
physical and cosmologically interpretable parameters $\mathbf{\Theta}_\mathrm{phys}^\mathrm{ShapeFit} = \left\lbrace \alpha_\parallel,\alpha_\perp, f\sigma_{s8}, m \right\rbrace$ and the 8 nuisance parameters $ \mathbf{\Theta}_\mathrm{nuis}^\mathrm{ShapeFit} = \left\lbrace b_1^\mathrm{N/S}, b_2^\mathrm{N/S}, \sigma_P^\mathrm{N/S}, A_\mathrm{noise}^\mathrm{N/S} \right\rbrace$ already introduced in section \ref{sec:theory_RSDcomp}. The fits are carried out in the same fashion as described therein, so $P_\mathrm{ShapeFit}(k,\mu)=P_\mathrm{RSD}(k,\mu)$ where the linear no-wiggle transfer function is modified via the shape parameter $m$.

\subsection{Reconstruction} \label{sec:theory_recon}
It is customary to use the technique of reconstruction \citep{Eis2007} to enhance the BAO peak detection within BAO fits.

The reconstructed catalogues are generated using the algorithm described by \cite{burden_efficient_2014,burden_reconstruction_2015} where the underlying dark matter density field is inferred from the actual galaxy field. This can be done efficiently only for  tracers with sufficient high-density of objects, in our case the LRG samples. During the reconstruction process, each galaxy position is displaced to the position where this galaxy would reside if there were no bulk flows. This process successfully removes most of the non-linear effects from the BAO feature and enhances the detection of the BAO peak. 

In this paper, as done in all similar SDSS analyses, we fit the reconstructed data with minimal information from the broadband clustering signal, attempting to isolate the signal of the BAO peak position along and across the LOS. This allows us to effectively constrain only $\alpha_\parallel$ and $\alpha_\perp$ from these catalogues. 

The reconstruction process effectively produces a new catalogue of galaxies which we refer to as the post-reconstructed (or post-recon) catalogue. Conversely, the original catalogue takes the name of pre-reconstructed (pre-recon) catalogue. 
We treat the pair, reconstructed and pre-reconstructed catalogs as two separate but correlated catalogs; as such, the data-vectors derived from  each can be combined using the appropriate correlation matrix, estimated from mock galaxy surveys as described in section \ref{sec:data-prepostcomb}.

 \subsection{Adopted naming convention for methodology and analysis approaches}
In the rest of this paper we adopt the following naming conventions.
We refer to BAO+RSD analyses as `classic' approach and often use these two names interchangeably. This can be seen as a data compression that extracts the BAO and RSD signature into three purely late-time physical parameters, or physical variables, per redshift bin: $\{\alpha_{\parallel}, \alpha_{\perp}, f\sigma_{s8}\}$
\footnote{The $f\sigma_{s8}$ convention was introduced by \cite{ShapeFit} only very recently, as it represents the quantity that the `classic' fixed-template approach actually measures instead of $f\sigma_8$. As ref.~\cite{ShapeFit} clearly explains, it is straightforward to convert between the two quantities at the cosmological interpretation step.}

The BAO signal is usually extracted from the reconstructed catalog. When this is not clear from the context we refer to this as BAO post-recon (as opposed to BAO pre-recon). The RSD analysis is always performed in the pre-recon catalogue.
The BAO+RSD `classic' analysis is extended by ShapeFit. We refer to this extended BAO+RSD+{\it Shape} as ShapeFit, interchangeably.
In this case the full data set is compressed into four physical parameters per redshift bin, the forth being the shape parameter $m$, which is however not purely a late-time parameter.
In the case that the shape parameter $m$ is varied during the template fit, but is \textit{not} used for the cosmological interpretation, i.e., $m$ is marginalized but only the compressed variables representing BAO+RSD information are interpreted, we refer to that as `classic' fit as well, because effectively the results are indistinguishable \cite{ShapeFit}. 
When the reconstructed catalogs are available, the full data set incorporates a stronger BAO signal by including a BAO post-recon analysis.
This represents a further improvement  to the original ShapeFit proposal and its applications to date \cite{ShapeFit,ShapeFitPT, BriedenPRL21}, where the BAO signal was extracted exclusively from the pre-recon catalog.

\section{Data}\label{sec:data}
We use the publicly available
data from the Sloan Digital Sky Survey-III \citep{adsabs:2011AJ....142...72E,boss:data} and -IV \citep{Blanton2017,eboss:data}, corresponding to the respective observation campaigns, BOSS \cite{2013AJ....145...10D} and eBOSS \cite{2016AJ....151...44D}. Both campaigns make use of two multi-object spectrographs \citep{Bolton12,Smee13} installed on the Apache Point Observatory 2.5-meter telescope located in New Mexico, USA \citep{Gunnetal2006} to carry out spectroscopic measurements from photometrically selected Luminous Red Galaxies (LRGs), Emission Line Galaxies (ELGs) and Quasar (QSO) samples, which have been used for both clustering and Ly-$\alpha$ studies. 

In this paper we focus on re-analyzing only the LRG \citep{Reid:2015gra,ebossLRG_catalogue} and quasar clustering \citep{ebossQSO_catalogue} catalogues. For simplicity we do not re-analyse the Main Galaxy Sample (MGS) \cite{2009ApJS..182..543A} and ELG samples \citep{ebossELG_catalogue}, neither the Ly-$\alpha$ forest studies \citep{dumasdesBorboux2020}, which would require an effort beyond the scope of this paper. The reason for not doing so is the complex treatment of systematics on the ELG sample, the low statistical power of the MGS, and the significantly different pipeline for analysing Ly-$\alpha$ forest data.

However, in our cosmology fits later in section \ref{sec:cosmo}, we incorporate the eBOSS DR16 Lyman-$\alpha$ BAO-only compressed variable results of \citep{dumasdesBorboux2020} obtained from their auto- (Ly$\alpha\times$Ly$\alpha$) and cross- (Ly$\alpha\times$QSO) spectra measurements at redshift $z_\mathrm{eff} = 2.33$. 
Collectively this data set probes the last 10 billion years of cosmic evolution through more than 2 million spectra.

To complement this `late-time' data-sample for the purpose of cosmological interpretation in section \ref{sec:cosmo}, we include additional `early-time' probes presented in section \ref{sec:data-early-time}.

\subsection{BOSS and eBOSS Data samples}
We analyze the power spectrum multipoles --monopole, quadrupole and hexadecapole-- of the catalogues listed in table \ref{tab:datastats}, consisting of a total number of 1,723,267 unique objects, covering a redshift range of $0.2<z<2.2$. The goal is to perform a consistent BAO, RSD and ShapeFit-type of analysis which does not make {\it a priori} assumptions about the true underlying cosmological model or family of models, yet at the same time maximizes the amount of inferred cosmological information.  
These catalogues were originally analyzed by the BOSS and eBOSS team, with a strong focus on BAO and RSD features: \citep{wedges,2012MNRAS.424..564R,Beutler:2016ixs,Beutler:2016arn,Vargas-magana18,Satpathyetal17,griebetal17} analyzed the BOSS DR12 samples, \citep{Gil-Marin:2020bct,Bautistaetal21,houetal2021,neveuxetal2020} analyzed the eBOSS DR16 samples. In addition, eBOSS also analyzed the ELG sample \citep{tamoneetal2020,demattia21,ebossELG_catalogue}, which we do not use in this paper. Finally, these measurements were consistently combined by the BOSS and eBOSS collaboration in \cite{alamdr12} and \cite{eboss_collaboration_dr16}, respectively. Additionally to the standard RSD and BAO analyses,  \cite{chuangetal16,pellejero-ibanezetal17} performed an analysis on BOSS data extracting also information on $\Omega_{\rm m} h^2$ from the shape of the power spectrum. Recently, further studies have been published focusing on analyzing BOSS and eBOSS data by fitting the full power spectrum to the prediction of a specific model (see for e.g., \cite{DAmico:2019fhj, Ivanov:2019pdj, 2020JCAP...05..032P,Trosteretal:2020,shi-fan, Semenaiteetal2021, Neveuxetal2021} just as few examples). Later in section~\ref{sec:cosmo-comparison-FM} we compare our results to their findings.

\begin{table}[htb]
    \centering
    \begin{tabular}{|c|c|c|c|c|c|}
    \hline
         Catalogue & tracer & range & patch & objects & Ref  \\
         \hline
         \hline
         BOSS DR12 & LRG & $0.2<z<0.5$ & north & 429,182 & \citep{Reid:2015gra} \\ 
         BOSS DR12 & LRG & $0.2<z<0.5$ & south & 174,819 & \citep{Reid:2015gra} \\ 
         \hline
         BOSS DR12 & LRG & $0.4<z<0.6$ & north & 500,872 & \citep{Reid:2015gra} \\ 
         BOSS DR12 & LRG & $0.4<z<0.6$ & south & 185,498 & \citep{Reid:2015gra} \\ 
         \hline
         BOSS DR12 + eBOSS DR16 & LRG & $0.6<z<1.0$ & north & 255,741 & \citep{ebossLRG_catalogue} \\ 
         BOSS DR12 + eBOSS DR16 & LRG & $0.6<z<1.0$ & south & 121,717 & \citep{ebossLRG_catalogue} \\ 
         \hline
         eBOSS DR16 & QSO & $0.8<z<2.2$ & north & 218,209 & \citep{ebossQSO_catalogue} \\ 
         eBOSS DR16 & QSO & $0.8<z<2.2$ & south & 125,499 & \citep{ebossQSO_catalogue} \\ 
         \hline
    \end{tabular}
    \caption{List of SDSS-III and -IV catalogues used in this paper and their number of objects. The total number of unique objects is 1,723,267, and the total effective volume is $2.82\,[{\rm Gpc}{h}^{-1}]^3$.}% or $9.13\,{\rm Gpc}^3$  }
    \label{tab:datastats}
\end{table}

Throughout this paper we always assume that the northern and southern hemispheres are statistically independent, as it is the common practice. In the same fashion, we consider that the different redshift bins are independent, unless they are overlapping. This is the case for the BOSS DR12 redshift bins at $0.2<z<0.5$ and $0.4<z<0.6$, for which the covariance is inferred from a suite of mock galaxy surveys, as described in section~\ref{sec:mocks}.
On the other hand  the eBOSS DR16 LRG and quasar sample do overlap in the redshift range $0.8<z<1.0$, but  their covariance can be neglected because of the low density of objects in this range (especially for quasars)
as motivated in section 3.1 of \cite{eboss_collaboration_dr16}.

\begin{figure}[tb]
    \centering
    \includegraphics[scale=0.25]{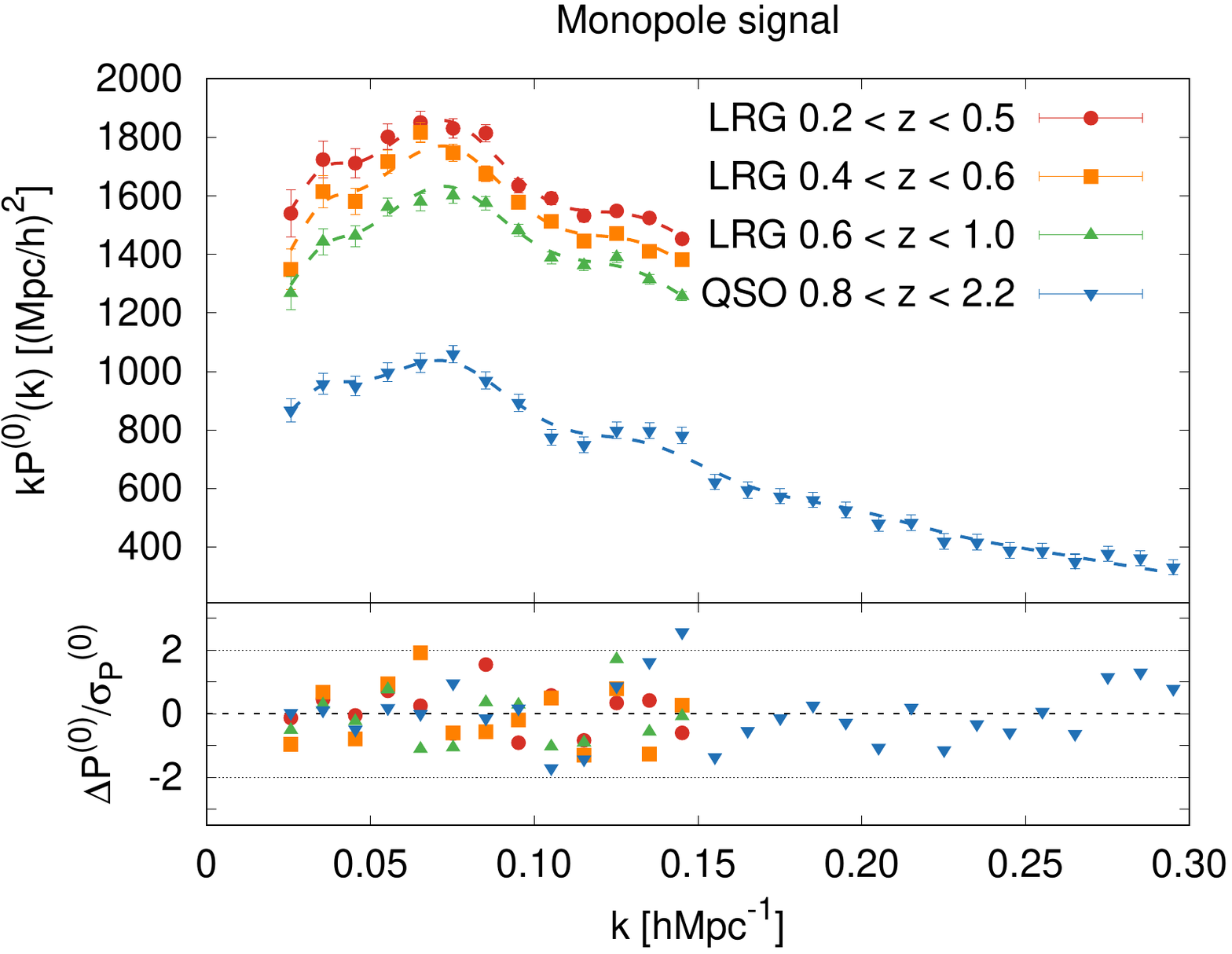}
    \includegraphics[scale=0.25]{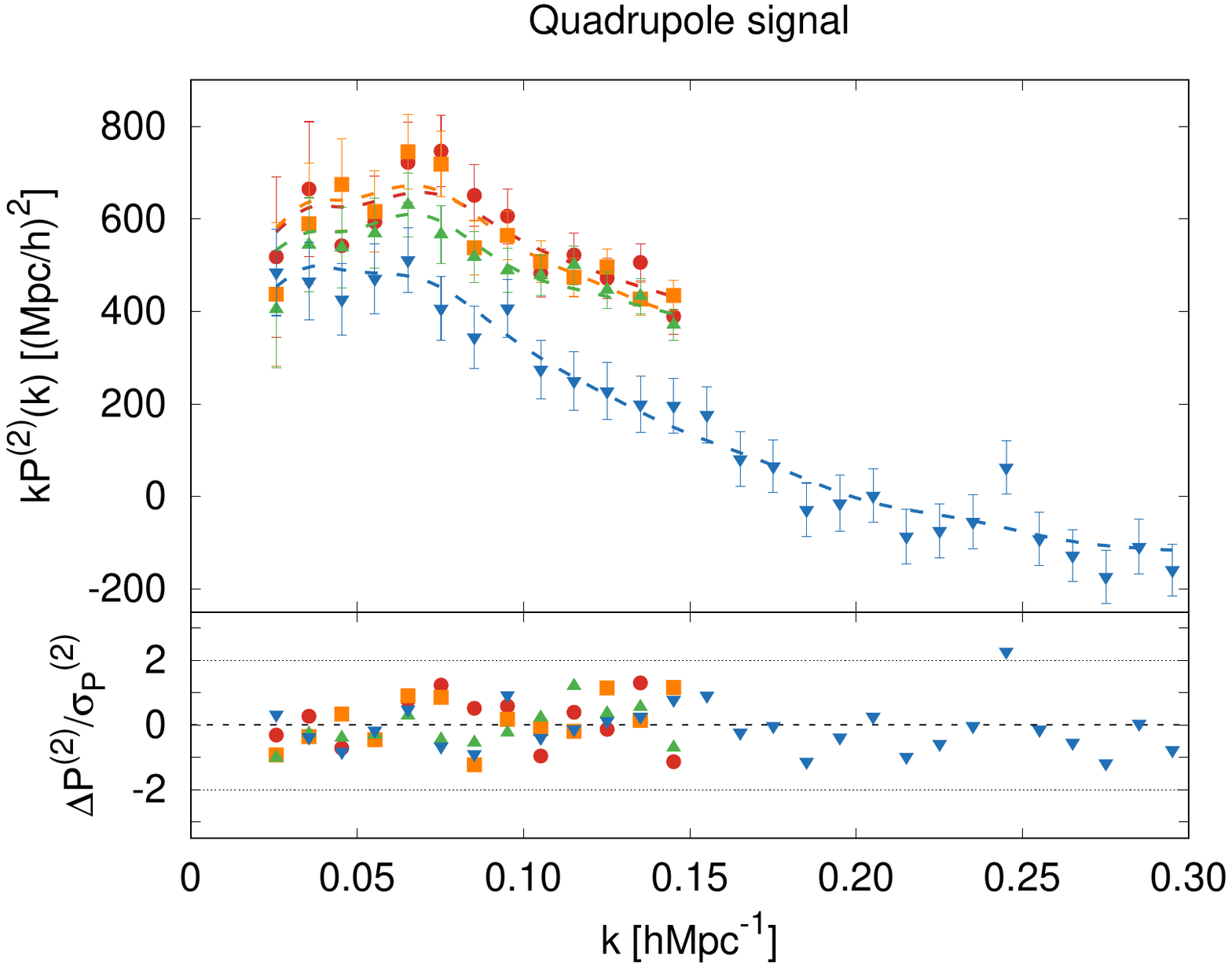}
    \includegraphics[scale=0.25]{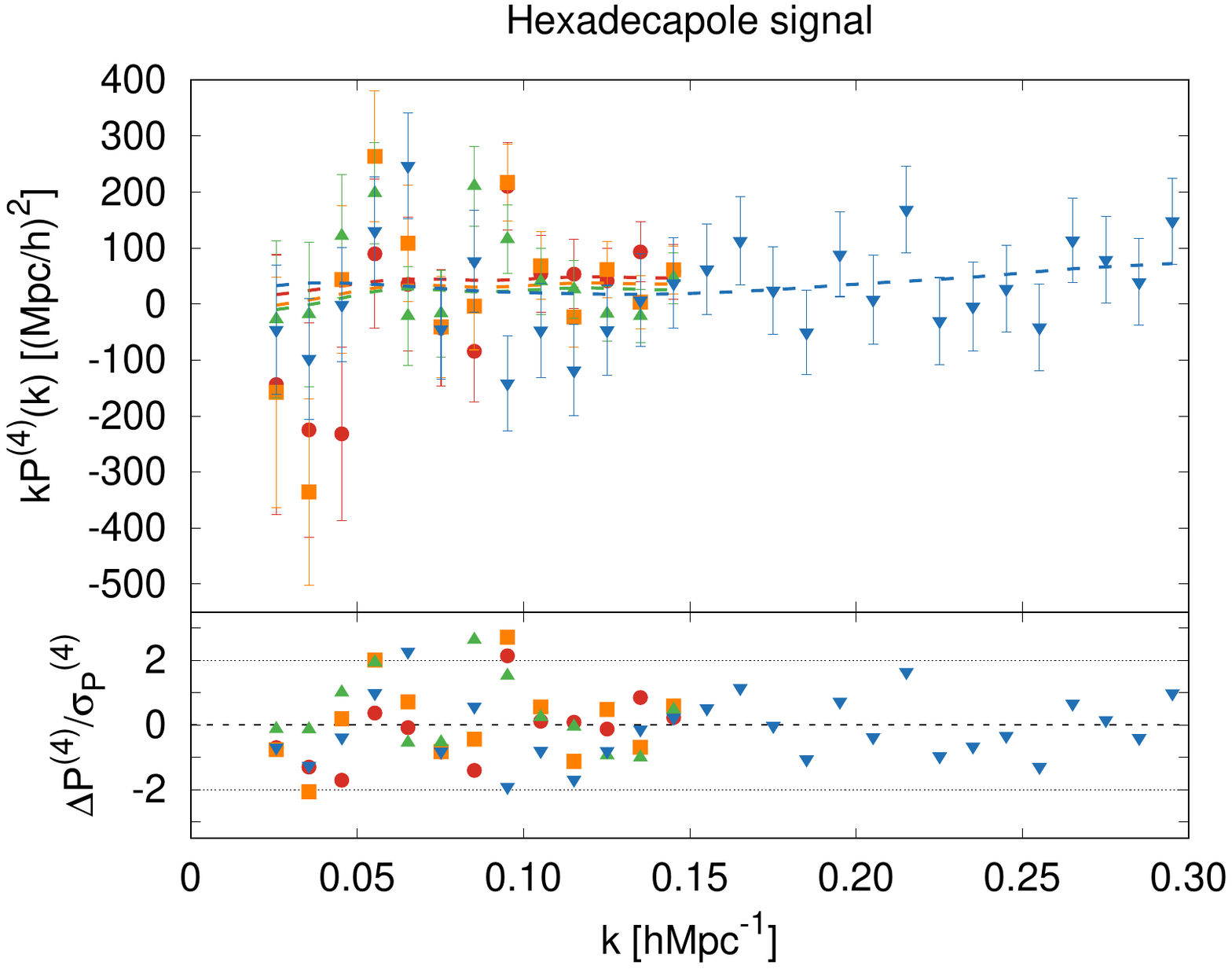}
    \includegraphics[scale=0.25]{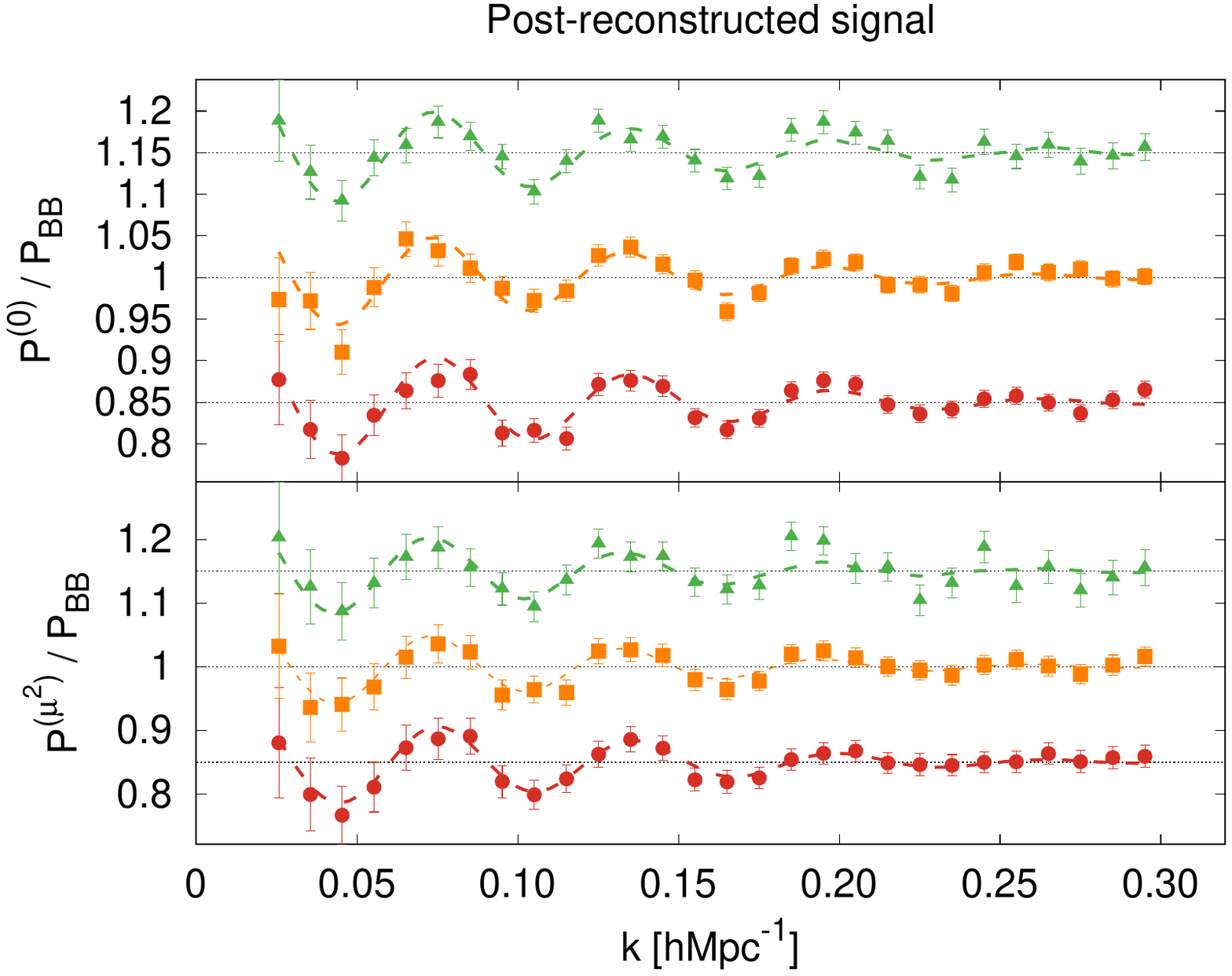}
    \caption{Measurements (points with error-bars) and best-fitting model (dashed lines with colors matching the data points). As labeled, the panels show the pre-reconstructed monopole (top-left), quadrupole (top-right) and hexadecapole (bottom-left) for the quasar sample and the three LRGs redshift bins. In each panel the bottom inset shows the residuals normalized by the {\it rms}. The bottom right panel shows the post-recon isotropic and anisotropic signals for the three LRGs redshift bins normalized by a smooth spectrum (broadband, BB) to highlight the BAO feature. Low (and high) LRG redshift bins are displaced vertically by $-(+)0.15$ for visibility.}
    \label{fig:dataP0}
\end{figure}

The power spectra multipole measurements for the pre-recon LRGs and QSOs are displayed in  figure~~\ref{fig:dataP0} along with the BAO post-recon signal in the three LRG redshift bins as points with error-bars. Colored dashed lines are the best-fits of the model (see section~\ref{sec:results}). In the pre-recon panels the lower insets show residuals with respect to this model, whereas for the post-recon panel the two insets show the BAO feature in the monopole (isotropic BAO, upper inset) and in the $\mu^2$-moment (anisotropic BAO, bottom inset), defined as, $P^{(\mu^2)}\equiv P^{(0)}+2/5P^{(2)}$.  Black dotted lines in the bottom right panel display the best-fit `mean' level for the broadband (no-wiggle) power. Note that the three LRG samples have been displaced vertically for visibility.

\subsection{Galaxy mocks}\label{sec:mocks}
Galaxy survey mocks are crucial to estimate the covariance matrices for the adopted data vectors. 
We employ a suite of galaxy mocks, matching the clustering properties and the sky-geometry  of the data samples presented in table~\ref{tab:datastats}.  These consist of $2\times2048$ realizations  of the Multi-Dark Patchy mocks \citep{kitaura_clustering_2016} for the northern and southern patches (hereafter Patchy mocks), for the two BOSS DR12 LRG samples. Additionally we consider $2\times1000$ realizations of the EZmocks \citep{Ezmocks} for the BOSS DR12 + eBOSS DR16 LRG sample, and for the eBOSS DR16 quasar sample, also for northern and southern patches.
The EZmocks  are generated from 5 different snapshots of large cubic periodic simulations based on the Zeldovich approximation \cite{Chuang15}, while the Patchy algorithm is based on 4 different snapshots of Augmented Lagrangian Perturbation Theory \cite{Kitaura13} and a bias scheme (hence the Patchy and EZmocks are often refereed to as fast mocks). Fast mocks are well suited for evaluating covariance matrices but their adoption to test or calibrate the accuracy of the adopted modeling of the signal require some care. 

\begin{table}[htb]
    \centering
    \begin{tabular}{|c|c|c|c|c|c|c|c|}
    \hline
         Cosmology & $\Omega_{\rm m}$ & $\Omega_{\rm b}$ & $h$ & $10\times\Omega_\nu$ & $n_s$ & $A_s\times10^9$ & $r_{\rm d}\,[{\rm Mpc}]$ \\
         \hline
         \hline
         Fiducial & $0.310$ & $0.0481$ & $0.676$ & $1.400$ & $0.97$ & $2.040$ & $147.78$ \\
         EZ & $0.307115$ & $0.048206$ & $0.6777$ & $0$ & $0.9611$ & $2.1151$ & $147.66$ \\
         Patchy & $0.307115$ & $0.048206$ & $0.6777$ & $0$ & $0.9611$ & $2.1476$ & $147.66$ \\
         \hline
    \end{tabular}
    \caption{List of cosmology models used in this paper: the reference or fiducial cosmology used for the fixed template (for convenience this is the same as the one used for the BOSS and eBOSS analyses in \cite{alam_clustering_2017,eboss_collaboration_dr16}) and the true underlying cosmology of the two set of mocks used in this paper.}
    \label{tab:cosmo}
\end{table}

In total we have  12,192 mock realizations of the pre-reconstructed catalogues. Additionally, we run the reconstruction algorithm introduced in the previous section on the LRG samples, obtaining an additional set of 10,192 realizations of the post-reconstructed mocks. Power spectrum multipoles are computed for each of these 22,384 mock realizations to extract a reliable power spectrum covariance, $C_{k,k'}^{\ell,\ell'}$, which allows us to individually fit each redshift-sample of both data catalogues and mock catalogues. The exact setup of these fits is described in section \ref{sec:data-prepostcomb}. 

The true underlying cosmology of these mocks and the fiducial cosmology used to analyse them can be found in table~\ref{tab:cosmo}. Additionally, in table \ref{tab:expected} we list the mocks's true underlying distance ratios, $D_H/r_{\rm d}$, $D_M/r_{\rm d}$, the expected values for the scaling parameters $\alpha_{\parallel,\,\perp}$, the true growth of structure parameter $f\sigma_{s8}$, and the expected shape parameter $m$ when these mocks are analyzed using the fiducial cosmology. Later in section \ref{sec:sys} we will show how the actual analyses on the mocks perform and how close they are to their expected values. 

\begin{table}[htb]
    \centering
    \begin{tabular}{|c|c|c|c|c|c|c|}
    \hline
         Redshift & $D_H/r_{\rm d}$ & $[\alpha_\parallel^{\rm exp}-1]\times10^2$ & $D_M/r_{\rm d}$ & $[\alpha_\perp^{\rm exp}-1]\times 10^2$ & $f\sigma_{s8}$ & $m^{\rm exp}\times10^2$  \\
         \hline
         \hline
          $0.38$ & $24.46$ & $-0.01$ & $10.35$ & $-0.09$ & $0.4736$ & $-1.15$ \\
         $0.51$ & $22.64$ & $0.03$ & $13.41$ & $-0.07$ & $0.4806$ & $-1.15$ \\ 
         $0.698$ & $20.21$ & $0.09$ & $17.43$ & $-0.04$ & $0.4659$ & $-1.15$\\
         $1.48$ & $12.92$ & $0.21$ & $30.11$ & $0.04$ & $0.3828$ & $-1.15$ \\
         \hline
    \end{tabular}
    \caption{True distances, expected dilations ($\alpha's$), growth rate and shape ($m$) when Patchy and EZ mocks are analyzed using the fiducial cosmology of table~\ref{tab:cosmo} as reference cosmology. Redshifts 0.38 and 0.51 correspond to the Patchy mocks true cosmology, whereas redshifts 0.698 and 1.48 to the EZmocks true cosmology. Note that the variables $D_H/r_{\rm d}$, $D_M/r_{\rm d}$ and $f\sigma_{s8}$ are not relative to the choice of the fixed-template used, and therefore do not have the index `exp'.}
    \label{tab:expected}
\end{table}

\subsection{Pre- and post-recon catalogue combination} \label{sec:data-prepostcomb}
The main results of this work (which are presented in section \ref{sec:results}) are the constraints on the compressed physical variables $\mathbf{\Theta}_\mathrm{phys}^\mathrm{combined} =  \{D_H/r_{\rm d}(z),\, D_M/r_{\rm d}(z),\,f\sigma_{s8}(z),\,m(z)\}^{\rm combined}$ obtained by consistently combining post-recon BAO and pre-recon ShapeFit results 
\begin{equation}
\begin{aligned}
&\mathbf{\Theta}_\mathrm{phys}^\mathrm{combined} = \mathbf{\Theta}_\mathrm{phys}^\mathrm{pre-recon} \join \mathbf{\Theta}_\mathrm{phys}^\mathrm{post-recon} = \\ 
 & \{D_H/r_{\rm d}(z),\, D_M/r_{\rm d}(z),\,f\sigma_{s8}(z),m(z)\}^{\rm pre-recon} \join \{D_H/r_{\rm d}(z),\, D_M/r_{\rm d}(z)\}^{\rm post-recon}.
\end{aligned}
\end{equation}

For a combined post-recon BAO + pre-recon ShapeFit analysis it is crucial to correctly incorporate the covariance between the compressed variables from both types of fits. Especially the scaling parameters are expected to show a strong correlation for each pair of pre-recon and post-recon catalog. It is highly non-trivial to model this correlation analytically due to {\it i)} the non-linear nature of the reconstruction scheme and {\it ii)} the evident differences in the BAO and ShapeFit underlying models (especially the no-wiggle power spectrum decomposition). 

One approach would be to infer the covariance matrix of the full combined data-vector $\langle P_\ell^{\rm pre-recon}(k,z)P_{\ell'}^{\rm post-recon}(k',z')\rangle$ and perform a simultaneous pre- and post-reconstruction fit directly from the multipoles. On the other hand, here we follow the approach taken by the BOSS and eBOSS team of combining the pre- and post-reconstruction results at the level of compressed variables: $\mathbf{\Theta}_\mathrm{phys}^\mathrm{pre-recon} \join \mathbf{\Theta}_\mathrm{phys}^\mathrm{post-recon}$, which has the advantage of dealing with a smaller covariance matrix than the previous approach: $12\times12$ elements for the BOSS LRG sample, and $6\times6$ for the eBOSS LRG sample compared to $380\times 380$ and $190 \times 190$ power spectrum elements, respectively. Recently, \cite{gil-marin22} showed how combining the pre- and post-recon information at the compressed variable stage only degrades the statistical precision on $5-10\%$ with respect to the simultaneous pre- and post-recon fits. 
 
From the pre-recon catalogues, we extract a set of compressed elements, $\mathbf{\Theta}_\mathrm{phys}^{\rm pre-rec}(z)$ for each of the 12,192 pre-recon mock realizations; and from the post-recon catalogues a compressed set of elements $\mathbf{\Theta}_\mathrm{phys}^{\rm post-rec}(z)$ from the 10,192 realizations. 

Using this information we are able to extract the block off-diagonal elements among different overlapping samples (for BOSS DR12 $0.2<z<0.5$ and $0.4<z<0.6$ samples), and among the pre- and post-reconstructed catalogues (for BOSS and eBOSS LRG samples). 

All pre-recon ShapeFit fits to the mocks and the data are carried out with the parameter and prior settings stated in table \ref{tab:results-priors}. For the post-recon BAO fits we use the same priors on the scaling parameters and uninformative uniform priors on the nuisance parameters. For all fits we follow as close as possible the configuration chosen within the official eBOSS BAO and RSD analyses. In particular, we choose $k_\mathrm{max}=0.15\,h\mathrm{Mpc}^{-1}$ for ShapeFit analyses of the LRG samples and $k_\mathrm{max}=0.30\,h\mathrm{Mpc}^{-1}$ for the ShapeFit analysis of the QSO sample and the BAO analyses of LRGs. In all cases, we apply a maximum scale cut at $k_\mathrm{min}=0.02\,h\mathrm{Mpc}^{-1}$ since larger scales are prone to observational systematics.  

\begin{table}[htb]
    \begin{minipage}{0.5\textwidth}
    \raggedleft
    \begin{tabular}{|c|c|}
      \hline
       Parameter (phys.) & Prior \\
       \hline
      $\alpha_\parallel$  & $[0.5, 1.5]$ \\
      $\alpha_\perp$ & $[0.5, 1.5]$ \\
      $f$ & $[0, 3]$ \\
      $m$ & $[-3, 3]$ \\
      \hline
    \end{tabular}
    \end{minipage}
    \begin{minipage}{0.5\textwidth}
    \raggedright
    \begin{tabular}{|c|c|}
    \hline
       Parameter (nuis.) & Prior \\
       \hline
      $b_1$ & $[0, 20]$ \\
      $b_2$ & $[\text{-} 20, 20]$ or $(5\pm2.5)$ \\
      %$b_{s2}$ & lag. \\
      %$b_{3\mathrm{nl}}$ & lag. \\
      $\sigma_P \, \Mpcoverh$ & $[0, 10]$ \\
      $A_\mathrm{noise}$ & $(1 \pm 0.3)$ \\
      \hline
    \end{tabular}
    \end{minipage}
    \caption{Baseline prior ranges for the physical (left) and nuisance (right) parameters used for \textit{ShapeFit}. Uniform priors are denoted as $[\mathrm{min},\mathrm{max}]$, Gaussian priors as $(\mathrm{mean} \pm \mathrm{std})$. The non-local bias parameters $b_{s2}$ and $b_{\rm 3nl}$ are assumed to follow the local-Lagrangian prediction as function of $b_1$ as described in section \ref{sec:shapefit}.}
    \label{tab:results-priors}
\end{table}
 
\subsection{Ancillary and external `early time' data} \label{sec:data-early-time}
Beside the large-scale structure datasets presented above we include the following complementary data, but only at the stage of interpreting the data in the light of cosmological models in section \ref{sec:cosmo}.

\begin{itemize}
    \item \textbf{BBN:} By measuring the light elements' abundances of distant absorption systems -- which serve as proxies for `primordial' times and early-time physics--  it is possible to infer the physical baryon energy density fraction $\omega_\mathrm{b}$ relying on our knowledge on nuclear reaction cross sections from solar observations, and our ability to correctly model the nuclear processes of Big Bang Nucleosynthesis (BBN) occurring only one second after the initial singularity. In this work we adopt the value $\omega_\mathrm{b} = 0.02235\pm0.00037$ from \cite{eboss_collaboration_dr16} (see also \cite{Cuceu20}) motivated by measurements of the relative deuterium to hydrogen abundance from \citep{Cooke:2018} and solar fusion cross sections derived by \citep{Adelberger:2010qa}.
    \item \textbf{Planck:} With its 2018 legacy data release \citep{Aghanim:2018eyx} the Planck satellite mission provided the most detailed temperature and polarization maps of the cosmic microwave background (CMB) radiation ever observed. This relic radiation with mean temperature $T_\mathrm{CMB}=2.7255K$ \citep{Fixsen:2009} was emitted when nuclei and electrons recombined $\approx$ 380,000 years after the Big Bang, at redshift $z_\mathrm{rec}=1090$.  We make use of the latest Planck data including the temperature and polarization auto and cross power spectra (TT, TE, EE, and lowE), as well as the Planck lensing measurements.  CMB lensing measurements are certainly  not early time, but  this  probe is used  only in section \ref{sec:cosmo} where model-dependence is re-introduced, so early-late separation is less important. 
\end{itemize}

\section{Model-Independent Results}\label{sec:results}

Here we present the main results  of this work, obtained from the ShapeFit analysis outlined in section \ref{sec:data-prepostcomb}. 
Our results are found in \ref{sec:results_all} and their comparison
to the official eBOSS results is in section \ref{sec:results_comparison}. 

\begin{figure}[t]
    \centering
    \includegraphics[scale=0.75]{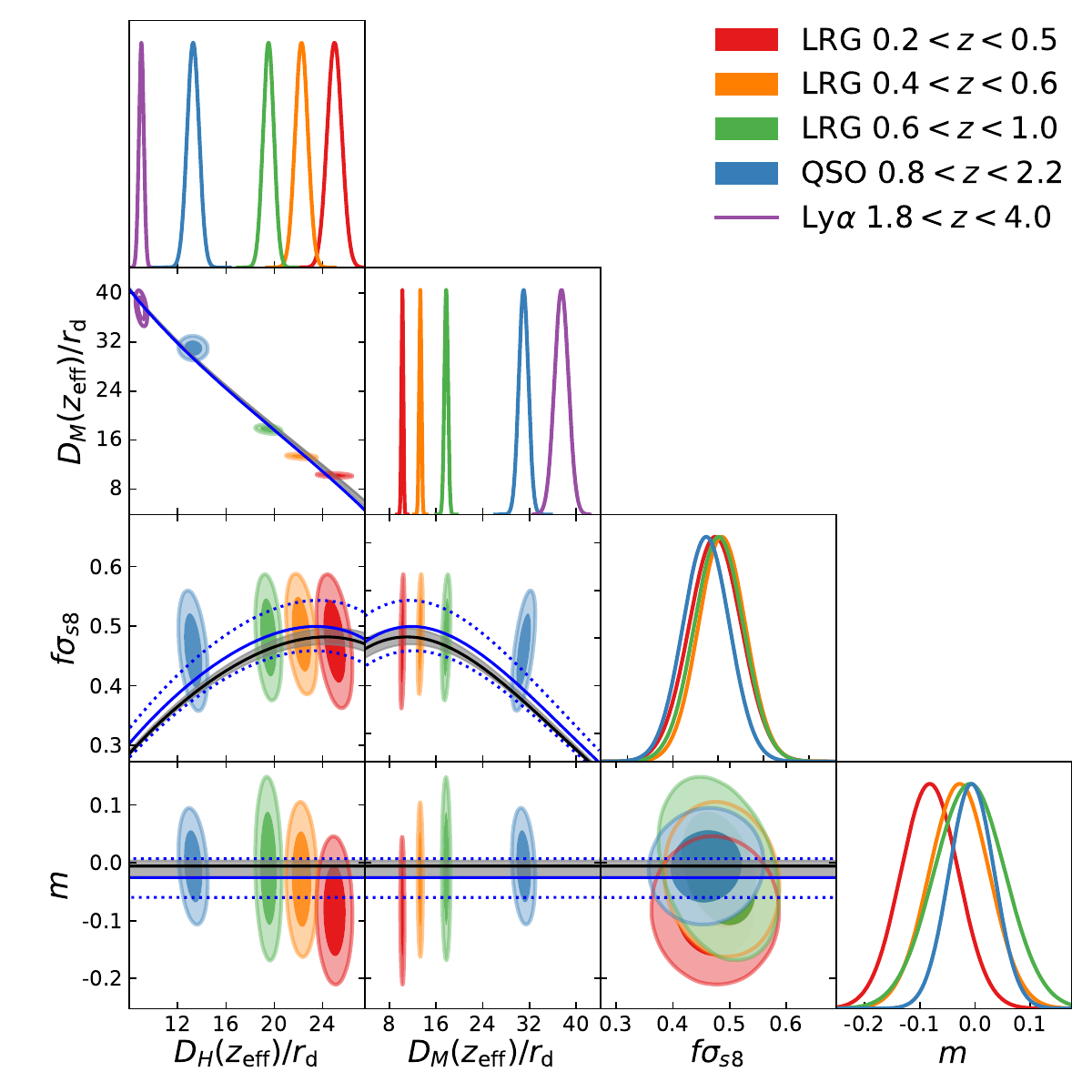}
    \caption{Main result of this paper: summary of the constraints on the compressed variables across all samples and redshifts. Filled, colored contours show the ShapeFit results. In the $D_H/r_\mathrm{d}-D_M/r_\mathrm{d}$ plane we also show the BAO-only result (empty, purple contour) obtained combining the Lyman-$\alpha$ auto- (Ly$\alpha\times$Ly$\alpha$) and cross- (Ly$\alpha\times$QSO) spectra measured at $z_\mathrm{eff}=2.33$ by \citep{dumasdesBorboux2020}. These constraints do not assume a $\Lambda$CDM model. 
    For comparison, we highlight the prediction for the Planck $\Lambda$CDM best-fit and allowed 2$\sigma$-range (black line, gray bands), and analogously the $\Lambda$CDM best-fit to the shown dataset including the corresponding 2$\sigma$-range (blue line, blue dotted lines),
    }
    \label{fig:results}
\end{figure}

\subsection{BAO, RSD and Shape evolution over 10 billion years of cosmic history} \label{sec:results_all}
ShapeFit results on the compressed parameters $\left\lbrace D_H/r_{\rm d}(z),\, D_M/r_{\rm d}(z),\,f\sigma_{s8}(z),\,m(z) \right\rbrace$ are presented jointly for all analysed redshift bins as filled colored contours in figure \ref{fig:results}. In addition, we show the BAO-only Lyman-$\alpha$ result of \citep{dumasdesBorboux2020} (purple empty contour), which is included in our baseline dataset for cosmological interpretation in section \ref{sec:cosmo}.    
The strength of the presented compressed variables constraints relies on their model-independence. As inherent to the fixed template fits described in section \ref{sec:theory_Pell}, throughout the fitting process no model assumptions or `internal model priors' based on $\Lambda$CDM or any extensions to it are applied. And that is what makes this compressed dataset such a unique and powerful probe of the underlying nature of the Universe. Let us briefly specify the importance of model-independence for the different pieces of information represented by these compressed variables.

\begin{itemize}
    \item \textbf{Geometry:} the geometrical information is traced via the AP anisotropy in units of the BAO scale, parameterized here by $D_H/r_{\rm d}(z)$ and $D_M/r_{\rm d}(z)$. Within any model, the parallel, $D_H(z)$, and perpendicular, $D_M(z)$, distances with respect to the LOS are directly linked to each other (see definition below eq.~\eqref{eq:thery_alpha_scaling}).
    By allowing the parameters $D_H/r_{\rm d}(z)$ and $D_M/r_{\rm d}(z)$ to vary freely, without any imposed correlation, we are able to cross-check whether our fundamental assumptions (FLRW metric, homogeneous and isotropic expansion, etc.) hold.
    \item \textbf{Growth:} the information on the history of structure growth is traced via RSD, parameterized here by the rescaled velocity fluctuation amplitude $f\sigma_{s8}(z)$. Within Einstein's theory of General Relativity, the redshift evolution of this quantity is directly linked to the matter density $\Omega_\mathrm{m}$, which also determines the geometry of the universe. By allowing $f\sigma_{s8}(z)$ to vary freely, on one hand we decouple these model-interdependencies between geometry and growth, and on the other hand are able to verify the validity of Einstein's theory in the first place. Note that the RSD compression provides a unique dataset on (still not sufficiently explored) large scales, that may give rise to the detection (or ruling out) of certain modified gravity models.
    \item \textbf{Shape:} the {\it Shape} information, parameterized by $m(z)$,
    % that we parameterize via $m(z)$ 
    incorporates a number of physical effects already described before (see section \ref{sec:theory_shapefit}). Most of these effects are of primordial, `early-time', origin, and are not expected to leave an imprint on the {\it Shape} that varies with redshift, so $m(z)=m$. However, by constraining this parameter independently for each redshift bin, we may be able to find hints for models that have a redshift dependent impact on the {\it Shape}, for example due to a primordial non-Gaussianity signal $f_\mathrm{NL}$, or use it as a flag for possibly unaccounted observational systematics (see \cite{BriedenPRL21} for more details).
\end{itemize}

It is important to note that geometry, growth and shape do not have the same degree of robustness when it comes to being affected by spurious signals (both coming from observations or theory modeling). In particular, the shape parameter is significantly affected by the assumptions  made when accounting for the imaging weights (see for e.g., fig. 3 of \cite{BriedenPRL21}). It is also sensitive to assumptions such as locality of the bias (see Appendix \ref{app:nonlocal_bias}.) On the other hand it is well known that the BAO feature as extracted by the template-based approach is very robust to these and other systematics.

Thanks to the compression technique used here we are able to disentangle different signatures of different physical origin, connecting them to an area of the spectral analysis (low-$k$, high-$k$, isotropic-anisotropic signal), and a specific compressed parameter. This allows us to study them individually and set an extra layer of diagnosis test (and perform a more robust analysis), which is not accessible in direct model fits, as the full modeling observables all contain mixed information across the analyzed spectral range.

We also note that ShapeFit by construction does not include features beyond the three quoted signatures. In principle the power spectrum does contain information also in the amplitude of the BAO wiggles, as well as in the shape of high-$k$ scales. However, this type of information is even less robust than the large-scale shape we include in this paper, and for this reason we have decided to not include it. However, we note that the ShapeFit formalism can be  extended to capture those signals if they become sufficiently robust in the future. This goes beyond the scope of the present paper.

Having said that, we begin by comparing these model-independent constraints to the standard cosmological model, the flat $\Lambda$CDM model. In figure \ref{fig:results} we show the $\Lambda$CDM best-fit to our BOSS+eBOSS dataset as blue solid line, with the allowed 2$\sigma$ region indicated via the blue dotted lines. We show the same (black line, grey bands) when considering Planck data only. See section \ref{sec:cosmo} for the exact setup of our cosmology fits. 

We can appreciate that the compressed constraints are in excellent agreement with the independent Planck-only $\Lambda$CDM best-fit. 
In particular, the model-independent BAO and AP constraints in the $D_H/r_\mathrm{d}-D_M/r_\mathrm{d}$ plane follow exactly the model prediction, which only allows a very tight relation. Therefore, there is no hint from this test of geometry that we would need to abandon our fundamental assumptions on the homogeneous and isotropic FLRW metric. The same holds for growth, for which the low redshift probes are in excellent agreement with the Planck $\Lambda$CDM prediction. However, as already noted by the eBOSS collaboration \cite{eboss_collaboration_dr16}, we observe a small excess of clustering of $1.5\sigma$ for the QSO sample at $z=1.48$. Although this is a rather mild anomaly (if any), we investigate the consistency between the LRG and QSO samples further in section \ref{sec:cosmo_highzvslowz}.
While our {\it Shape} measurements are all consistent with Planck's $\Lambda$CDM prediction within 1$\sigma$, we note a subtle tendency of decreasing $m$ with decreasing redshift.

In summary, the model-independent analysis of BOSS and eBOSS galaxies and Lyman-$\alpha$ delivers a unique cross-check of our fundamental assumptions and provides further powerful confirmation of 
the standard, flat $\Lambda$CDM model.

\begin{figure}[htb]
    \centering
    \includegraphics[trim=0 180 0 0, clip=true,scale=0.5]{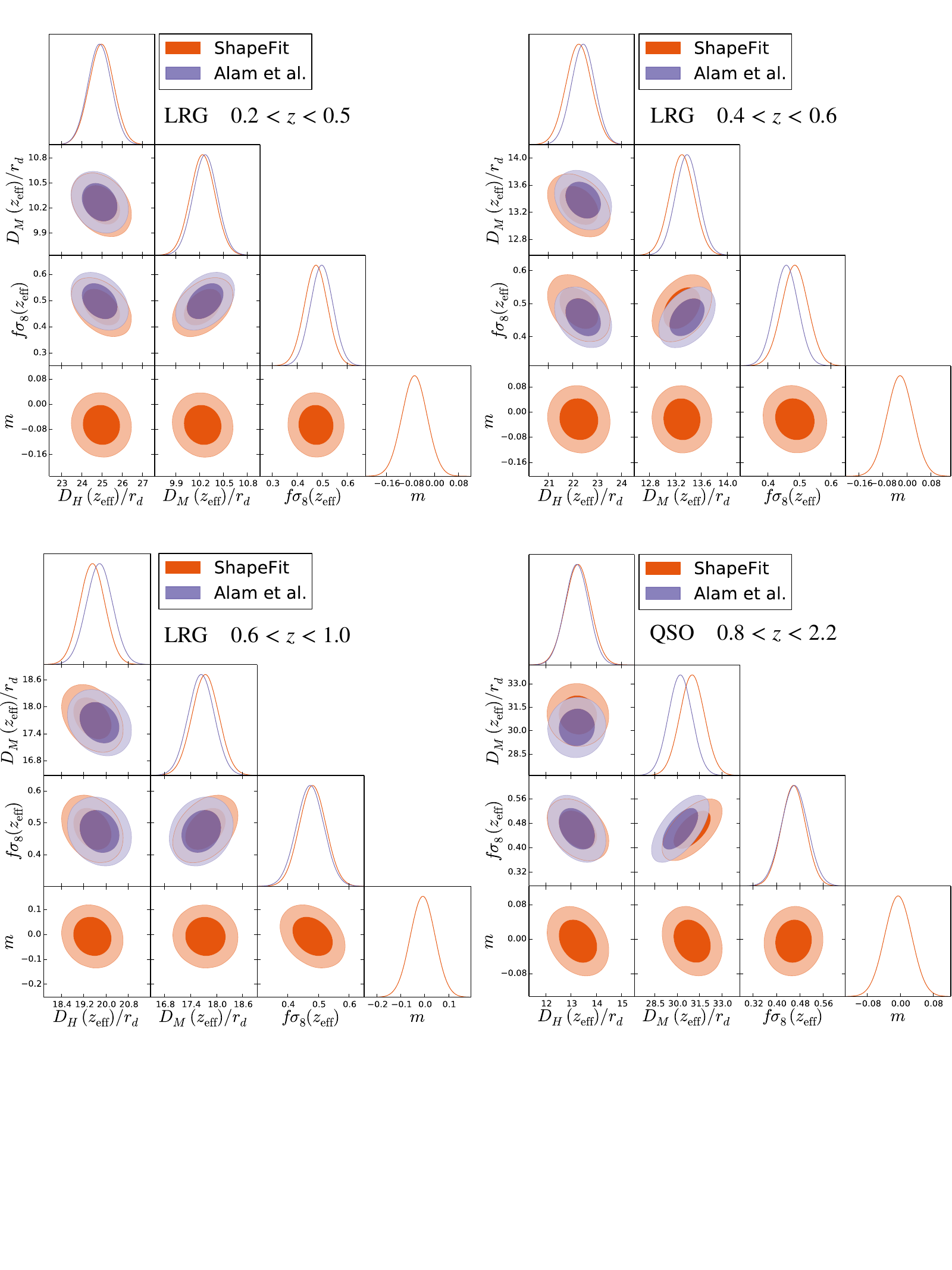}

    \caption{Comparison between the compressed variables inferred in this paper using ShapeFit (orange contours) and those from the official BOSS and eBOSS papers \cite{alam_clustering_2017,eboss_collaboration_dr16} (purple contours) for all individual samples. ShapeFit constraints come from only using the Fourier space signal (the power spectrum), whereas the Alam et al. contours display the consensus between Fourier and configuration space. For the LRG panels both approaches (orange and purple contours) display the joint analysis of the full shape pre-recon and BAO post-recon signals. The numerical results are provided in table~\ref{tab:boss_compare}.}
    \label{fig:boss_compare}
\end{figure}

\begin{table}[htb]
    \centering
    \begin{tabular}{|c|c|c|c|c|c|}
    \hline
    Sample ($z_{\rm eff}$) & method. & $D_H/r_{\rm d}$ & $D_M/r_{\rm d}$ & $f\sigma_8$ & $m$  \\
    \hline
    \hline
       LRG ($0.38$) & Alam et al. & $24.89\pm0.58$ & $10.27\pm 0.15$ & $0.497\pm0.045$ & $-$ \\ 
       LRG ($0.38$) & {\it ShapeFit} & $24.98 \pm 0.61$ & $10.24 \pm 0.16$ & $0.462 \pm 0.045$ & $-0.066\pm 0.042$ \\ %lrg geocorr
       \hline
       LRG ($0.51$) & Alam et al. & $22.43\pm0.48$ & $13.38\pm0.18$ & $0.459\pm0.038 $ & $-$ \\ 
       LRG ($0.51$) & {\it ShapeFit} & $22.26\pm 0.53$ & $13.30\pm 0.19$ & $0.482 \pm0.041$ & $-0.023\pm 0.044$ \\ % LRG geocorr
       \hline
       LRG ($0.698$) & Alam et al. & $19.77\pm0.47$ & $17.65\pm0.30$ & $0.473\pm0.044$ & $-$ \\ 
       LRG ($0.698$) & {\it ShapeFit} & $19.54\pm 0.45$ & $17.70\pm 0.31$ & $0.478\pm 0.043$ & $-0.008\pm 0.052$ \\ % lrg geocorrected
       \hline
       QSO ($1.48$) & Alam et al. & $13.23\pm0.47$ & $30.21\pm0.79$ & $0.462\pm0.045$ & $-$ \\ 
       QSO ($1.48$) & {\it ShapeFit} & $13.27 \pm 0.50$ & $31.01\pm 0.82$ & $0.458\pm 0.041$ & $-0.005\pm 0.033$ \\ %% qso geocorrected
       \hline
    \end{tabular}
    \caption{Comparison of  constraints on  compression parameters quoted in Alam et al.\cite{eboss_collaboration_dr16} using the classic method, and the compression proposed by ShapeFit. For Alam et al. the results include both power spectrum and correlation function, whereas for ShapeFit only the power spectrum signal is used. In both cases the pre- and post-recon signals have been combined if available. Figure~\ref{fig:boss_compare} displays the triangle plots for the same samples. Note that the LRG samples at $z_{\rm eff}=$ 0.38 and 0.51 are correlated. The full covariance matrices for all these samples can be found in Appendix \ref{app:datavectors}.}
    \label{tab:boss_compare}
\end{table}

\subsection{Comparison with official BOSS and eBOSS results} \label{sec:results_comparison}

As a next step we compare our ShapeFit constraints with the official BOSS and eBOSS results from \cite{alam_clustering_2017,eboss_collaboration_dr16}. 
We compare to their consensus results obtained from Fourier and configuration space for pre-recon and post-recon catalogs where available, while our combined pre- and post-recon constraints are obtained from Fourier Space only. 

The compressed variables constraints are shown in figure \ref{fig:boss_compare} and table \ref{tab:boss_compare} for each individual sample for ShapeFit (orange contours) and for the classic approach used by BOSS and eBOSS (purple contours). We find that both approaches are in excellent agreement only showing small deviations of order $0.2\sigma$ and at most $1\sigma$ in $D_M/r_{\rm d}$ for the quasars. We see that the shape parameter is nearly uncorrelated with the other parameters, therefore we expect its effect on their error bars to be negligible. The ShapeFit error bars are in very good agreement with the official reported errors, although these are constructed in different way as described above: for the official BOSS and eBOSS results there is an information gain coming from the correlation function signal, although this might be partially compensated by the inclusion of a systematic error contribution at the level of the compressed variables, which we do not consider here.\footnote{We explore the potential systematic error contribution of ShapeFit in section~\ref{sec:sys}.}
However, note that most of the systematic error budget arises from modeling the two-point statistics and the choice of fiducial cosmology (see for e.g., fig. 14 of \cite{smithetal20}), which in ShapeFit is already accounted for via the shape parameter $m$. 

We conclude that our ShapeFit constraints on $\left\lbrace D_H/r_{\rm d}, D_M/r_{\rm d}, f\sigma_8 \right\rbrace$ are consistent with the official results and can be safely used for cosmological parameter estimation. In this work we are interested in using this set of parameters together with the shape $m$ for cosmological interpretation, see section \ref{sec:cosmo}. For further details on the ShapeFit systematic budget including the systematic error on $m$, see section \ref{sec:sys_budget}.  

\section{Re-introducing model-dependence: Cosmology Interpretation} \label{sec:cosmo}
The advantage of parameter compression methods such as ShapeFit is that the whole power spectrum analysis presented before, from the performance on mocks towards the systematic budget determination, is performed only once, without the need to repeat it for every cosmological model in consideration. Therefore, once the three (four) compressed variables are obtained with the classic fit (ShapeFit), their cosmological interpretation is much more streamlined than for FM Fits. 

In this section we are interested in the cosmological implications for selected cosmological models, of the ShapeFit results for the BOSS and eBOSS LRGs and QSOs samples described before, with particular focus on the information gain provided by the new shape parameter $m$ with respect to the classic BAO+RSD approach. 

The models considered, the varied parameters and prior ranges for all the cosmological models considered are provided in table \ref{tab:cosmo-priors}. These should be seen as few examples, chosen to compare more directly the ShapeFit performance with that of other approaches.

In order to connect the compressed set of parameters of the four redshift bins with the parameters of the model we run a Monte Carlo Markov Chain employing CLASS \cite{2011JCAP...07..034B} as a Boltzmann code for generating the linear power spectrum. We follow the standard approach for connecting the scaling parameters and the growth of structure with the parameters of the model. A key point, and a novelty of ShapeFit is how to connect $m$ to the relevant parameters of the model. This is done by computing the smooth linear power spectrum (see Appendix \ref{app:smoothBAO}) and applying it to eq.~\eqref{eq:theory_m}. A more detailed explanation of this process is provided in \cite{ShapeFit}, and in particular in fig.~5.

We start by considering the baseline $\Lambda$CDM model in section \ref{sec:cosmo-baseline} and proceed with the extended cosmologies in section \ref{sec:cosmo-ext}. We investigate the cosmological implications of the individual LSS tracers (LRGs, QSOs and Ly-$\alpha$) in section \ref{sec:cosmo_highzvslowz} and compare our results to other fitting approaches in section \ref{sec:cosmo-comparison}. 

\begin{table}[htb]
\centering
    \begin{tabular}{|c|c|c|c|}
      \hline
      Type & Parameter (phys.) & Prior & Usage\\
       \hline
   \multirow{6}{*}{Baseline $\Lambda$CDM}  & $\omega_\mathrm{b}$  & $[0.005, 0.04]$ & Always\\
     & $\omega_\mathrm{cdm}$  & $[0.001, 0.99]$ & Always \\
     & $h$ & $[0.2, 2]$ & Always \\
     & $\ln\left(10^{10}A_s\right)$ & $[0.1, 10]$ & Always \\
     & $n_s$  & $[0.5, 1.5]$ & With Planck\\
     & $\tau_\mathrm{reio}$  & $[0.004, 1.0]$ & With Planck\\
      \hline
    \multirow{5}{*}{Extensions to $\Lambda$CDM} & $\Sigma m_\nu \,(\mathrm{eV})$ & $[0.0,1.0]$ & Section \ref{sec:cosmo-mnu} \\
    & $N_\mathrm{eff}$ & $[0.0,9.0]$ & Section \ref{sec:cosmo-neff} \\
    & $\Omega_\mathrm{k}$ & $[-0.8,0.8]$ & Section \ref{sec:cosmo-Ok}, \\
    & $w_0$ & $[-1.5,-0.0]$ & Section \ref{sec:cosmo-w0}\\
    & $w_0+w_a$ & $[-5.0,-0.0]$ & Section \ref{sec:cosmo-w0} \\
      \hline
    \end{tabular}
    \caption{Prior ranges for the cosmological parameters. Whenever the extended parameters are not varied, they are fixed to $\Sigma m_\nu = 0.06 \, \mathrm{eV}$, $N_\mathrm{eff}=3.046$, $\Omega_\mathrm{k}=0$, $w_0 = -1$, $w_a = 0$. The reionisation optical depth $\tau_\mathrm{reio}$ and scalar tilt $n_s$ are only varied when Planck data is included in the chains, otherwise the scalar tilt is fixed to the fiducial value $n_s^\mathrm{fid}=0.97$}
    \label{tab:cosmo-priors}
\end{table}

\subsection{Baseline \texorpdfstring{$\Lambda$}{L}CDM} \label{sec:cosmo-baseline}

\begin{figure}[htb]
    \centering
    \includegraphics[width=0.49\textwidth]{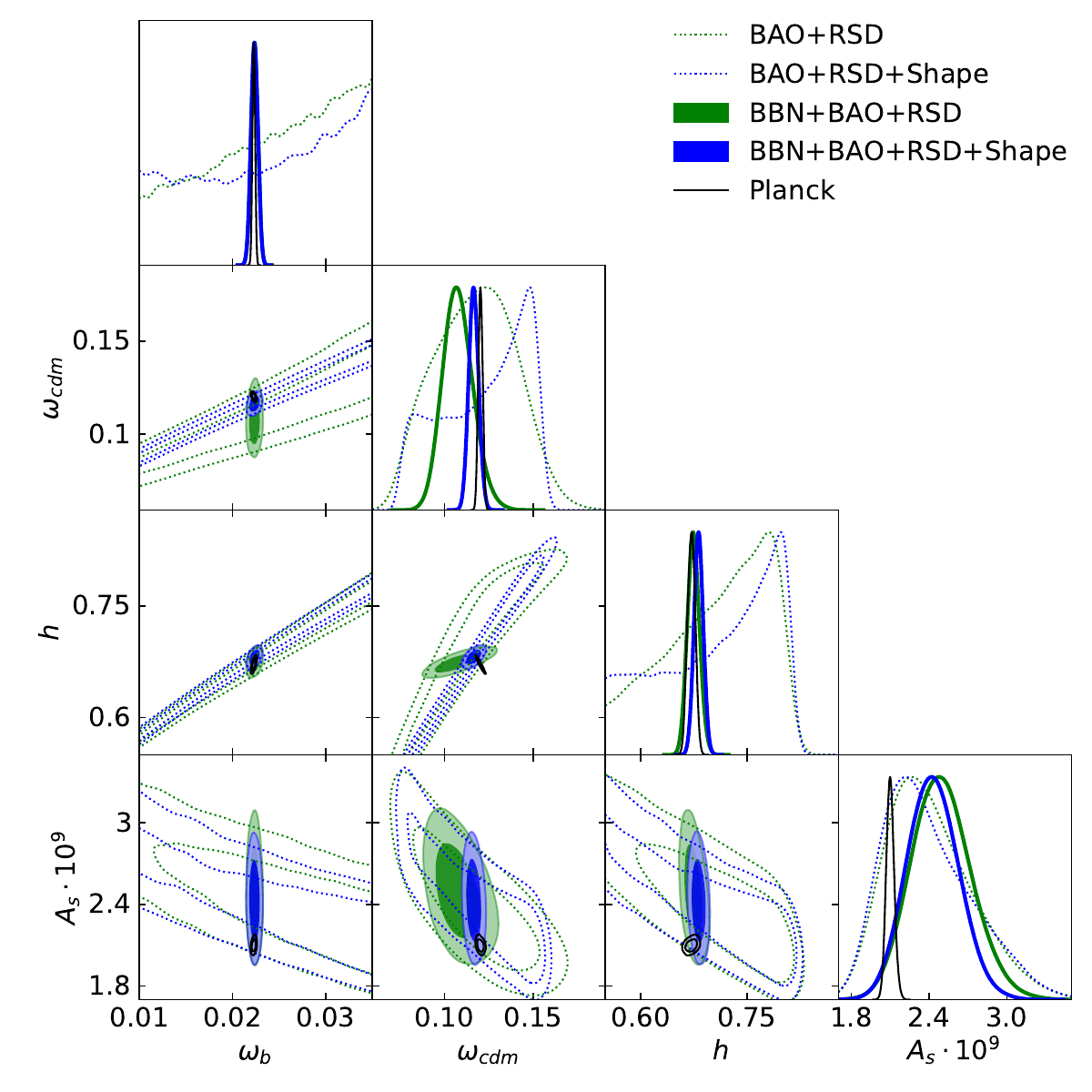}
        \includegraphics[width=0.49\textwidth]{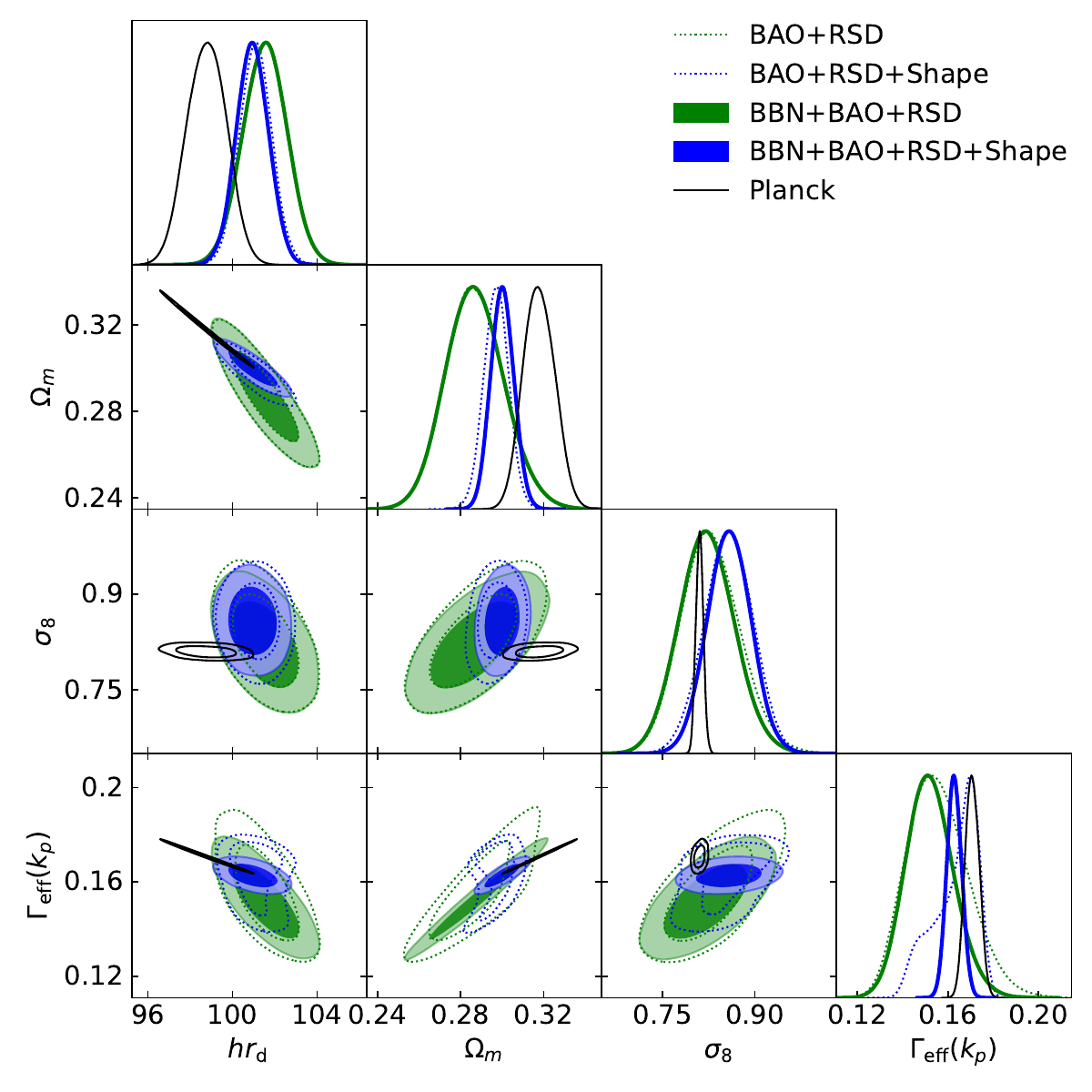}
    \caption{Baseline $\Lambda$CDM results for the classic fit (BAO+RSD information, green contours), ShapeFit (BAO+RSD+{\it Shape} information, blue contours) in comparison with Planck (black contours). The filled contours include a BBN-motivated Gaussian prior on $\omega_\mathrm{b}$, while the dotted contours span the full uniform prior range of the baryon density. The left panel shows the results in the basis of the varied parameters, while the right panel shows the derived cosmological parameters, that are more naturally constrained by our LSS dataset. As a result, the parameters on the right panel are (almost) insensitive to the BBN prior, which is not the case for those on the left panel.
    }
    \label{fig:cosmo}
\end{figure}

We show our baseline results for the classic fit (green), that includes the BAO+RSD information, and for ShapeFit (blue), which adds the {\it Shape} information, in figure \ref{fig:cosmo}. The cosmological constraints from the BOSS+eBOSS surveys alone are shown as empty, dotted contours, while the filled, continuous contours include the BBN-motivated Gaussian prior on $\omega_\mathrm{b}$ introduced in section \ref{sec:data-early-time}. For comparison, we show the $\Lambda$CDM constraints from Planck alone (empty, black contours), which are in good agreement with the LSS
ones.

The two panels of figure \ref{fig:cosmo} correspond to the same cosmological runs, but for different parameter bases. On the left, we show the basis of varied parameters, while on the right we show those derived parameters that are more closely related to the physical parameters our LSS dataset is sensitive to, presented in section \ref{sec:results_all}. Strikingly, the left panel parameters are quite unconstrained from LSS data alone. Both for the classic fit and for ShapeFit, there is a perfect degeneracy between $\omega_\mathrm{b}, \omega_\mathrm{cdm}$ and $h$ that can only be broken via the BBN prior or other early-time information. On the other hand, the right panel's parameters are almost insensitive to the BBN prior, which can be understood as follows. 

The sound horizon scale in units of the Hubble constant today, $hr_\mathrm{d}$, is measured from the isotropic BAO information. Therefore, its constraints are nearly identical for the classic fit and ShapeFit. This is not the case for $\Omega_\mathrm{m}$, which is determined within the classic approach from the anisotropic BAO and AP effect alone, whereas for ShapeFit there is additional information coming from the shape $m$. As argued in \cite{BriedenPRL21} this parameter effectively constrains the combination $\Gamma_0 = \Omega_\mathrm{m} h$, which is also known as `shape parameter' \cite{BBKS, BondSzalay83,EH_TransferFunction}. However, this parameter combination does not take into account the shape sensitivity to $\omega_{\rm b}$ due to the baryon suppression \cite{Sugiyama95}. Therefore, in the right panel we show the more complex, scale-dependent `effective shape parameter' $\Gamma_\mathrm{eff}(k)$ defined in eq. (30) of \cite{EH_TransferFunction} evaluated at the pivot scale $k_p$ introduced in section \ref{sec:theory_shapefit}. This parameter is constrained very well by ShapeFit, which propagates into an improvement of $\Omega_\mathrm{m}$ constraint with respect to the classic fit, even without imposing the BBN prior. Interestingly, we do not observe the same for the Hubble constant $h$, which for LSS data alone remains unconstrained even after adding $m$. Finally, the matter fluctuation amplitude $\sigma_8$ is well determined by both the classic fit and ShapeFit through our RSD measurement of the velocity fluctuation amplitude $f\sigma_{s8}$. Note that this constraint is completely independent of the BBN prior, whereas the constraint on the primordial fluctuation amplitude $A_s$ shows a certain $\omega_\mathrm{b}$-dependence. This is due to the fact that our LSS maps are sensitive to the total matter power spectrum amplitude and are not able to disentangle whether the amplitude is of primordial origin from inflation, from early time evolution of the transfer function (for example related to the baryon suppression at the time of photon decoupling) or attributed to the late-time growth of structures. Therefore, $\sigma_8$ is the natural variable to express the net clustering amplitude.   

Another interesting aspect, related to the two parameter bases shown in the left and right panel of figure \ref{fig:cosmo} respectively, is that for Planck alone the situation is inverted. While Planck constraints on physical densities  $\omega_\mathrm{b}, \omega_\mathrm{cdm}$ and the Hubble parameter $h$ are much tighter than in the LSS case, they are of comparable size when considering the absolute matter density $\Omega_\mathrm{m}$ and the sound horizon in units of the Hubble constant $hr_\mathrm{d}$, which are strongly degenerate in the Planck case. It is interesting to note that the constraint on $\Omega_m$ from LSS alone (with a very weak $n_s$ prior, see appendix \ref{app:prior}) is more stringent than that from Planck, yet well in agreement. This demonstrates the complementary nature between the shown early-time and late-time datasets.

We conclude that the shape $m$ delivers a significant piece of information leading to a strong improvement in $\Lambda$CDM parameter constraints. This is related to the findings of \cite{Philcox_Sherwin_Farren_Baxter_21} who obtain a constraint on $H_0$ from the galaxy power spectrum  marginalizing over the sound horizon scale. Here, for example, for $\Omega_\mathrm{m}$ we find an improvement of factor 2 when including the {\it Shape}. A breakdown of the combined constraining power shown here into the contributions from individual tracers can be found in section \ref{sec:cosmo_highzvslowz}. The exact numbers and how they compare to other approaches can be found in table \ref{tab:other_results} of section \ref{sec:cosmo-comparison}.

\subsection{Extensions to the baseline \texorpdfstring{$\Lambda$}{L}CDM model} \label{sec:cosmo-ext}

We consider a variety of extensions to the baseline $\Lambda$CDM model in a similar way as presented in the official BOSS and eBOSS cosmology papers \cite{BOSSDR12,eboss_collaboration_dr16}. Similar to those works, we focus on models that involve neutrino physics (sections \ref{sec:cosmo-mnu} and \ref{sec:cosmo-neff}) and models that change the geometry and growth history of the universe, such as curvature (section \ref{sec:cosmo-Ok}) and varying dark energy (section \ref{sec:cosmo-w0}). 

\subsubsection{Massive neutrinos} \label{sec:cosmo-mnu}

The measurement of the sum of neutrino masses $\Sigma m_\nu$ is of major interest for the scientific community and within reach for upcoming (and ongoing) cosmological surveys. The presence of massive neutrinos, or any relativistic, weakly-interacting particle species, that becomes non-relativistic once the temperature of the universe drops below its mass (also referred as `warm dark matter') leaves a unique imprint on cosmological observables. If correctly modeled and verified for different probes, these features (described below in more detail) can lead towards the measurement of the sum of neutrino masses.

We know that neutrinos must possess a non-zero mass from neutrino oscillation experiments, that measure the mixing angle between neutrino flavours from different neutrinos sources, such as from solar and atmospheric \cite{Fukuda98}, reactor \cite{Abe12} 
and accelerator \cite{Abe14} 
origins. The measured mixing angles can be translated into squared mass differences between the flavours. Although these measurements do not probe the absolute mass scale of neutrinos, they can be used to construct the minimal sum of neutrino masses allowed by the oscillation data given that the lightest neutrino has a mass of zero. Since the oscillation data is primarily sensitive to the \textit{squared} mass differences (and not their sign), there are two possible mass hierarchies: either the smaller mass split occurs between the lightest and the second-to-lightest neutrino (normal hierarchy) or it occurs between the heaviest and the second-to-heaviest neutrino (inverted hierarchy). For the normal (inverted) hierarchy, the minimum neutrino mass sum consistent with the oscillation data is $\Sigma m_\nu > 0.0588 \,\mathrm{eV} $ ($\Sigma m_\nu > 0.0995 \,\mathrm{eV} $) \cite{Esteban19}. 

While from particle physics experiments it is very challenging to measure the absolute neutrino mass scale -the strongest and most recent $95\%$ upper limit of $\Sigma m_\nu < 2.4 \,\mathrm{eV}$ comes from the KATRIN experiment \cite{KATRIN2022}- cosmological surveys are currently beating this limit by a an order of magnitude. Intriguingly, the state-of-the-art $95\%$ upper limit on the sum of neutrino masses coming from the combination of Planck with BOSS+eBOSS BAO and RSD data of $\Sigma m_\nu < 0.102 \,\mathrm{eV}$ \cite{eboss_collaboration_dr16} is already very close to the lower limit predicted by neutrino oscillation experiments assuming the inverted hierarchy. Upcoming surveys will most probably either exclude the inverted hierarchy, detect a non-zero neutrino mass sum or even discriminate between neutrino masses of the individual flavours. However, any potential finding in this direction depends on the exact choice of underlying model. Therefore, it is crucial to verify the implications of non-zero neutrino masses for independent probes with robust control over any systematic effects that may arise from each probe. In this work, we introduce the shape of the power spectrum, parameterized through $m$, as an additional observable that may serve (among others) as a smoking gun towards a non-zero neutrino mass detection.

The implications of non-zero neutrino masses for cosmology are manifold and rather complex. Their subtle impact on cosmological observables has been reviewed in \cite{Lesgourgues:2006nd}. In a nutshell, there are three main effects. First, the transition of neutrinos from relativistic to non-relativistic species leads to a change in geometry. Depending on whether the total matter density today $\omega_\mathrm{m}=\omega_\mathrm{cdm}+\omega_\mathrm{b}+\omega_\nu$ or the cold dark matter + baryon density $\omega_\mathrm{cb}=\omega_\mathrm{cdm}+\omega_\mathrm{b}$ is kept fixed, either the early-time scale of equality between matter and radiation $k_\mathrm{eq}$ changes (in the former case) or the late-time distance ladder parameterised for example through the Hubble parameter $h$ changes (in the latter case). This effect can be measured by combining CMB with BAO data for example. Second, since the massive (but still very light) neutrinos have much higher thermal velocities than ordinary matter, they do not cluster on scales smaller than their free-streaming scale. As a consequence, they wash out small-scale perturbations and slow down the growth of structures. This has a measurable impact on the redshift dependence of the growth rate, assessed by RSD data. The third effect is related to the second one. Since massive neutrinos do not only change the redshift-dependence of the growth rate, but also its scale dependence, they induce a step-like suppression of the matter power spectrum at their free streaming scale. This has a measurable impact on the shape parameter $m$. 

So far, this third effect has not been taken into account within the classic approach, for a number of reasons. As shown in figure 1 of \cite{ShapeFit}, the step-like suppression induced by massive neutrinos is very degenerate with the other $\Lambda$CDM parameters, in particular combinations of $\omega_\mathrm{cdm},n_s,\sigma_8$, at least within the restricted wavenumber range of $0.02<k \hoverMpc<0.15$ usually investigated in galaxy clustering. Moreover, the marginalization over bias parameters, the FoG effect, etc., make it even harder to extract robust neutrino mass information from the data. Finally, the BAO+RSD information is considered more robust than the {\it Shape} information, which is subject to systematic uncertainties of observational (see section \ref{sec:sys}) and modeling nature, such as unaccounted for scale dependent bias, relativistic effects or primordial non-gaussianity (see \cite{BriedenPRL21} for the latter case). Using the ShapeFit framework, where the shape $m$ is measured directly with nuisance parameters already marginalized out and the possibility to apply to it any custom systematic error budget, it is more convenient and intuitive to test the neutrino-mass-induced shape dependence of the power spectrum.

We assume three degenerate neutrinos with effective number of neutrino species $N_\mathrm{eff}=3.046$, and with the mass split equally among them. As in the classic BAO+RSD approach by \cite{eboss_collaboration_dr16}, we compare our $f\sigma_{s8}$ measurements to the model prediction of the `cold' velocity fluctuation amplitude $f\sigma_8^\mathrm{cb}$, obtained from the cold dark matter + baryon power spectrum $P_\mathrm{cb}$ instead of the total matter power spectrum $P_\mathrm{m}$. This quantity has been shown to represent the galaxy clustering amplitude in a more universal way, since galaxies are tracers of the cold+baryon density field, that excludes the neutrino perturbations \cite{Raccanelli:2017kht,Vagnozzi:2018pwo}. Accordingly, we obtain theoretical predictions for the shape $m$ consistent with the `cb' prescription by using the cold+baryon transfer function in the numerator of eq.~\eqref{eq:theory_m}.

\begin{figure}[htb]
    \centering
    \includegraphics[width=0.49\textwidth]{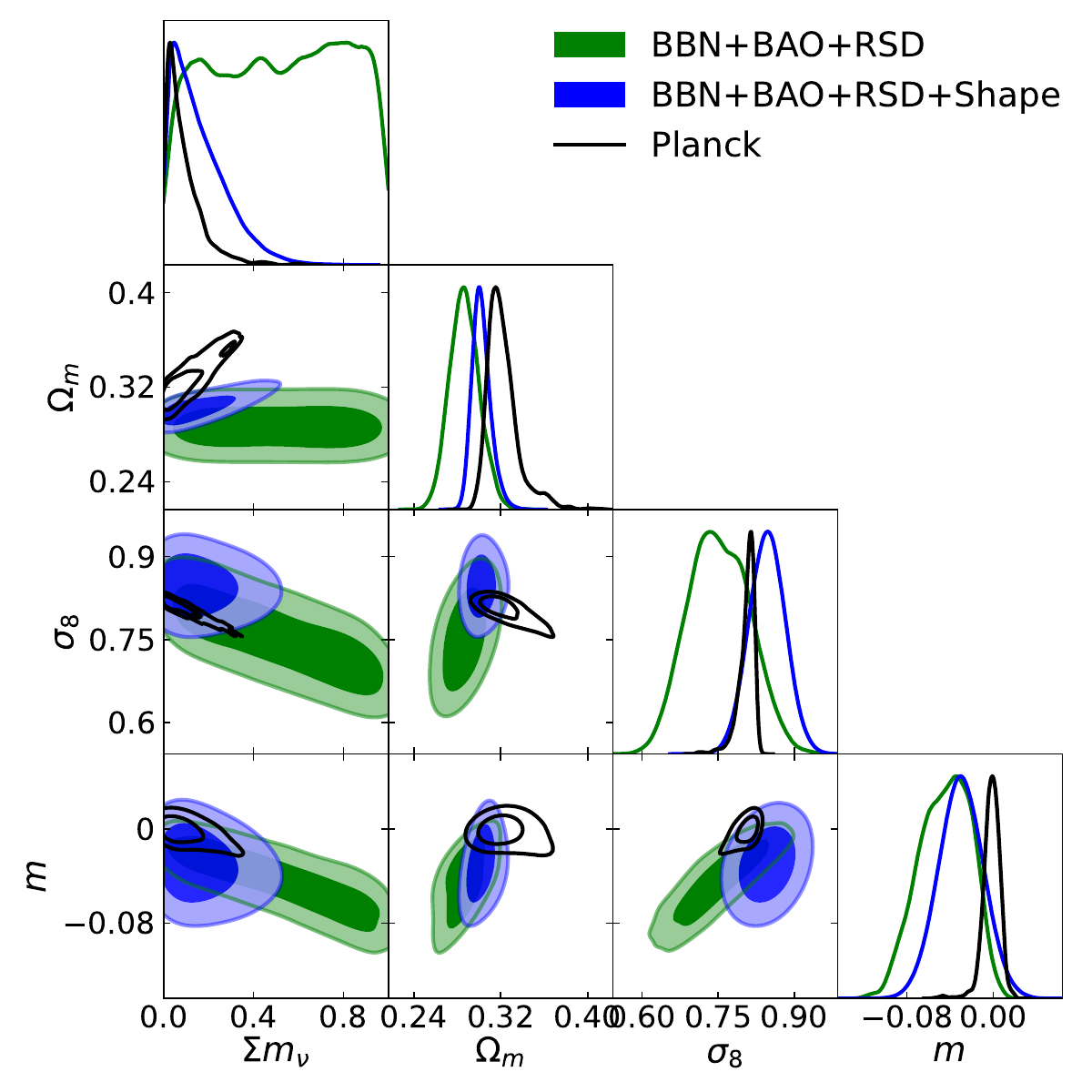}
        \includegraphics[width=0.49\textwidth]{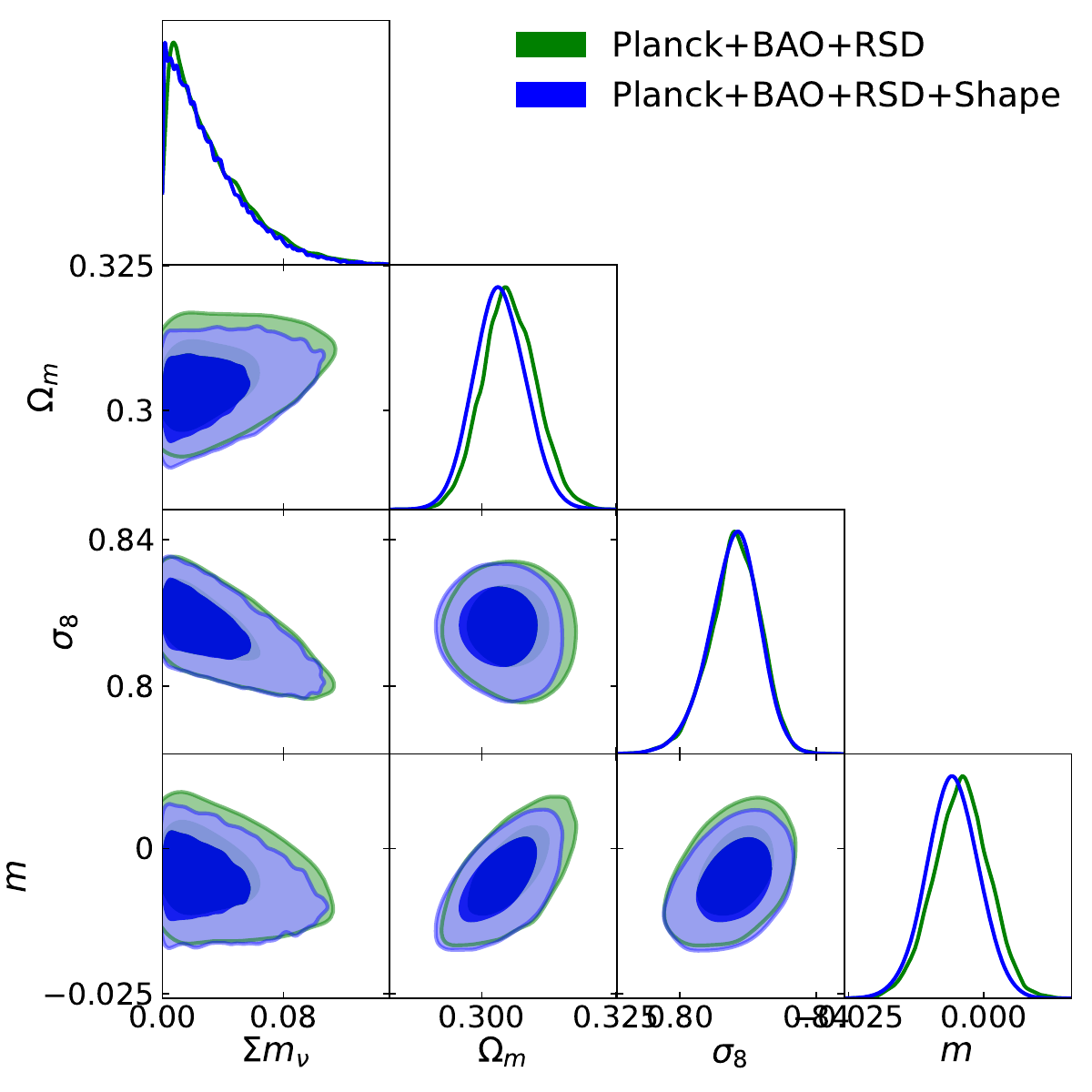}
    \caption{Neutrino mass,$\Sigma m_\nu$, bounds for different data combinations and  $\Sigma m_\nu$  degeneracy with $\Omega_\mathrm{m}$, $\sigma_8$, and the power spectrum slope at the pivot scale, $m$. Results obtained with the classic BAO+RSD Fit are shown in green, while the ShapeFit results are shown in blue. In the left panel we compare both types of fit when combined with the BBN-motivated prior on $\omega_{\rm b}$ and with primordial tilt $n_s$ fixed to the reference value. For comparison, we also show the constraints obtained from Planck alone in black. In the right panel we show the results combined with Planck, allowing $\omega_{\rm b}$ and $n_s$ to vary freely.}
    \label{fig:cosmo-Mnu}
\end{figure}

We show our constraints on $\Sigma m_\nu$ and its correlation with $\Omega_\mathrm{m}$, $\sigma_8$, and the shape $m$ in figure \ref{fig:cosmo-Mnu}. Again, green and blue contours represent the classic fit and ShapeFit respectively. Parameter constraints derived from Planck alone are shown in black. The left panel shows parameter constraints for LSS+BBN data, while in the right panel LSS data is combined with Planck. 

From the left panel we see that the BBN+BAO+RSD data alone can not constrain neutrino mass, due to its degeneracy with fluctuation amplitude $\sigma_8$. However, this degeneracy is broken by the slope $m$. Hence, ShapeFit is able to provide an upper 95\% confidence level bound on the neutrino mass of $\Sigma m_\nu < 0.40 \, \mathrm{eV}$, which is the tightest neutrino mass bound ever obtained from LSS data (in combination with the BBN prior). From Planck alone we find $\Sigma m_\nu < 0.24 \, \mathrm{eV}$, consistent with \cite{Aghanim:2018eyx}. 

Note that the ShapeFit constraint relies on a fixed fiducial value of the primordial tilt $n_s$. Letting it vary freely would completely degrade the ShapeFit neutrino mass constraint, since the slope variation induced by $\Sigma m_\nu$ is degenerate with $n_s$ on the scales
we consider here. Planck on the other hand provides the angular power spectrum shape on a huge range of scales, so that the scale-independent tilt can be inferred with high precision. 

In the right panel we show the case where Planck data and the BOSS+eBOSS data are fitted jointly. In this case the neutrino mass constraints become much tighter due to the complementary information Planck provides with respect to the LSS surveys. For our Planck+BAO+RSD dataset we find $\Sigma m_\nu < 0.085 \, \mathrm{eV}$. Note that this is slightly smaller than the value reported in \cite{eboss_collaboration_dr16}, we have verified that this is due to the fact that we exclude the MGS and ELG samples from our dataset. 

Interestingly, this upper bound barely changes when adding the shape $m$, yielding $\Sigma m_\nu < 0.082 \, \mathrm{eV}$, as the additional information within the LSS shape is superseded by the Planck data. This can also be seen from figure \ref{fig:results} and has already been shown in figure 2 of \cite{ShapeFit}. We conclude that for this specific extension to the $\Lambda$CDM model the shape information from our BOSS+eBOSS dataset does not add much to the information that Planck provides. 
This also holds for all the other extended models analysed in this section. Therefore, in what follows, we focus on the cosmological constraints obtained from our dataset without Planck.

Nevertheless, $\Sigma m_\nu < 0.082 \, \mathrm{eV}$ at 95\% C.L. implies that the minimum mass for the inverted hierarchy is excluded at 98\% C.L. (assuming Gaussian errors).

\subsubsection{Varying effective number of neutrino species} \label{sec:cosmo-neff}

Another degree of freedom related to neutrinos is the effective number of neutrino species $N_\mathrm{eff}$. Given the standard model of particle physics and the measurement of Z-boson decay,  we know that three neutrino species exist. However, we can not exclude that extra neutrino species (or other weakly interacting particles) existed in the early universe, when the temperature was higher than the energy range probed by laboratory experiments.

In this case the radiation density comprised of photons and neutrinos would change as
\begin{equation}
    \omega_r =\omega_\gamma + \omega_\nu = \left[ 1 + \frac{7}{8} \left( \frac{4}{11} \right)^{4/3} N_\mathrm{eff} \right] \omega_\gamma~. 
\end{equation}
Varying the parameter $N_\mathrm{eff}$ thus induces a change in the sound horizon $r_\mathrm{d}$ and the scale of matter and radiation equality $k_\mathrm{eq}$ while keeping the photon density (and therefore the CMB temperature fixed). At the background level, this effect is completely degenerate with the Hubble parameter $h$. Hence, extra relativistic degrees of freedom could help to reconcile the Hubble tension. However, this degeneracy is broken at the perturbation level, where $N_\mathrm{eff}$ has a variety of subtle effects. Current CMB observations from Planck strongly disfavor deviations from 3 neutrino species, from our Planck dataset we find $N_\mathrm{eff} = 2.941 \pm 0.45$ (95\%) in agreement with \cite{Aghanim:2018eyx}.

We investigate whether ShapeFit is able to track the effect of $N_\mathrm{eff}$ on the matter power spectrum, which are \textit{i)} a smooth tilt in the transition region between the small and the large scale limit and \textit{ii)} a modulation of the BAO amplitude. ShapeFit is sensitive to effect \textit{i)} through $m$, but not to effect \textit{ii)}. 

From the left panel of figure \ref{fig:cosmo-NeffOk} we see that including $m$ reduces the cosmological parameter space allowed by LSS data. But neither the classic fit nor ShapeFit (both including the BBN-motivated prior on $\omega_\mathrm{b}$) are able to constrain $N_\mathrm{eff}$ due to its degeneracy with $h$. In fact, effect \textit{i)} is a pure background effect coming from the shift in $k_\mathrm{eq}$. To capture effect \textit{ii)} as well we would need to extract the BAO amplitude from the data with ShapeFit, which is a challenging task due to non-linear effects, such as the BAO wiggle suppression, that need to be modeled accurately and cross-checked to be free of observational systematic effects. Note that we do not consider here the dependence of the baryon density $\omega_\mathrm{b}$ and the Helium fraction $Y_\mathrm{He}$ on $N_\mathrm{eff}$ within the theoretical BBN prediction. We leave this, and a more complete fitting procedure including the BAO wiggle amplitude for future work.

When combining our LSS dataset with Planck we obtain $N_\mathrm{eff} = 3.16 \pm 0.41$ (95\%)  for the classic fit and $N_\mathrm{eff} = 3.12 \pm 0.38$ (95\%)  for ShapeFit, both consistent with the official Planck+BAO constraint $N_\mathrm{eff} = 2.99 \pm 0.33$ (95\%)  from \cite{Aghanim:2018eyx}.

\subsubsection{Curvature} \label{sec:cosmo-Ok}

\begin{figure}[tb]
    \centering
    \includegraphics[width=0.49\textwidth]{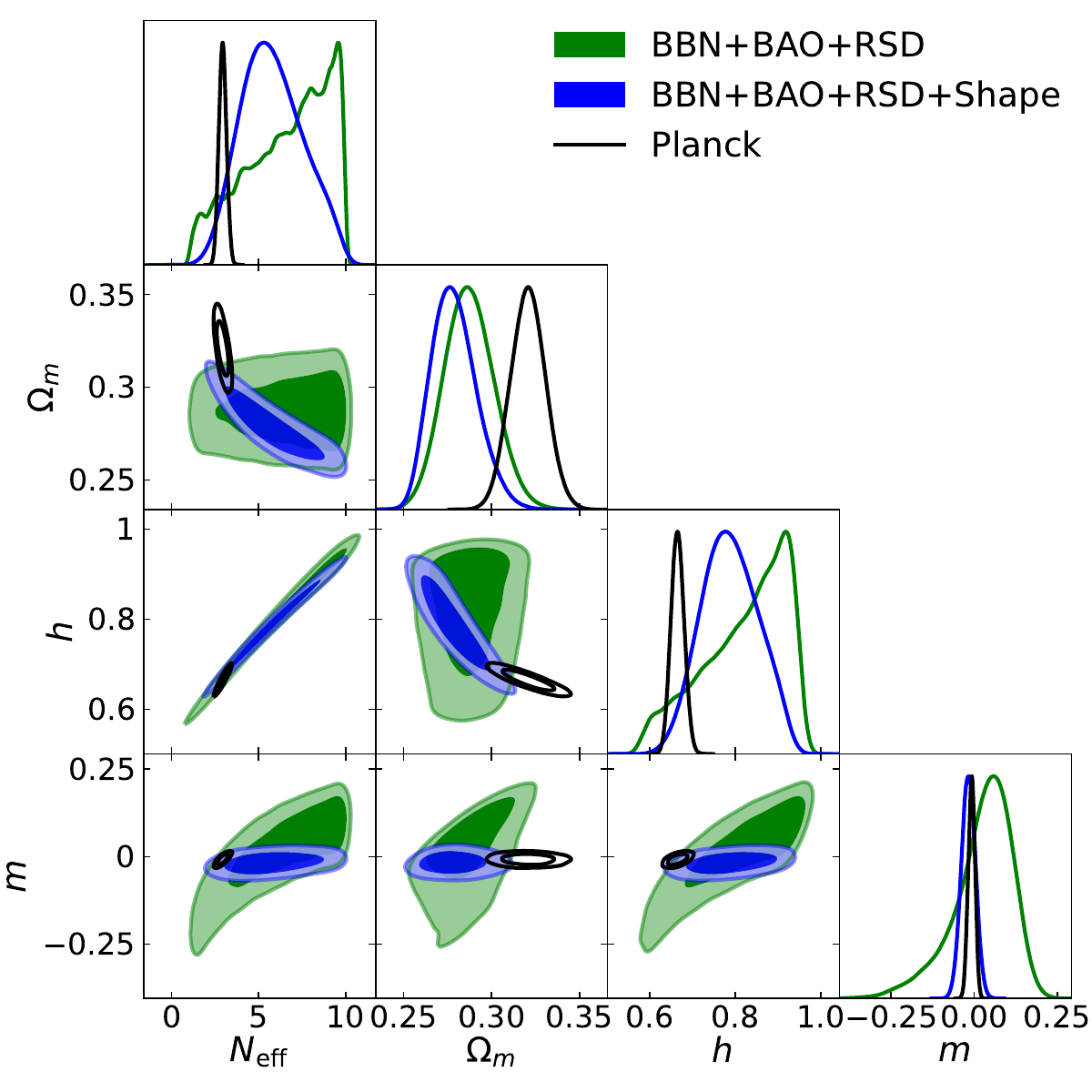}
        \includegraphics[width=0.49\textwidth]{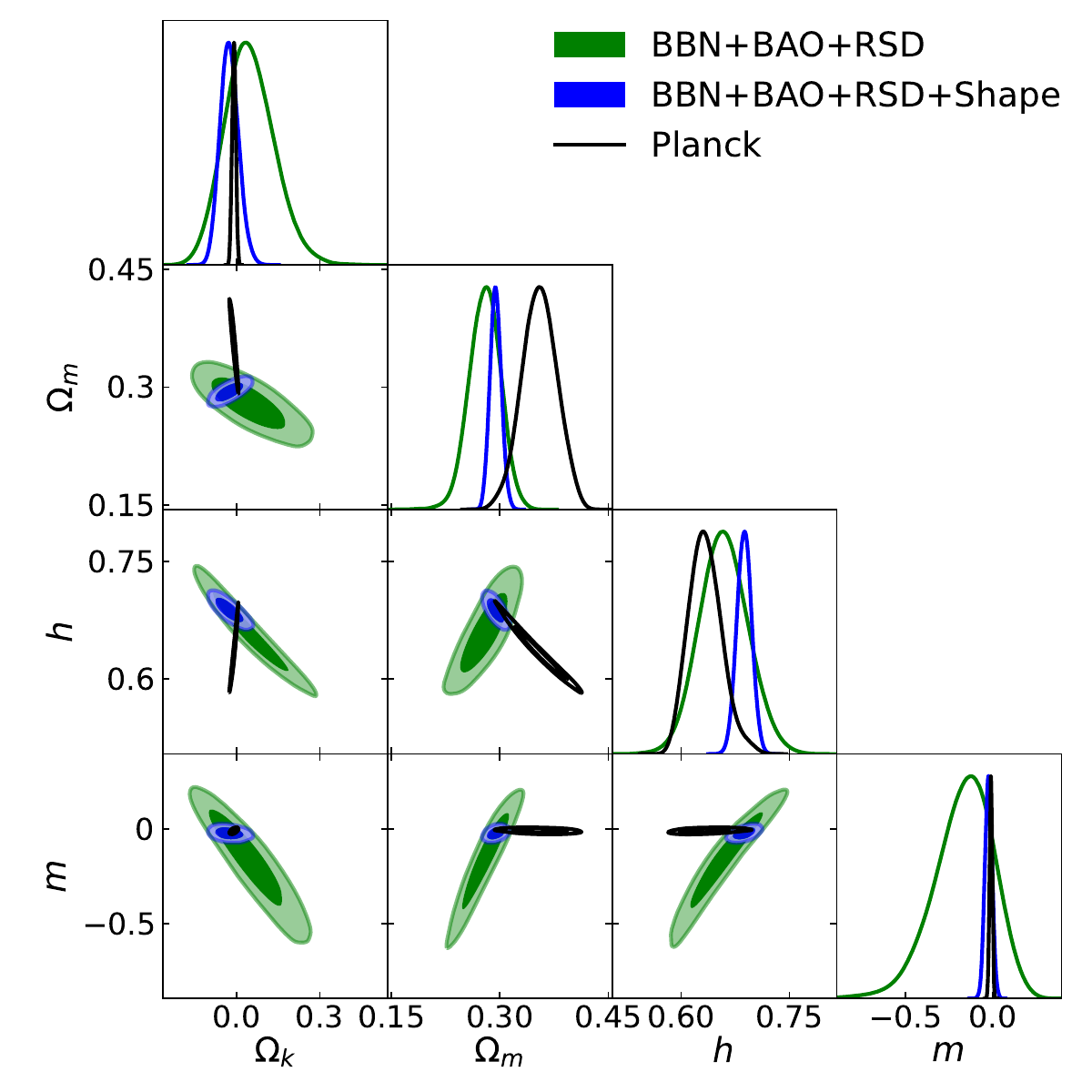}
    \caption{Left panel: Cosmological constraints in the case of allowing free number of effective degrees of freedom. Right panel: Cosmological constraints in the case of allowing for curvature.}
    \label{fig:cosmo-NeffOk}
\end{figure}

Spatial curvature is usually parameterised by the curvature energy density parameter $\Omega_k$ today entering the Friedmann equation and Hubble expansion rate with a redshift dependence proportional to $(1+z)$. 
Also, non-zero curvature changes the geometry and as such the comoving angular diameter distance $D_M^{\Omega_k=0}$ (already defined in section \ref{sec:theory_Pell} below eq.~\eqref{eq:thery_alpha_scaling}) as

\begin{equation}
\begin{aligned}
    D_M^{\Omega_k} = \frac{c}{H_0} S_k \left( \frac{D_M^{\Omega_k=0}}{c/H_0} \right), \qquad S_k(x) = \begin{cases}
    \sin\left(\sqrt{-\Omega_k}x/ \sqrt{-\Omega_k}\right) \quad &\Omega_k<0, \\
    x \quad &\Omega_k =0,\\
    \sinh\left(\sqrt{\Omega_k}x/ \sqrt{\Omega_k}\right) \quad &\Omega_k>0 ~. \\
    \end{cases}
\end{aligned}
\end{equation}

While, when combining LSS data with Planck, the evidence for a flat universe with $\Omega_k=0$ is striking \cite{Aghanim:2018eyx,eboss_collaboration_dr16} (see also table \ref{tab:comparison-with-classic}), here we are particularly interested in the constraining power of our LSS data set on  $\Omega_k$ %=0$ %from our LSS dataset alone 
and whether the shape $m$ helps to break its degeneracy with $\Omega_\mathrm{m}$ and $h$.

In the right panel of figure \ref{fig:cosmo-NeffOk} we show the constraints on these cosmological parameters along with their degeneracy with $m$ for the classic fit (green), ShapeFit (blue) and Planck only (black). The latter provides the tightest constraint on curvature of $\Omega_k=-0.0104\pm0.0067$. Note that we also include the Planck lensing signal here, which is in agreement with a flat universe and strongly improves constraints with respect to considering the Planck temperature and polarization spectra only leading to $\Omega_k=-0.043 \pm 0.017$. For the classic fit we find $\Omega_k=0.047_{-0.099}^{+0.083}$ and for ShapeFit
$\Omega_k=-0.027_{-0.037}^{+0.032}$ delivering an improvement in constraining power of a factor $\sim 2.7$. As can be seen in the figure, this improvement comes from the measurement of the shape $m$. 

All these results are consistent with zero curvature and their combination (either $\mathrm{Planck} + \mathrm{Classic}$ or $\mathrm{Planck} + \mathrm{ShapeFit}$) gives $\Omega_k=-0.0015\pm0.0016$. So, the Planck {\it Shape} information dominates over the LSS {\it Shape} constraint and the BAO+RSD measurements alone are sufficient due to their high degree of complementarity to Planck enabling to lift most of the parameter degeneracies. 

Note that curvature only affects the geometry and growth of the universe, it does not leave any imprint on the galaxy power spectrum slope. Still, the slope measurement matters, due to its sensitivity to $\Omega_\mathrm{m}$, which breaks the degeneracy with $\Omega_k$. But once Planck is added, the shape $m$ does not further improve constraining power. 

\subsubsection{Varying Dark Energy} \label{sec:cosmo-w0}

The most important science driver behind any state-of-the-art  spectroscopic survey is to unravel the nature of dark energy, which delivers the force behind the accelerated expansion of the universe observed by many disparate probes. Currently, this accelerated expansion is in accordance with the General Relativity prediction of Einstein's cosmological constant $\Lambda$, but a microscopic explanation for this constant term in the Friedman equation, leading to a constant expansion rate, does not exist yet. 

The most general macroscopic description for this term relies in the assumption of a fluid called dark energy with negative equation of state paramter $w = p_\mathrm{DE}/\rho_\mathrm{DE}$ relating the dark energy pressure $p_\mathrm{DE}$ and density $\rho_\mathrm{DE}$. The dark energy equation of state parameter $w$ governs the evolution 
of the universe at late times. In general, the scale-factor dependence of the energy density of any fluid with constant equation of state is $\rho(a) \propto a^{-3(1+w)}$. For $w<-1/3$, 
it describes a fluid giving rise to accelerated expansion.
For $w=-1$ in particular, the dark energy density $\rho_\mathrm{DE}$ remains constant yielding the same expansion rate as predicted by a cosmological constant $\Lambda$.

In this section we consider three cases for the evolution of the equation of state $w(a)$ with scale factor $a$,
\begin{equation}
\begin{aligned}
    w(a) = 
    \begin{cases}
    -1 \qquad & (\Lambda\mathrm{CDM}), \\
    w_0 \qquad & (w\mathrm{CDM}), \\
    w_0 + w_a(1-a) \qquad & (w_0 w_a\mathrm{CDM}), \\
    \end{cases}
\end{aligned}
\end{equation}
where the first case is consistent with $\Lambda$CDM, the second exhibits an equation of state constant with cosmic time and the third allows for a time dependence phenomenologically motivated by \cite{Chevallier_Polarski01,Linder03}.

\begin{figure}[htb]
    \centering
    \includegraphics[width=0.49\textwidth]{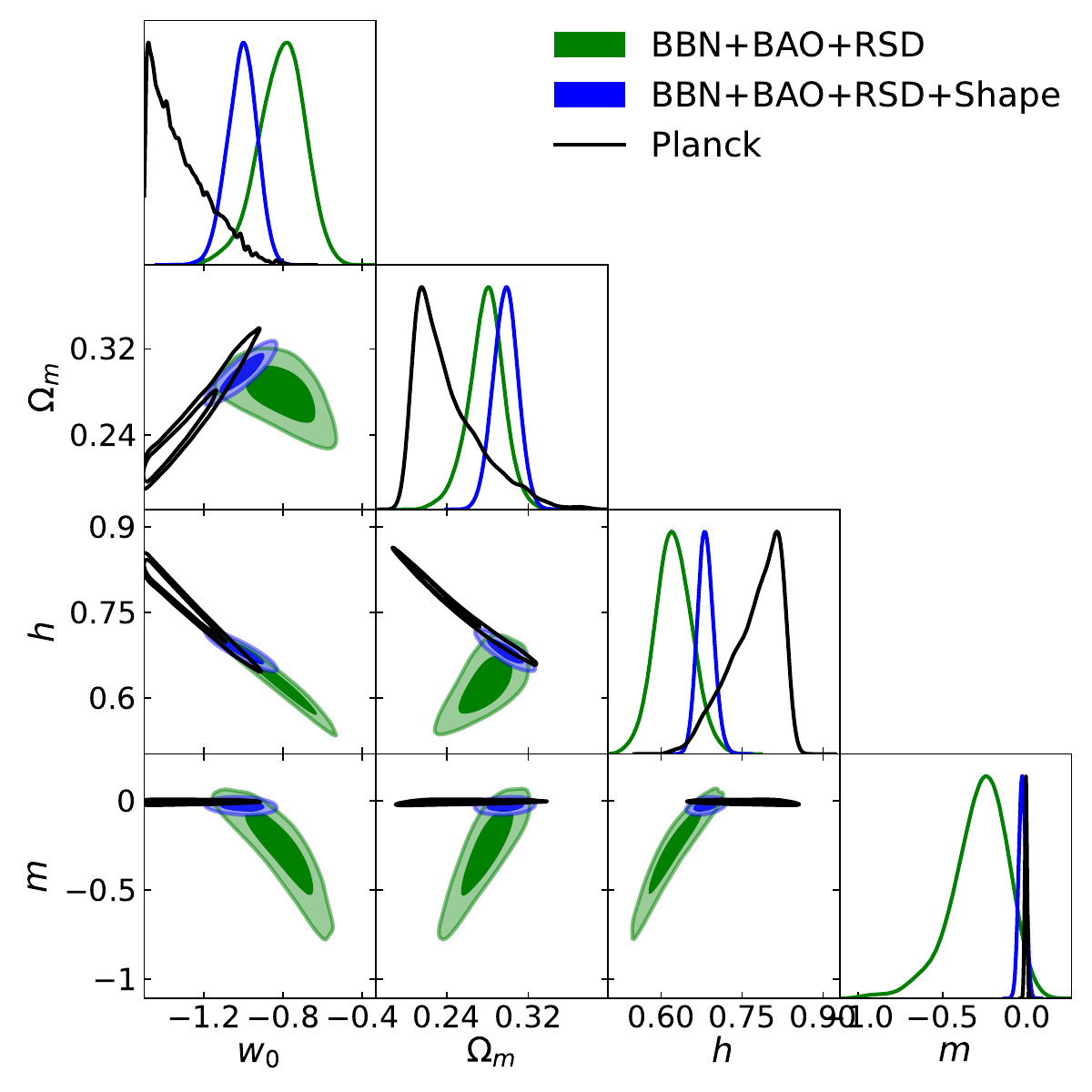}
        \includegraphics[width=0.49\textwidth]{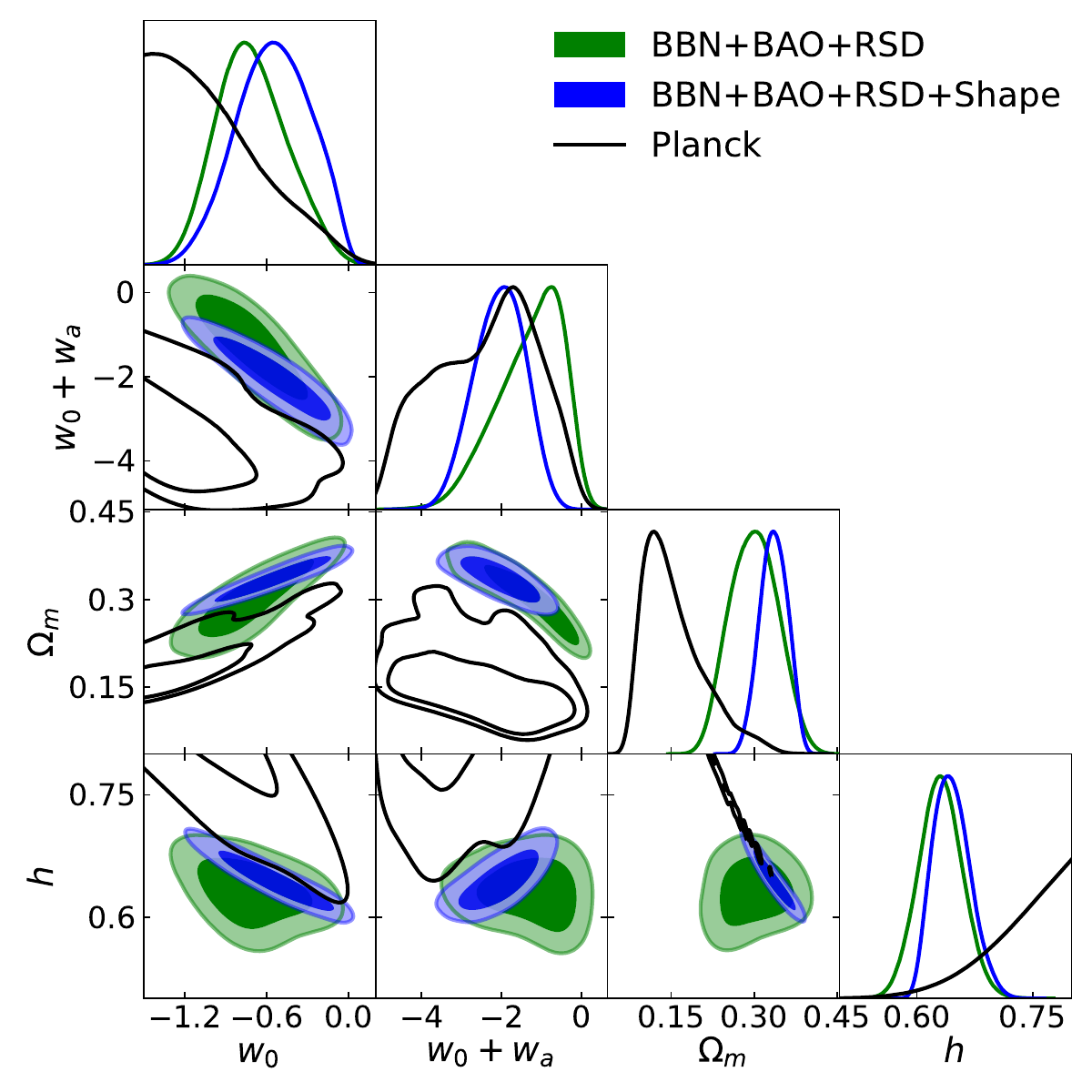}
    \caption{Left panel: Cosmological constraints within the $w_0$CDM model. Right panel: Cosmological constraints within the $w_0 w_a$CDM model.}
    \label{fig:cosmo-w0wa}
\end{figure}

We show the parameter constraints for the $(w\mathrm{CDM})$ and $(w_0 w_a\mathrm{CDM})$ cases in the left and right panel of figure \ref{fig:cosmo-w0wa}, respectively. Again, we display the results of the classic fit in green, ShapeFit in blue and Planck in black, where for Planck we include the full temperature, polarization and lensing data, as described in section \ref{sec:data-early-time}. 

Our findings are analog to the case of allowing for curvature in section \ref{sec:cosmo-Ok}. Although varying the dark energy equation of state does not lead to a change in galaxy power spectrum shape,\footnote{In fact, varying dark energy does change the shape of the power spectrum at the scale of equality between matter and dark energy. However this scale is of order $k\lesssim 10^{-3}\,h/$Mpc and is thus not observable.} the measurement of $m$ breaks the degeneracy of $w_0$ with $\Omega_\mathrm{m}$ and $h$. Hence, with  LSS-only information, 
ShapeFit yields  a constraint on the equation of state of $w_0 = -1.007_{-0.073}^{+0.083}$, is nearly as precise as that from the combination of Planck with the classic BAO and RSD ($w_0 = -1.090_{-0.041}^{+0.050}$), and a factor $\sim 1.5$ improvement with respect to the classic fit result of $w_0 = -0.81 \pm 0.12$. 

The same observation holds for the $(w_0 w_a\mathrm{CDM})$ case. The exact numbers are reported in table~\ref{tab:comparison-with-classic}. There, we also show the results after combining Planck with our LSS dataset, finding that Planck dominates over the {\it Shape} constraints similar to the case of varying curvature. We conclude that, due to their high degree of complementarity to Planck, the information contained in BAO and RSD is sufficient to constrain cosmological parameters globally, i.e., by combining disparate probes of the universe.

\begin{figure}[htb]
    \centering
    \includegraphics[width=\textwidth]{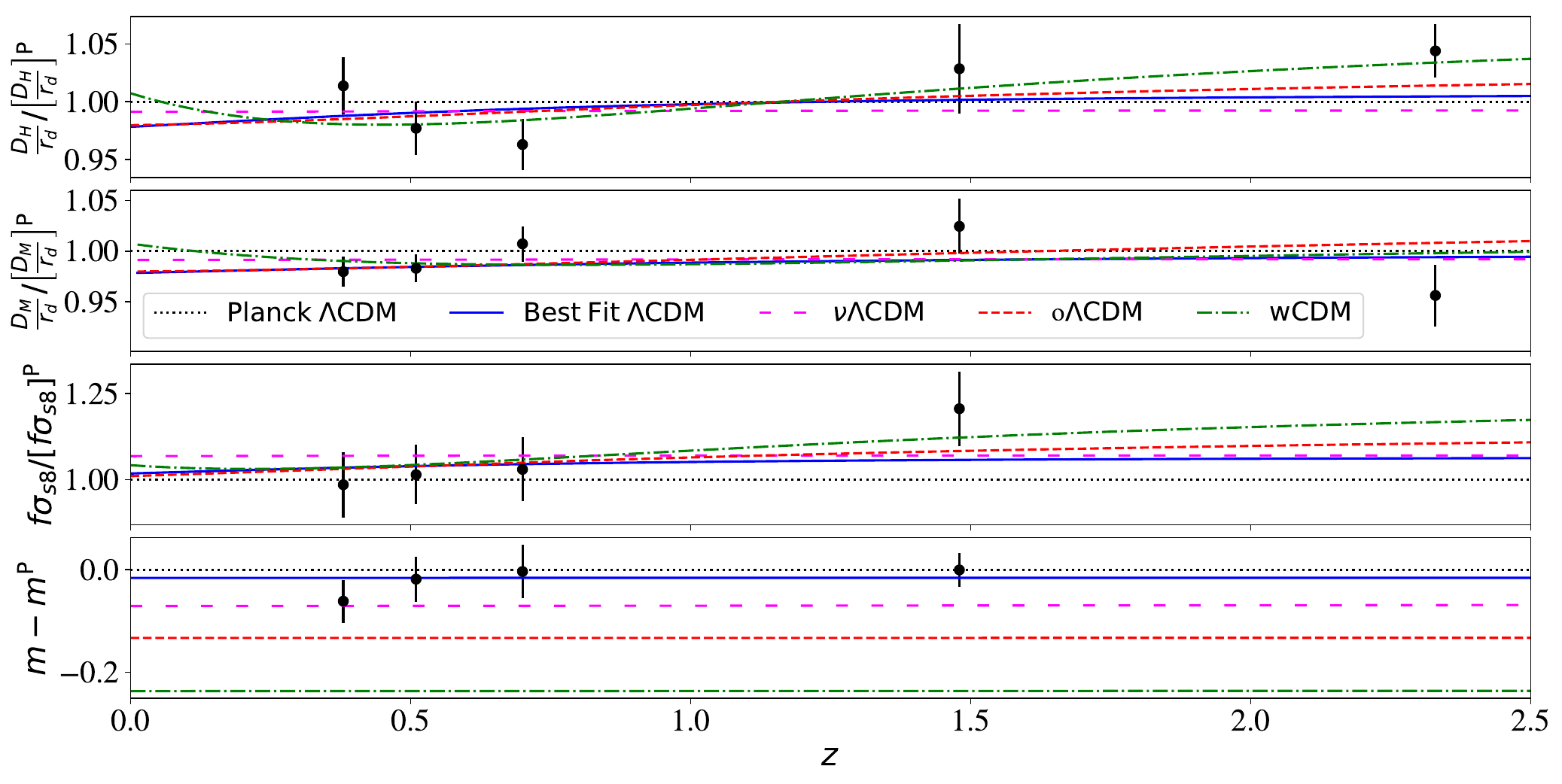}
    \caption{Comparison of several models with respect to the compressed dataset. Each panel shows the redshift dependence (colored lines) and measurements (black data points) of the compressed variables rescaled to the Planck (labeled `P') $\Lambda$CDM prediction (black dotted lines). The blue, continuous lines represent the best-fit $\Lambda$CDM prediction to the ShapeFit dataset. The magenta, sparsely dashed lines correspond to a $\nu\Lambda$CDM model with $\Sigma m_\nu = 0.40\,\mathrm{eV}$ that is excluded by the ShapeFit dataset at $2\sigma$. Finally, we show the o$\Lambda$CDM (red, dashed lines) and wCDM (green, dash-dotted) models, that deliver the best-fit to the classic dataset, but are vastly excluded by the measurement of the shape $m$.}
    \label{fig:all_extended}
\end{figure}

Finally, we visualize the main conclusion of this section in figure \ref{fig:all_extended}, where we directly compare the most important extended models investigated in this section  to the compressed data set and their $\Lambda$CDM predictions.

Each subpanel of figure \ref{fig:all_extended} shows one of the compressed, physical parameters with respect to the Planck $\Lambda$CDM prediction displayed via the black dotted lines as a function of redshift. The measurements presented in section \ref{sec:data} and figure \ref{fig:results} are displayed as black data points and the blue continuous line represents the $\Lambda$CDM best-fit to the data set. Note that the information contained here so far is identical to figure \ref{fig:results}. In addition, we show the theoretical predictions from three extensions to the baseline $\Lambda$CDM model, selected in the following way:

\begin{itemize}
    \item $\boldsymbol{\nu\Lambda}$\textbf{CDM} (magenta, sparsely dashed lines): We show the model corresponding to a neutrino mass of $\Sigma m_\nu = 0.4\,\mathrm{eV}$, which is excluded by ShapeFit at the 
    95\% confidence level. This corresponds to the $2\sigma$ edge of the blue contours in the left panel of figure \ref{fig:cosmo-Mnu}.
    \item $\boldsymbol{o \Lambda}$\textbf{CDM} (red, dashed lines): We show the best-fit model to the classic data set, consisting of BAO+RSD only, without the {\it Shape} information, when allowing $\Omega_k$ to vary freely. This corresponds to the best-fit of the green contours in the right panel of figure \ref{fig:cosmo-NeffOk}.  
    \item $\boldsymbol{w}$\textbf{CDM} (green, dash-dotted lines): Again, we show the best-fit model to the classic data set, but when allowing $w_0$ to vary freely. This corresponds to the best-fit of the green contours in the left panel of figure \ref{fig:cosmo-w0wa}.
\end{itemize}

Figure \ref{fig:all_extended} explicitly shows that -using LSS data alone-  ShapeFit  constrains models that leave in imprint on the power spectrum slope, such as in the $\nu\Lambda$CDM case. In addition, ShapeFit helps to constrain models by lifting parameter degeneracies, even if the parameter extensions themselves do not change the power spectrum slope, such as the $o\Lambda$CDM and $w\Lambda$CDM models. 
For these models in particular, by focusing on the red and green lines on figure~\ref{fig:all_extended}, we can appreciate that the classic parameters are fitted equally well as the concordance $\Lambda$CDM, but these deviations from  $\Lambda$CDM, deliver a  {\it Shape} prediction in strong disagreement with the data. 
Therefore, the shape $m$ is a powerful probe when constraining models using LSS data alone. 

Note that figure \ref{fig:all_extended} is similar to figures 2 and 7 of the eBOSS cosmological results paper \cite{eboss_collaboration_dr16}, but complementary in the parameter selection of the  extended models shown. While their parameter choices are tuned to deliver a `good fit' to Planck, but a `bad fit' to their presented dataset, here we tune the parameters of $\Sigma m_\nu$, $\Omega_k$ and $w_0$ (and the remaining $\Lambda$CDM parameters) the other way around. As mentioned before, we select them such that the BAO and RSD compressed variables are fit well, but are in vast disagreement with our {\it Shape} measurement, (which is in agreement with Planck). 

\subsection{Consistency between individual tracers}
\label{sec:cosmo_highzvslowz}

\begin{figure}[htb]
    \centering
    \includegraphics[width=0.6\textwidth]{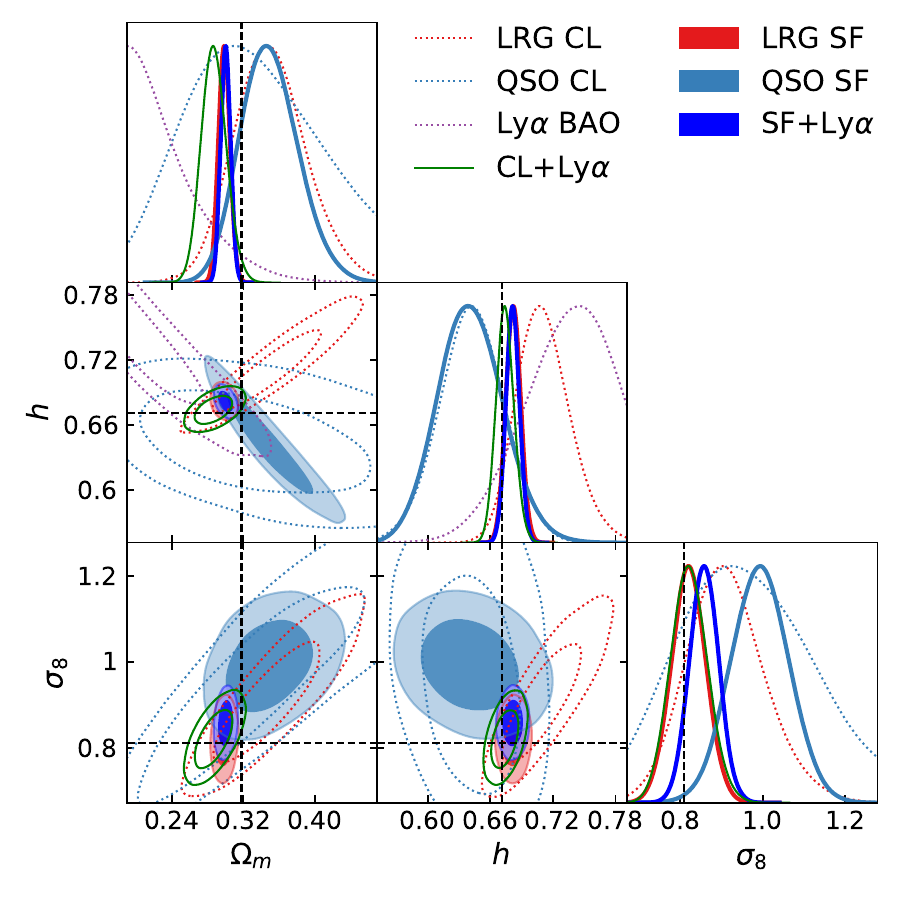}
    \caption{Comparison of cosmological constraints for different tracers. Empty contours show the classic (CL) analysis (BAO+RSD for galaxies, BAO for Lyman-$\alpha$)   $\Lambda$CDM results for LRGs (red), QSOs (light blue), Ly-$\alpha$ (purple) and all combined (green). Filled contours show the corresponding ShapeFit (SF) $\Lambda$CDM results via the same color scheme, but with combined results shown in blue. Note that the green and blue contours here represent the same cases as in figure \ref{fig:cosmo}. In all cases a BBN prior on $\omega_\mathrm{b}$ is assumed. Dashed black lines indicate the Planck best-fit cosmology. }
    \label{fig:highzlowz}
\end{figure}

We investigate the consistency between the different BOSS and eBOSS tracers (LRG's, QSO's, Ly$\alpha$) within the baseline $\Lambda$CDM model.

In figure \ref{fig:highzlowz} we show cosmological constraints from the three LRG samples at redshifts $0.2<z<1.0$ (red), the QSO sample at $0.8<z<2.2$ (light blue) and the Ly$\alpha$ forest at $1.8<z<4.0$ (purple) each, including a BBN prior. The results are shown both for the classic fit (empty dotted contours, labeled `CL') and ShapeFit (filled contours, labeled `SF') for comparison. Additionally, we show the cosmological constraints when combining all samples (green and dark blue contours) already presented in section \ref{sec:cosmo-baseline}. The best-fit Planck cosmology is indicated by black dashed lines.

%From the figure 
We can appreciate the strong tightening of constraints due to the {\it Shape} for the LRG and QSO samples, especially in the $\Omega_\mathrm{m}-h$ plane. As expected, this effect is stronger for the individual samples than for the combined ones, because already in the classic case the individual samples are highly complementary in their cosmological parameter degeneracy directions. Thus the {\it Shape} becomes less important the larger the analysed redshift range is. Also note that the Ly$\alpha$ forest plays a crucial role breaking the degeneracies in the $\Omega_{\rm m}-h$ plane for the classic case. This is not the case for ShapeFit where the degeneracy is broken even across a smaller redshift baseline (see the consistency between red and dark blue contours).

In the $\Omega_\mathrm{m}-h$ plane the consistency between LRG's and QSO's for ShapeFit is remarkable. However,  the $\sigma_8$ parameter recovered by ShapeFit LRG's and QSO's reveals a discrepancy of $2.1 \sigma$. This should be compared to the $3.6 \sigma$ tension between the same samples reported in Neveux et al. \cite{Neveuxetal2021}, who find,
\begin{align}
\sigma_{8,z=0}^\mathrm{LRG} = 0.760 \pm 0.046 \qquad \sigma_{8,z=0}^\mathrm{QSO} = 1.12 \pm 0.10 \qquad [\mathrm{Neveux}~\mathrm{et}~\mathrm{al.}]~,
\end{align}
whereas (using the same Gaussian priors on $\omega_\mathrm{b}$ and $n_s$ as in \cite{Neveuxetal2021}) we find,

\begin{align}
\sigma_{8,z=0}^\mathrm{LRG} = 0.814 \pm 0.043 \qquad \sigma_{8,z=0}^\mathrm{QSO} = 0.993 \pm 0.071 \qquad [\mathrm{this} ~\mathrm{work}]~.
\end{align}

While the LRG results of both methods can not be directly compared to each other, since \cite{Neveuxetal2021} do not include the BOSS medium redshift sample at $z_\mathrm{eff}=0.51$ and neither the BAO post-recon information, the $1\sigma$ discrepancy in $\sigma_8$ for QSO alone is surprising. We suggest that it is related to the differences between the FM approach of \cite{Neveuxetal2021} and ShapeFit, coming from the lack of constraining power of the quasar sample alone. Nevertheless, after combining the samples and covering a larger redshift range, our results are well in agreement with \cite{Neveuxetal2021}, as shown in table \ref{tab:other_results}.

\subsection{Comparison with other approaches} \label{sec:cosmo-comparison}
In this section we compare the cosmological results presented in sections \ref{sec:cosmo-baseline} and \ref{sec:cosmo-ext} to the official eBOSS collaboration results \cite{eboss_collaboration_dr16} and the results obtained by other, independent groups, who have used the same BOSS+eBOSS dataset (or a subset of it). As the initial motivation of ShapeFit has been to provide a bridge between fixed-template (classic) and varying-template (full modeling) approaches \cite{BriedenPRL21}, how ShapeFit compares to each of these very different ways of interpreting spectroscopic galaxy surveys is of particular interest. 

\subsubsection{Full modeling approaches}
\label{sec:cosmo-comparison-FM}

We compare our main results in  light of the standard flat-$\Lambda$CDM model with three free parameters, $\{\Omega_{\rm m},\,H_0,\,A_s\}$ with the most recent studies applied to the BOSS and eBOSS samples using different full modelling approaches. The first two rows of table~\ref{tab:other_results} display the best-fitting $\Omega_{\rm m},\,H_0,\,A_s$,\footnote{The $\sigma_8$ is  a derived parameter, obtained from  the values of the $\{\Omega_{\rm m},\,H_0,\,A_s\}$ parameters. However, since some authors choose to show it (instead of reporting $A_s$) we decide to display it as well.} values for the ShapeFit approach, with and without assuming a BBN prior on $\omega_{\rm b}$, and with a fixed value of $n_s$ to its reference value as described in table~\ref{tab:cosmo-priors}. This is followed by the results of full modelling studies, of which four have been applied to the BOSS DR12 LRG data, and two to the QSO and LRG eBOSS DR16 data (which contains the BOSS DR12 data). In addition, we also include the parameters derived from applying ShapeFit to these samples. We briefly describe below the main assumptions of these different analyses and compare them to our findings.

\begin{table}[htb]
    \centering
    %\small
    \footnotesize
    \begin{tabular}{|c|c|c|c|c|c|c|}
    \hline
        Sample & Priors & Method / Ref. & $\Omega_{\rm m}$ & $H_0\mathrm{\left[\frac{km/s}{Mpc}\right]}$ & $A_s\cdot10^9$ & $\sigma_8$  \\
        \hline
        \hline

  \multirow{2}{*}{Full} & $n_s$ &  \textbf{ShapeFit}  &  $0.2971 \pm 0.0061$ & 
  %$0.7125$ (unc.) &
  $-$ & $2.39_{-0.43}^{+0.24}$ & $0.857 \pm 0.040$ \\
  & $\omega_\mathrm{b},n_s$ &  \textbf{ShapeFit}  & $0.3001 \pm 0.0057$ & $68.16\pm0.67$ & $2.43\pm0.20$ & $0.858\pm0.036$ \\
 \hline 
 & $\frac{\omega_\mathrm{b}}{\omega_\mathrm{m}}, n_s$ &   D'Amico \citep{DAmico:2019fhj} & $0.309 \pm 0.010$ & $68.5 \pm 2.2$ & $1.52 \pm 0.84$ & $-$ \\
 & $\omega_\mathrm{b},n_s$ &   Ivanov \citep{Ivanov:2019pdj} & $0.295 \pm 0.010$ & $67.9 \pm 1.1$ & $-$ & $0.721\pm 0.043$ \\
LRG & $\omega_\mathrm{b},n_s$ &    Philcox \citep{2020JCAP...05..032P} & $0.2962^{+0.0082}_{-0.0080}$ & $67.81^{+0.68}_{-0.69}$ & $-$ & $0.739^{+0.040}_{-0.041}$ \\
DR12 & $\omega_\mathrm{b}$ &  Tr\"oster \citep{Trosteretal:2020} & $0.317^{+0.015}_{-0.019}$ & $70.4\pm2.4$ & $-$ & $0.71\pm0.049$ \\
      & $\omega_\mathrm{b},n_s$ &  Chen \citep{shi-fan} & $0.3030\pm0.0082$ & $69.23\pm0.77$ & $-$ & $0.733\pm0.047$ \\
   & $n_s$ &  ShapeFit \cite{BriedenPRL21} & $0.295 \pm 0.014$ & $-$ & $2.56\pm0.51$ & $0.806\pm0.065$ \\
   %& $\omega_\mathrm{b},n_s$ &  ShapeFit \cite{BriedenPRL21} & $0.297 \pm 0.014$ & $66.0\pm1.7$ & $2.63\pm0.46$ & $0.806\pm0.063$ \\
   \hline
 LRG  & $\omega_\mathrm{b},n_s$ &  Neveux \citep{Neveuxetal2021}  & $0.315\pm0.013$ & $66.9 \pm 1.9$ & $-$ & $0.763\pm0.046$ \\
 DR16 & $\omega_\mathrm{b},n_s$ &  \textbf{ShapeFit}  & $0.2984 \pm 0.0066$ & $68.20\pm0.73$ & $2.24\pm0.24$ & $0.820\pm0.043$ \\
  \hline
 QSO  & $\omega_\mathrm{b},n_s$ &  Neveux \citep{Neveuxetal2021}  & $0.321\pm0.016$ & $65.1 \pm 1.9$ & $-$ & $1.12\pm 0.10$ \\
 DR16 & $\omega_\mathrm{b},n_s$ &  \textbf{ShapeFit}  & $0.350 \pm 0.033$ & $64.1\pm3.1$ & $3.25\pm0.47$ & $0.993\pm0.072$ \\
  \hline
 LRG  & $\omega_\mathrm{b},\omega_\mathrm{cdm},n_s$ &   Semenaite \citep{Semenaiteetal2021} & $0.3037\pm 0.0081$ & $68.55^{+0.84}_{-0.94}$ & $-$ & $0.800\pm 0.039$ \\
 +QSO  & $\omega_\mathrm{b},n_s$ &   Neveux \citep{Neveuxetal2021}  & $0.308\pm0.010$ & $66.4 \pm 1.4$ & $-$ & $0.869\pm0.046$ \\
 DR16 & $\omega_\mathrm{b},n_s$ &  \textbf{ShapeFit}  & $0.3012 \pm 0.0057$ & $68.24\pm0.67$ & $2.42\pm0.20$ & $0.860\pm0.036$ \\
  \hline

    \end{tabular}
    \caption{Comparison of ShapeFit results with other full modelling approaches when analyzing BOSS and eBOSS samples. The first column shows the used data sets in each approach, where `Full' refers to the final release of the BOSS+eBOSS dataset shown in figure \ref{fig:results} (LRG+QSO+Ly$\alpha$); LRG DR12 refers to the BOSS LRG sample, $0.2<z<0.75$; LRG DR16 to the whole BOSS and eBOSS LRG samples, $0.2<z<1.0$; and QSO DR16 to the whole eBOSS quasar sample, $0.8<z<2.2$. The second column shows which parameters have either been fixed or varied within a Gaussian prior in each analysis (see text for details). The first two rows (and also those rows labeled in bold) display the best-fitting parameters of this work either with or without the prior on $\omega_{\rm b}$, motivated by BBN. Additionally, we display the results of other studies analyzing part of the BOSS and eBOSS data described here. All the quoted analyses assume a flat-$\Lambda$CDM model, but D'Amico et al, Ivanov et al. and Philcox et al. (along with our results) have fixed the $n_s$ parameter to Planck best-fit finding, and Tr\"oster et al., Semenaite et al. and Neveux et al., allow $n_s$ to vary within a certain range (see text). This difference may considerably enlarge the error-bars on some of the $\Lambda$CDM parameters.}
    \label{tab:other_results}
\end{table}

In D'Amico et al. \cite{DAmico:2019fhj} the authors employ the EFT \cite{Baumanetal12,Carrascoetal12} approach to model the power spectrum monopole and quadrupole signals of the BOSS DR12 LRG sample split in two redshift bins: the so called LOWZ sample, $0.15<z<0.43;\,z_{\rm eff}=0.32$ and the CMASS sample, $0.43<z<0.70;\,z_{\rm eff}=0.57$, considering the scales $k\leq0.20\,{\rm Mpc}^{-1}h$ for CMASS and $k\leq0.18\,{\rm Mpc}^{-1}h$ for LOWZ. For CMASS, both northern and southern samples are considered, whereas for the LOWZ sample only the northern cap is analyzed. The $\omega_{\rm b}/\omega_{\rm m}$ ratio and $n_s$ index are assumed fixed at Planck's best-fit values. 

Ivanov et al. \cite{Ivanov:2019pdj} also use EFT to model the power spectrum monopole and quadrupole signals of BOSS DR12 applying the following redshift binning, $0.2<z<0.5;\,z_{\rm eff}=0.38$ and $0.5<z<0.75;\,z_{\rm eff}=0.61$; and considering the scales $k\leq0.25\,{\rm Mpc}^{-1}h$, using both northern and southern galactic caps. An informative prior on $\omega_{\rm b}$ is employed motivated Planck observations, and $n_s$ is fixed to Planck's best-fit value. The neutrino mass is varied within the narrow range of (0.06-0.18) eV. 

Compared to Ivanov et al. and D'Amico et al. results, our value for $A_s$ (or $\sigma_8$) is significantly larger. This is likely due to the fact that in their analysis the (necessary) re-normalization of the window function is ignored. We find, however, a very good agreement for $\Omega_{\rm m}$ as well as for $H_0$, which are less affected by this effect. Our analysis returns significantly smaller error-bars: a factor 1.8 smaller for $\Omega_{\rm m}$, a factor of 1.6 smaller for $H_0$ compared to  Ivanov et al., a factor of 3.2 smaller for $H_0$ compared to D'Amico et al., because our analysis employs a larger sample, $0.2<z<4.0$, and it employs pre- and post-recon catalogues coherently, when available. The weaker constrain on $H_0$ from D'Amico et al. \cite{DAmico:2019fhj} with respect to Ivanov et al. \cite{Ivanov:2019pdj} is due to the type of prior used:  D'Amico et al. fix the $\omega_{\rm b}/\omega_{\rm m}$ ratio, whereas Ivanov et al. the baryon density $\omega_{\rm b}$ (see the effect of this prior choice in Appendix B of \cite{ShapeFitPT}).

Philcox et al. \citep{2020JCAP...05..032P} use the same BOSS DR12 sample, and the same piors on $w_b$, $\Sigma m_{\nu}$ and $n_s$, and a similar EFT model as described in Ivanov et al., but additionally include the signal from the post-reconstructed catalogues which improves the BAO constraint. This allows them to better constrain $H_0$ and $\Omega_{\rm m}$ with respect to the corresponding pre-reconstructed study of Ivanov et al.. Compared to Philcox et al., ShapeFit with a BBN prior yields error-bars tighter by a factor of 1.4 for $\Omega_{\rm m}$, equal for $H_0$, and tighter by a factor of $1.1$ for $\sigma_8$. As before the disagreement in the best-fit value for $\sigma_8$ is caused by the (non) re-normalization of the window in their analysis. In general the agreement with the obtained $\Omega_{\rm m}$ and $H_0$ best-fit values is very good. 

Tr\"oster et al. \cite{Trosteretal:2020} use the same BOSS DR12 samples as in Ivanov et al., but instead of employing the power spectrum monopole and quadrupole signals they choose to model three LOS-wedges of the correlation function. They employ a perturbation theory model, gRPT, for describing the real-space statistics, combined with the TNS model from \cite{Taruya:2010mx}, and assume the usual local Lagrangian relations for the non-local biases. They use an informative but wider-than-usual prior on $\omega_{\rm b}$ (about 10 times wider than the usual BBN prior from \cite{Cooke:2018}). This choice impacts the error-bars on $H_0$, which are much wider than the above studies. They also choose to use an uninformative wide prior on $n_s$ which also affects $H_0$, as well as $\Omega_{\rm m}$ and $\sigma_8$. Similarly to the above studies, Tr\"oster et al. does not account for the necessary normalization of the window which impacts the recovered value of $\sigma_8$.

Chen et al. \cite{shi-fan} use the same BOSS DR12 samples as in Ivanov et al., and as Philcox et al. they combine both pre- and post-recon catalogue information by simultaneously fitting the pre-recon power spectrum and post-recon correlation function. They fix $n_s\,\,\omega_{\rm b},\,\Sigma m_\nu$ to fixed values consistent with Planck results. Unlike the previous works they employ an implementation of the Lagrangian Perturbation theory \cite{matsubara2007,Bernardeauetal2002} to model both two-point statistics multipoles, and properly account for the normalization of the window of the power spectrum. We find that our results are in very good agreement for $\Omega_{\rm m}$ and $H_0$, whereas their $\sigma_8$ best-fit value is $\sim 1\sigma$ lower than ours for this sample. 

For completeness we also include the results derived from applying ShapeFit to the BOSS DR12 LRG sample, divided in two bins as in Ivanov et al, as described in \cite{BriedenPRL21}. Unlike the other studies, we do not apply any prior on $\omega_{\rm b}$ which only allows us to report constraints on $\Omega_{\rm m}$ and $A_s$. The difference in $\sigma_8$ is partially (although not completely) given by the normalization of the survey window, and the agreement for $\Omega_{\rm m}$ with the other studies is very good.
In addition, we note that our value for $\sigma_8$ for the full sample is fully consistent with the one found by the BOSS and eBOSS collaborations, as reported later in Table~\ref{tab:comparison-with-classic-lcdm}. We do not further discuss the low values of $\sigma_8$ reported by some of the references reported in  Table~\ref{tab:other_results}, as it goes beyond the scope of this analysis.

Neveux et al. \citep{Neveuxetal2021} use the power spectrum monopole, quadrupole and hexadecapole signals of the BOSS DR12 LRG sample between $0.2<z<0.5$, the BOSS and eBOSS LRG samples between $0.6<z<1.0$ and the eBOSS quasar sample between $0.8<z<2.2$. Unlike our main result, they do not employ any reconstruction data, nor the Ly-$\alpha$ measurements. They describe the data using the perturbation theory implementation RegPT \cite{Taruyaetal12}, along with the TNS model \cite{Taruya:2010mx} under the assumption of local Lagrangian for the non-local biases. We choose to display their results with priors on $\omega_{\rm b}$ and $n_s$, to make the analysis setting closer to ours, and also display the results for the LRG sample and QSO sample alone. For comparison we also report our results on the same samples (see section~\ref{sec:cosmo_highzvslowz} for a comparison of $\sigma_8$ under the same prior conditions).

For the whole LRG+QSO sample we find good consistency with our results on $\Omega_{\rm m}$ and $\sigma_8$. However, Neveux et al. reports a value for $H_0$ which is about $1\sigma$ smaller than our finding. This difference can be caused by the slightly different choice of priors, as well as the difference in the data-catalogue selection: we employ the post-reconstructed catalogues, as well as the galaxies between $0.5<z<0.6$. The impact of this can be more clearly seen on the size of the error-bars inferred from the LRG DR16 sample alone, where ours are significantly smaller. The most striking result arise when the QSO sample alone is considered, as the errorbars on $\Omega_{\rm m}$ and $H_0$ are substantially smaller in Neveux et al. compared to our approach: a factor of 2 and a factor of 1.6, respectively. On the other hand the value on $\sigma_8$ found by Neveux et al. is 1.4 times looser than ours. These differences deserve a more careful investigation we may address in future work. 

Semenaite et al. \cite{Semenaiteetal2021} use the data from BOSS DR12 as in Ivanov et al, along with the eBOSS quasar sample between $0.8<z<2.2$. They do not use any post-reconstructed catalogue, nor eBOSS LRG data, nor Lyman-$\alpha$ data. They model three LOS-wedges of the correlation function using a perturbation theory implementation (\textsc{Respresso} \cite{respresso,Taruyaetal12}), where the velocity power spectra are given by \cite{Beletal19} and the redshift-space distortions are modelled according to the TNS model \cite{Taruya:2010mx}. Instead of using the local Lagrangian bias relations they choose to use another approach motivated by the findings in \cite{Sheth_Chan_Scoccimarro13}. We choose to display their results when Gaussian priors are used around $\omega_{\rm b}$, $\omega_c$ and $n_s$, as this is the closest choice to our type of analysis. We find consistent results on $\Omega_{\rm m}$ and $H_0$, but a $\sim1\sigma$ discrepancy for $\sigma_8$. Their error-bars are larger than ours, partly because of not fixing $n_s$, but also because not employing the reconstructed catalogues, nor using the eBOSS LRG or Ly-$\alpha$ datasets.

\subsubsection{Classic approaches}
\label{sec:cosmo-comparison-classic}
The comparison  with other analyses that used the classic approach, and in particular the final cosmological analysis of eBOSS \cite{eboss_collaboration_dr16} serves two main purposes. We aim to \textit{i)} show that interpreting our classic dataset (without the {\it Shape}) delivers the same cosmological results as in \cite{eboss_collaboration_dr16}, \textit{ii)} quantify the power of {\it Shape} in constraining cosmological models more in detail than done in sections \ref{sec:cosmo-baseline} and \ref{sec:cosmo-ext} and \textit{iii)} quantify the differences among the compression methods after combining with the Planck likelihood introduced in section \ref{sec:data-early-time}.

All results of this paper are shown in comparison to the official BOSS+eBOSS counterpart (labeled `(e)BOSS') in table \ref{tab:comparison-with-classic-lcdm} for the $\Lambda$CDM model and  \ref{tab:comparison-with-classic} for $\Lambda$CDM extensions,  in the same format as table 4 of \cite{eboss_collaboration_dr16}. Note that for the fits without Planck, \cite{eboss_collaboration_dr16} use only the BAO signal, while for our classic fits we always include the BAO and RSD signals. However, for the runs labeled `Planck+(e)BOSS' we use publicly available fits from \cite{eboss_collaboration_dr16} labeled `CMBLens+BAORSD' therein, which include both the BAO and RSD signals. Also, note that all results presented in this table do not rely on the BBN prior, which is the reason why $H_0$ remains unconstrained for all runs that do not include the Planck likelihood.

Concerning \textit{i)}, we have already shown in section \ref{sec:results_comparison} that the agreement with \cite{eboss_collaboration_dr16} at the level of compressed variables is very good. However,  we investigate whether the small residual differences leak into a measurable difference at the level of cosmological model parameters. Note that the dataset we use in this paper is slightly different from that in \cite{eboss_collaboration_dr16}, where they also include isotropic BAO and RSD (assuming fixed BAO) measurements from the MGS sample and the anisotropic BAO+RSD measurement of the ELG sample at effective redshifts $z=0.15$ and $z=0.85$ respectively. We have monitored the effect of this deviation for the $\Lambda$CDM and $\nu\Lambda$CDM models. We find that by including the ELG and MGS data in our pipeline we recover exactly the same $\Omega_\mathrm{m}$ and $\Sigma m_\nu$ constraints as \cite{eboss_collaboration_dr16}. We therefore conclude that the main discrepancy between the official (e)BOSS and our classic results, which always remain well within $1\sigma$, are attributed to this slightly different choice.

About \textit{ii)}, in general, we see that ShapeFit tends to yield slightly larger values of $\Omega_\mathrm{m}$ than the classic fit. Also, the constraining power improves significantly when adding the shape $m$, by a factor of about $2.5, 3, 1.5 $ for $\Omega_\mathrm{m}$, $\Omega_\mathrm{k}$ and $w_0$, respectively. In this way, ShapeFit delivers competitive evidence for a flat universe ($\Omega_\mathrm{k}=0$) and for a standard cosmological constant ($w_0=-1$). In the case of $\nu\Lambda$CDM, ShapeFit has the ability to constrain the sum of neutrino masses, which the classic fit is not sensitive to.

\begin{table}[htb]
    \centering
    \begin{tabular}{|L|C|C|C|C|}
 \hline
 & \Omega_\mathrm{m} & H_0\mathrm{\left[\frac{km/s}{Mpc}\right]} & A_s\times 10^9 & \sigma_8 \\  \hline \hline
 \mathrm{(e)BOSS ~ BAO} & 0.299\pm 0.016 & - & - & -\\
 \mathrm{Classic} ~ \mathrm{[this ~ work]} & 0.287\pm 0.014 & - & 2.41_{-0.41}^{+0.26} & 0.828\pm0.048 \\
 \mathrm{ShapeFit} ~ \mathrm{[this ~ work]} & 0.2971\pm 0.0061 & - & 2.39_{-0.43}^{+0.24} & 0.857\pm0.040  \\
 \hline
 \mathrm{BBN+(e)BOSS ~ BAO} & 0.299\pm 0.016 & 67.35\pm0.97 & - & -\\
 \mathrm{BBN+(e)BOSS} & 0.297^{+0.014}_{-0.016} & 67.23\pm0.95 & 2.57^{+0.29}_{-0.34} & 0.850\pm0.033\\
 \mathrm{BBN+Classic} & 0.287\pm 0.014 & 67.42_{-0.91}^{+0.84} & 2.50_{-0.25}^{+0.22} & 0.822\pm0.044 \\
 \mathrm{BBN+ShapeFit} & 0.3001\pm 0.0057 & 68.16\pm0.67 & 2.43\pm0.20 & 0.858\pm0.036  \\ 
 \hline
 \mathrm{Planck} & 0.3178\pm 0.0079 & 67.13\pm 0.56 & 2.1_{-0.032}^{+0.028} & 0.8101_{-0.0061}^{+0.0062}  \\
% \mathrm{Planck + (e)BOSS ~ BAO} & 0.3109\pm 0.0053 & 67.68\pm 0.40 & - & -  \\
  \mathrm{Planck + (e)BOSS} & 0.3109\pm 0.0054 & 67.68\pm 0.40 & 2.113^{+0.027}_{-0.031} & 0.8117^{+0.0055}_{-0.0060}  \\
 \mathrm{Planck + Classic} & 0.3081\pm 0.0050 & 67.83\pm 0.38 & 2.117_{-0.033}^{+0.029} & 0.8089_{-0.0064}^{+0.0060} \\
 \mathrm{Planck + ShapeFit} & 0.3067\pm 0.0047 & 67.94 \pm 0.36 & 2.121_{-0.033}^{+0.030} & 0.8091_{-0.0065}^{+0.0057}  \\ \hline 
    \end{tabular}
    \caption{$\Lambda$CDM model: Comparison of our Classic and ShapeFit cosmological constraints with the official BOSS+eBOSS results that include either the BAO only signal (labeled `(e)BOSS BAO')  or the full BAO+RSD signals (labeled `(e)BOSS'). We present mean values with 68\% C.L.}
    \label{tab:comparison-with-classic-lcdm}
\end{table}

\begin{landscape}
\begin{table}[t]
    \centering
    \small
    \makebox{
    \begin{tabular}{|L|L|C|C|C|C|C|C|C|}\hline
& &  \Omega_\mathrm{m} & H_0\mathrm{\left[\frac{km/s}{Mpc}\right]} &\Sigma m_\nu\,[\mathrm{eV}] & N_\mathrm{eff} & \Omega_k & w_0 & w_a \\
\hline
 \hline
 \multirow{6}{*}{$\nu \Lambda \mathrm{CDM}$} & \mathrm{ShapeFit}~ \mathrm{[this ~ work]} & 0.300_{-0.011}^{+0.008} & - & <0.54 & - & - & - & -\\
 & \mathrm{BBN} + \mathrm{ShapeFit}& 0.302_{-0.010}^{+0.007} & 68.03\pm0.68 & <0.40 & - & - & - & -\\
 & \mathrm{Planck} & 0.321_{-0.015}^{+0.009} & 66.95_{-0.68}^{+1.1} & <0.26 & - & - & - & -\\
 & \mathrm{Planck + (e)BOSS} & 0.3089\pm 0.0058 & 67.87\pm 0.45 & <0.10  & - & - & - & -\\
 & \mathrm{Planck + Classic} & 0.3052\pm 0.0052 & 68.14\pm 0.40 & <0.085 & - & - & - & -\\
 & \mathrm{Planck + ShapeFit} & 0.3034\pm 0.0049 & 68.28\pm 0.39 & <0.082 & - & - & - & -\\
 \hline
  %& \mathrm{ShapeFit} & 0.701\pm 0.016 & - & - & - & - & - & -\\
\multirow{3}{*}{$N_\mathrm{eff} \Lambda \mathrm{CDM}$} & \mathrm{Planck} & 0.321\pm 0.011 & 66.5^{+1.4}_{-1.7} & - & 2.94_{-0.24}^{+0.21} & - & - & -\\
 & \mathrm{Planck + Classic} & 0.3066 \pm 0.0060 & 68.5 \pm 1.3 & - & 3.16 \pm 0.22 & - & - & -\\
 & \mathrm{Planck + ShapeFit} & 0.3053 \pm 0.0056 & 68.44 \pm 0.12 & - & 3.12 \pm 0.19 & - & - & -\\
 \hline
\multirow{9}{*}{$\mathrm{o}\Lambda\mathrm{CDM}$} & \mathrm{(e)BOSS ~ BAO} & 0.285 \pm 0.023 & - & - & - & 0.078^{+0.086}_{-0.099}  & - & - \\
 & \mathrm{Classic}~ \mathrm{[this ~ work]} & 0.276_{-0.019}^{+0.021} & - & - & - & 0.054_{-0.092}^{+0.079}  & - & - \\
 & \mathrm{ShapeFit}~ \mathrm{[this ~ work]} & 0.2943_{-0.0092}^{+0.0080} & - & - & - & -0.022_{-0.038}^{+0.032}  & - & - \\
 & \mathrm{BBN} + \mathrm{Classic}~  & 0.279_{-0.021}^{+0.023} & 65.9 \pm 3.5 & - & - & 0.047_{-0.099}^{+0.083}  & - & - \\
 & \mathrm{BBN} + \mathrm{ShapeFit} & 0.2942_{-0.0085}^{+0.0078} & 68.8 \pm 1.1 & - & - & -0.027_{-0.037}^{+0.032}  & - & - \\
 & \mathrm{Planck} & 0.355\pm 0.025 & 63.4^{+2.6}_{-2.1} & - & - & -0.0104 \pm 0.0067 & - & - \\
 & \mathrm{Planck + (e)BOSS} & 0.3105\pm 0.0056 & 67.75\pm 0.56 & - & - & 0.0003\pm 0.0017  & - & - \\
 & \mathrm{Planck + Classic} & 0.3077 \pm 0.0052 & 68.10\pm 0.51 & - & - & 0.0014 \pm 0.0017  & - & - \\
 & \mathrm{Planck + ShapeFit} & 0.3058 \pm 0.0047 & 68.25\pm 0.49 & - & - & 0.0015 \pm 0.0016  & - & - \\
 \hline
\multirow{7}{*}{$w\mathrm{CDM}$} & \mathrm{(e)BOSS ~BAO} & 0.271_{-0.017}^{+0.038} & -  & - & - & - & -0.69\pm 0.15 & -\\
 & \mathrm{Classic}~ \mathrm{[this ~ work]} & 0.279_{-0.016}^{+0.018} & -  & - & - & - & -0.81_{-0.11}^{+0.13} & -\\
 & \mathrm{ShapeFit}~ \mathrm{[this ~ work]} & 0.296 \pm 0.013 & -  & - & - & - & -0.998_{-0.073}^{+0.085} & -\\
%  & \mathrm{BBN+Classic} & 0.279_{-0.015}^{+0.018} & -  & - & - & - & -0.81_{-0.11}^{+0.12} & -\\
 & \mathrm{BBN+ShapeFit} & 0.298 \pm 0.013 & 68.23\pm1.6  & - & - & - & -1.007_{-0.073}^{+0.083} & -\\
% & \mathrm{Planck} & 0.2415_{-0.04}^{+0.015} & - & - & - & - &  -  & - \\
 & \mathrm{Planck   + (e)BOSS} & 0.3039\pm 0.0092 & 68.6\pm 1.0 & -  & - & - & -1.037\pm 0.039  & -\\
 & \mathrm{Planck   + Classic} & 0.2928 \pm 0.0093 & 69.9 \pm 1.2 & -  & - & - & -1.090_{-0.041}^{+0.050}  & -\\
 & \mathrm{Planck   + ShapeFit} & 0.2906 \pm 0.009 & 70.1 \pm 1.2 & -  & - & - & -1.093_{-0.044}^{+0.048}  & -\\
 \hline
\multirow{5}{*}{$w_0w_a\mathrm{CDM}$} & \mathrm{BBN+Classic} & 0.300_{-0.051}^{+0.041} & -  & - & - & - & -0.70_{-0.31}^{+0.23} & -0.58_{-0.76}^{+1.3}\\
 & \mathrm{BBN+ShapeFit}& 0.335 \pm 0.027 & -  & - & - & - & -0.55_{-0.27}^{+0.30} & -1.50_{-0.92}^{+0.96}\\
% & \mathrm{Planck} & 0.1564_{-0.074}^{+0.025} & - & - & - & - &  -  & - \\
 & \mathrm{Planck   + (e)BOSS} & 0.329\pm 0.017 & 66.1\pm 1.7 & -  & - & - & -0.70 \pm 0.19  & -0.99_{-0.52}^{+0.62} \\
 & \mathrm{Planck   + Classic} & 0.333 \pm 0.025 & 65.7^{+2.2}_{-2.6} & -  & - & - & -0.63 \pm 0.26  & -1.29_{-0.69}^{+0.79}\\
 & \mathrm{Planck   + ShapeFit} & 0.330 \pm 0.023 & 66.0^{+2.2}_{-2.5} & -  & - & - & -0.64 \pm 0.25  & -1.27_{-0.65}^{+0.78}\\
 \hline
\end{tabular}}
    \caption{$\Lambda$CDM extensions: Comparison of our Classic and ShapeFit cosmological constraints with the official BOSS+eBOSS results that include the BAO only signal (labeled `(e)BOSS BAO') without Planck and the full BAO+RSD signals (labeled `(e)BOSS') in the fits combined with Planck. We present mean values with 68\% C. L., only in the case of $\Sigma m_\nu$ we present upper limits at 95\% confidence level. We do not display the cases, for which the extended parameters are unconstrained.}
    \label{tab:comparison-with-classic}
\end{table}
\end{landscape}

Regarding \textit{iii)}, when adding Planck data there is very good agreement between our classic and the officially-reported eBOSS results. We have verified that our own Planck-only constraints are in very good agreement with the official Planck results from \cite{Aghanim:2018eyx}. As already anticipated, for the $\nu\Lambda$CDM case in the right panel of figure \ref{fig:cosmo-Mnu}, the improvement of ShapeFit with respect to the classic fit is very modest once Planck is included because the shape information within Planck dominates over our constraint on $m$ from the LSS maps.

Still, this comparison indicates that for upcoming galaxy survey data, for example DESI \cite{aghamousa_desi_2016}, which will measure the {\it shape} more accurately, $m$ might play a significant role for constraining cosmology, even when including Planck.

\section{Systematic checks and  performance on synthetic catalogues} \label{sec:sys}
We aim to validate the robustness of the results presented in section~\ref{sec:cosmo} with respect to variations of the chosen baseline data pipeline, and to quantify the systematic error-budget of the used approach.
As the results are insensitive to reasonable deviations from the adopted assumptions (see Appendix \ref{app:nonlocal_bias}), this section is likely of interest for experts, other readers may omit it at a first sitting. 
  
For the specific performance of the perturbation theory model used we refer to the original ShapeFit paper \cite{ShapeFit} and PT challenge ShapeFit paper \cite{ShapeFitPT}. Both of them present the comparison of the model with respect to full N-body mocks, the Nseries mocks, and the PT challenge mocks, and both samples consist of haloes catalogues which have been populated with galaxies following a Halo Occupation Distribution (HOD) consistent with the LRG catalogues from BOSS data.

\subsection{Galaxy fast-mocks} \label{sec:sys_mocks}

We start by running the same baseline setup used for the data on section~\ref{sec:cosmo}, on the 2048 and 1000 realizations of the Patchy and EZmocks, respectively. These mocks were originally produced for inferring the covariance matrix of the power spectra of the data, the compressed-parameter matrix needed to combine pre- and post-recon parameters, and the correlation among the two overlapping redshift bins. However, we need to bear in mind that these mocks rely on approximate fast techniques and are not full N-body. Consequently, they do not require such high computational resources as a full N-body simulation, at the expense of not being fully accurate when describing the clustering. Therefore, they should not be used to determine the systematic error budget of our approach. Yet, we run the full data pipeline on them, rather than to validate the pipeline, to validate the mocks, as well as to check how consistent data and mocks are in terms of the inferred errorbars.

\begin{figure}[htb]
    \centering
    \includegraphics[scale=0.25]{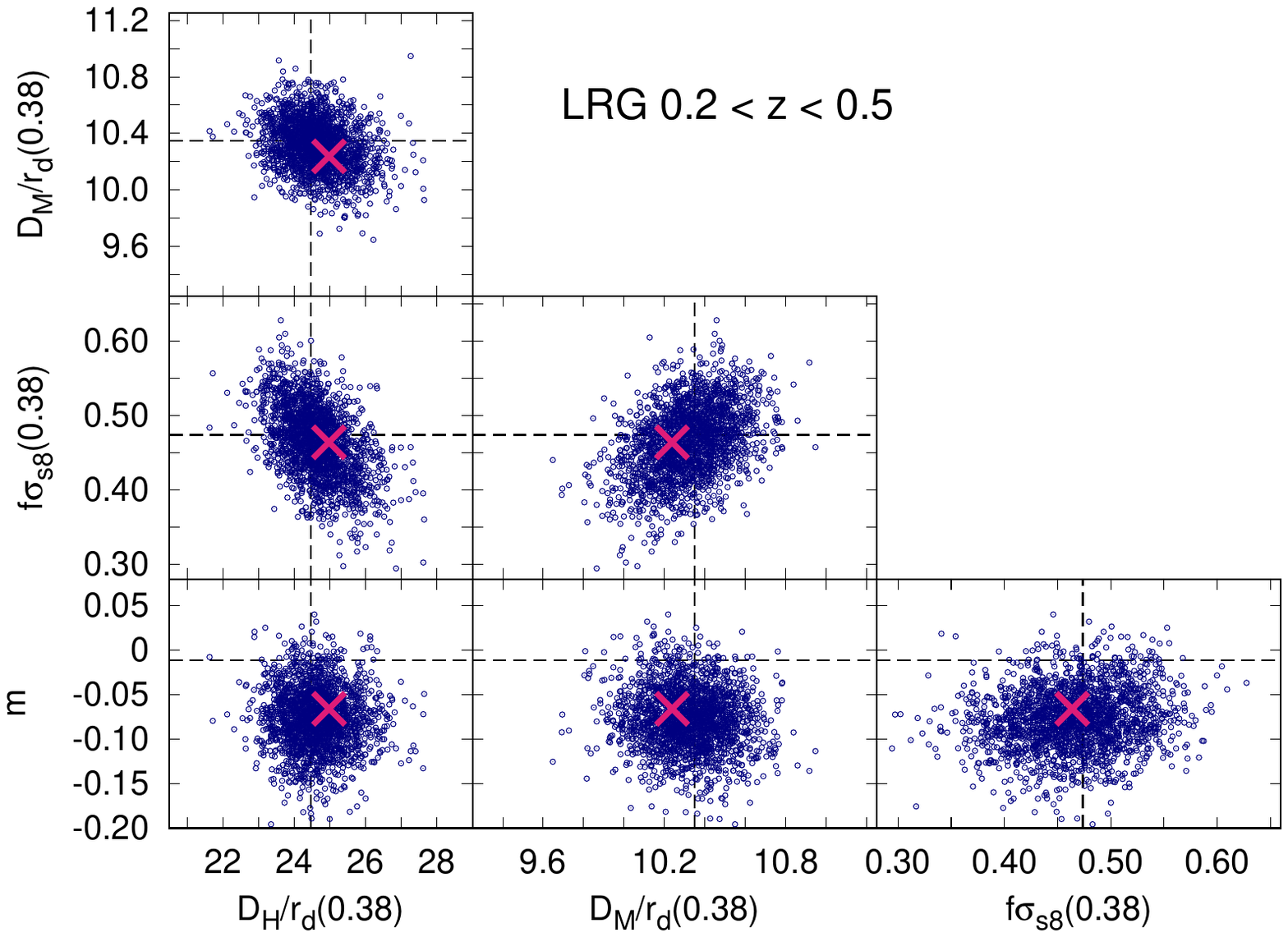}
        \includegraphics[scale=0.25]{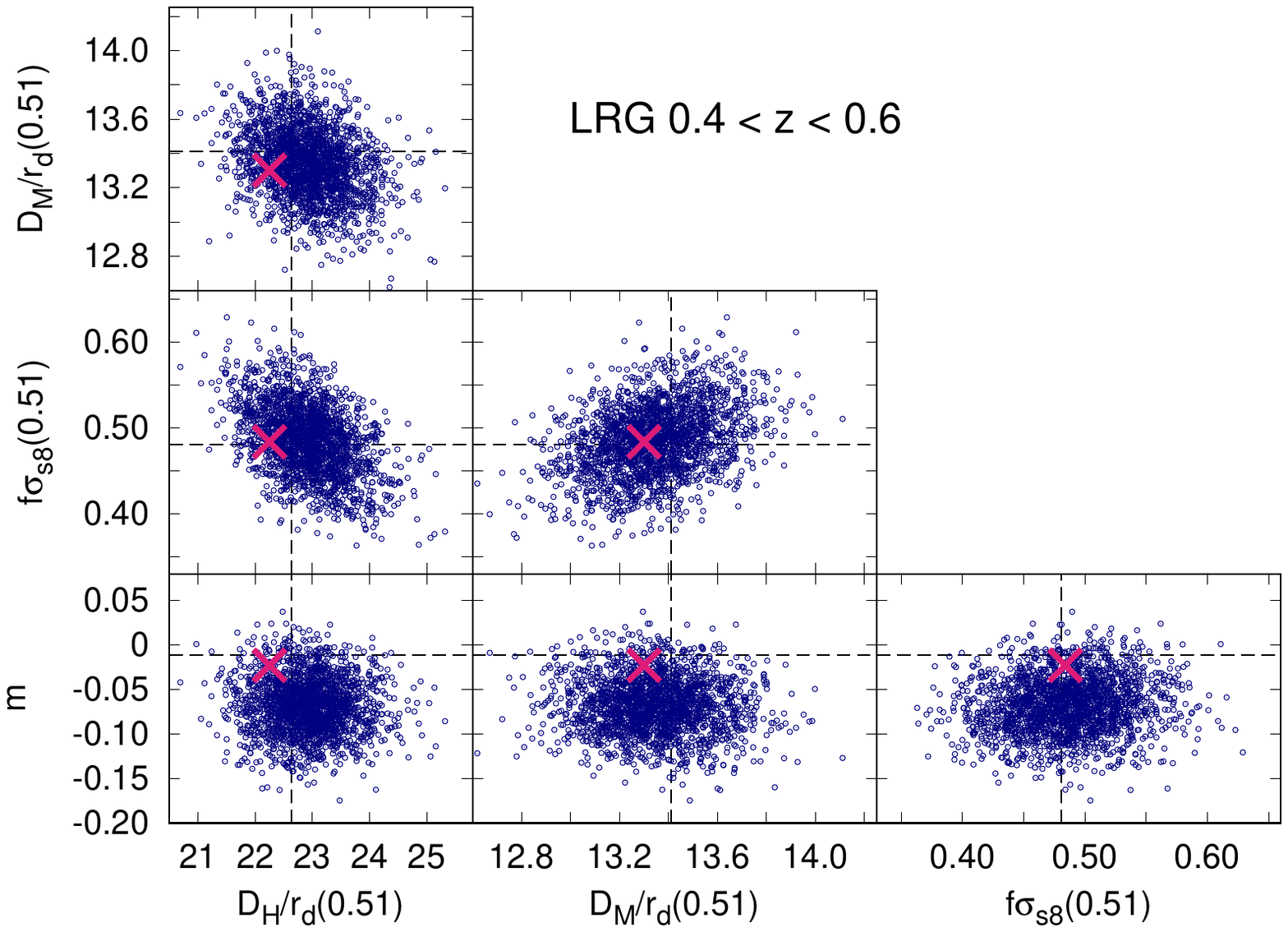}
    \includegraphics[scale=0.25]{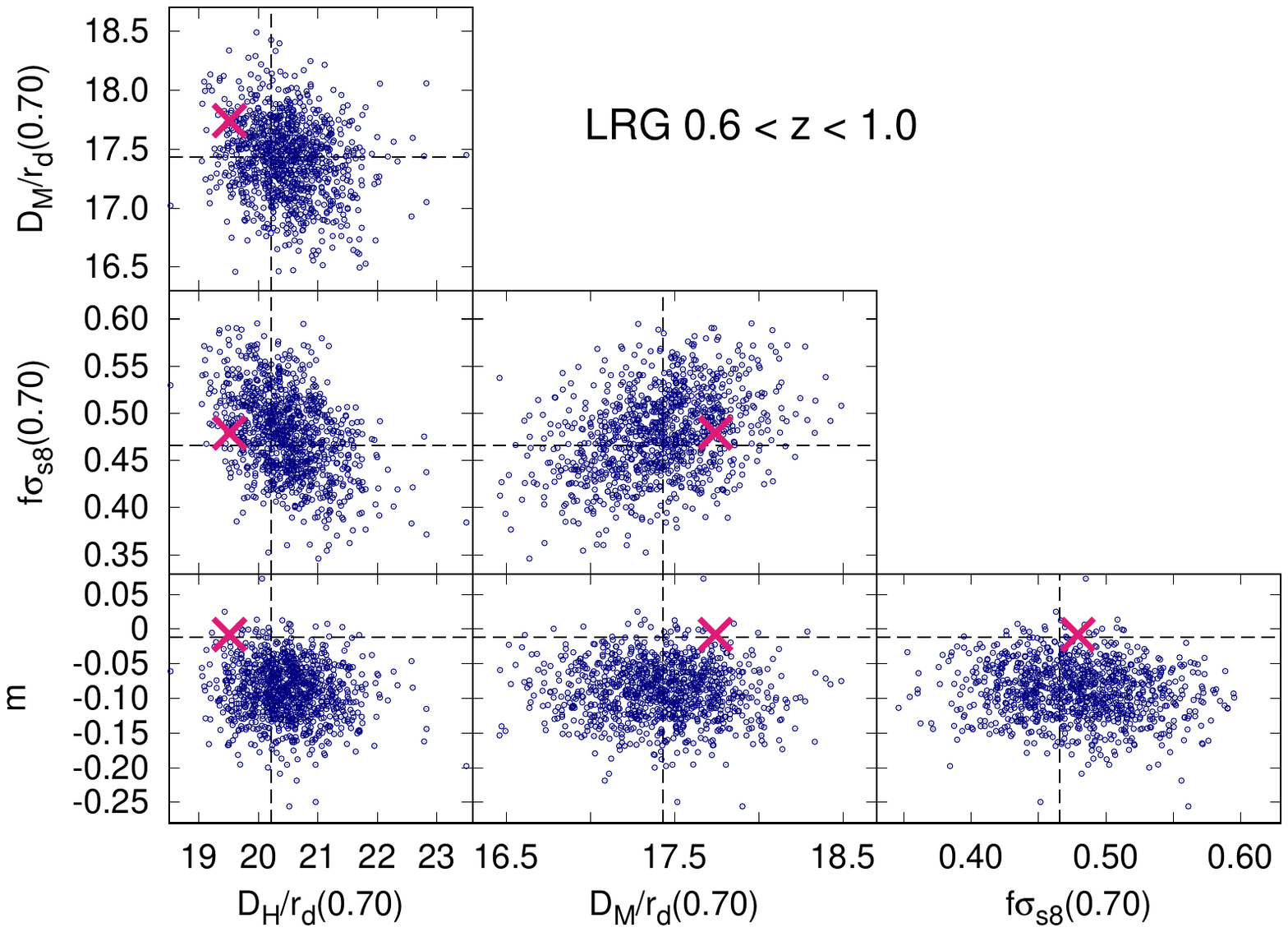}
        \includegraphics[scale=0.25]{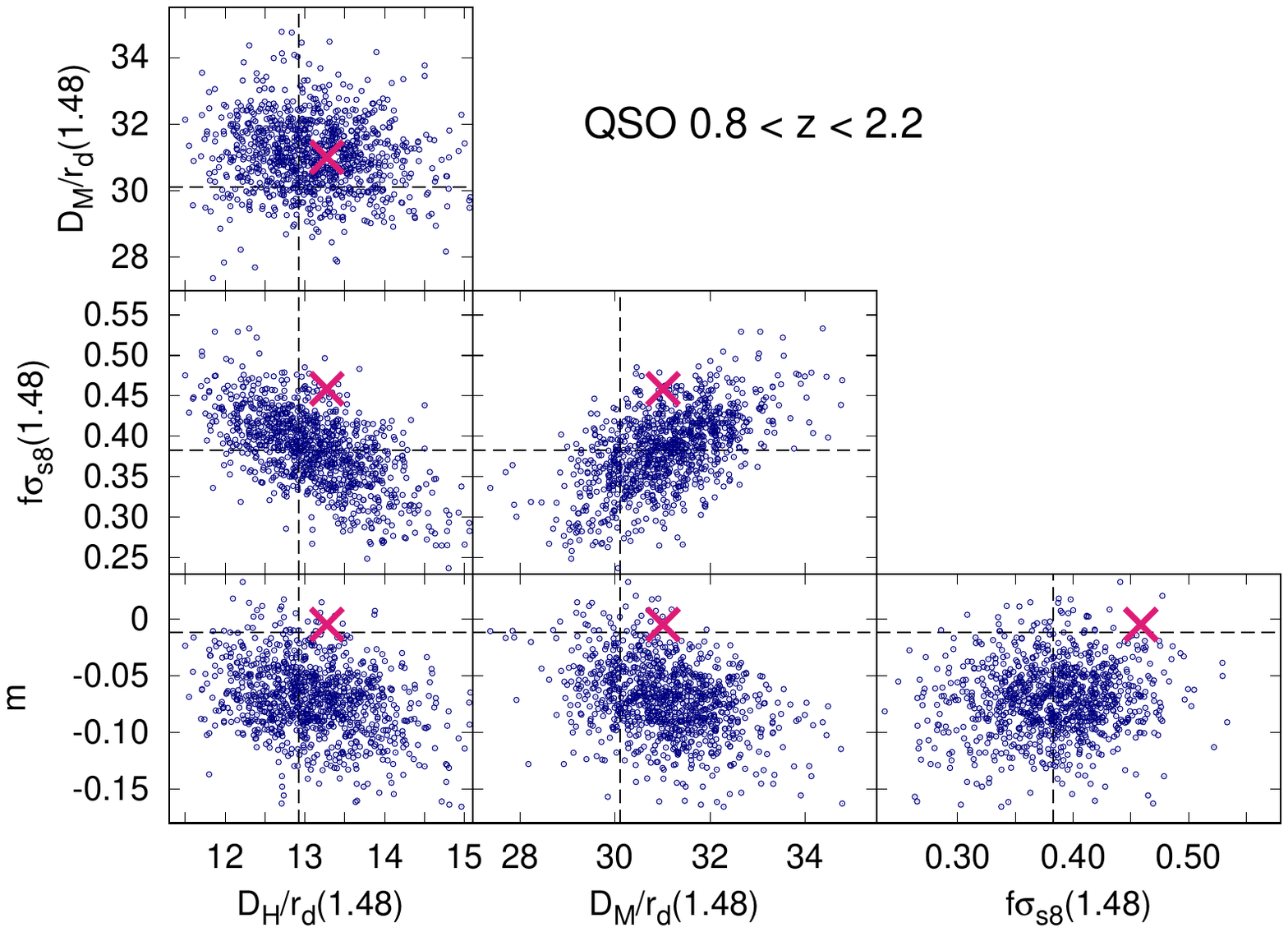}
    \caption{Triangle plots for the BOSS and eBOSS samples when the ShapeFit pipeline is applied.  The four physical parameters of interest are shown in each panel: the BAO longitudinal and transverse distances, $D_H/r_{\rm d}$ and $D_M/r_{\rm d}$, respectively; the growth of structure $f$ times $\sigma_{s8}$; and the shape parameter $m$. Each panel displays one of the four studied BOSS/eBOSS samples, as indicated. The blue dots display the best-fits for these parameters in each of the mock realizations: 2048 for the Patchy mocks corresponding to the LRG samples $0.2<z<0.5;\,(z_{\rm eff}=0.38)$ and $0.4<z<0.6;\,(z_{\rm eff}=0.51)$; 1000 realizations for the EZmocks corresponding to LRG sample $0.6<z<1.0;\,(z_{\rm eff}=0.70)$, and the QSO sample $0.8<z<2.2;\,(z_{\rm eff}=1.48)$. The red cross represents the result for the actual data, as it is presented in table~\ref{tab:boss_compare}. The dotted black lines represent the expected values for the cosmology of the mocks (see table~\ref{tab:cosmo}). All cases display the inferred parameters from both northern and southern patches. For the LRG samples, the pre-recon full shape information has been consistently combined with the post-recon catalogue BAO signal as described in section~\ref{sec:data-prepostcomb}.}
    \label{fig:signal}
\end{figure}

Figure~\ref{fig:signal} displays the triangle plot for the four physical parameters of interest, and for the four studied galaxy samples. The blue dots represent the result of the 2048 and 1000 realizations on the Patchy (for LRG $0.2<z<0.6$) and EZmocks (for LRG $0.6<z<1.0$ and QSO $0.8<z<2.2$), respectively. The red cross displays the performance of the data, as presented in table~\ref{tab:boss_compare}. The horizontal and vertical dotted black lines display the expected values given the true cosmology of the mocks (see table~\ref{tab:cosmo}). 

 We see that best-fit values of the data lie within the scatter cloud of the best-fitting values for the mocks. This is indeed expected as the mocks were designed to reproduce the clustering of the data. However, we do not necessarily expect that both data and mocks share the same cosmology, even within the statistical uncertainty of the samples. When comparing the performance of the mocks with their expected values (marked by the black dashed lines), we find that for the $D_H/r_{\rm d}$, $D_M/r_{\rm d}$ and $f\sigma_{s8}$ parameters, the agreement is excellent. Indeed, the mocks were designed with the aim of being able to reproduce both the BAO and the anisotropic clustering signal, as these are the two main scientific goals of the BOSS and eBOSS programs. Additionally, we find that the mocks tend to have a low value of the shape parameter, $m$, with respect to its expected value, with an offset of around $-0.05$ to $-0.10$, which corresponds to $1-2\sigma$ statistical of the BOSS/eBOSS data sample. We explore this particular feature more in detail in section~\ref{sec:sys_budget}, using full N-body mocks data (but see also Appendix \ref{app:nonlocal_bias} for the impact of the local bias assumption on the shape parameter for these mocks). 
In short, this observed behaviour corresponds to a limitation of the mocks (rather than to systematics of the model) due to the clustering of the mocks in the scale range of $0.05<k\,[h{\rm Mpc}^{-1}]<0.10$. In particular, pre-virialization terms (as for example those captured by 1-loop corrections) contribute to the clustering signal on the scales where $m$ is measured. But pre-virialization  is not fully taken into account by the Zeldovich approximation, which these mocks rely on (see also Appendix \ref{app:nonlocal_bias}). As a result the model fitting procedure compensates by artificially lowering the best-fitting value of $m$. 

 This limitation has not been explicitly reported before because it did not impact how well the BAO and $f\sigma_{s8}$ parameters could be recovered from these mocks, as $m$ is very uncorrelated with them. 

\begin{figure}[htb]
    \centering
    \includegraphics[scale=0.25]{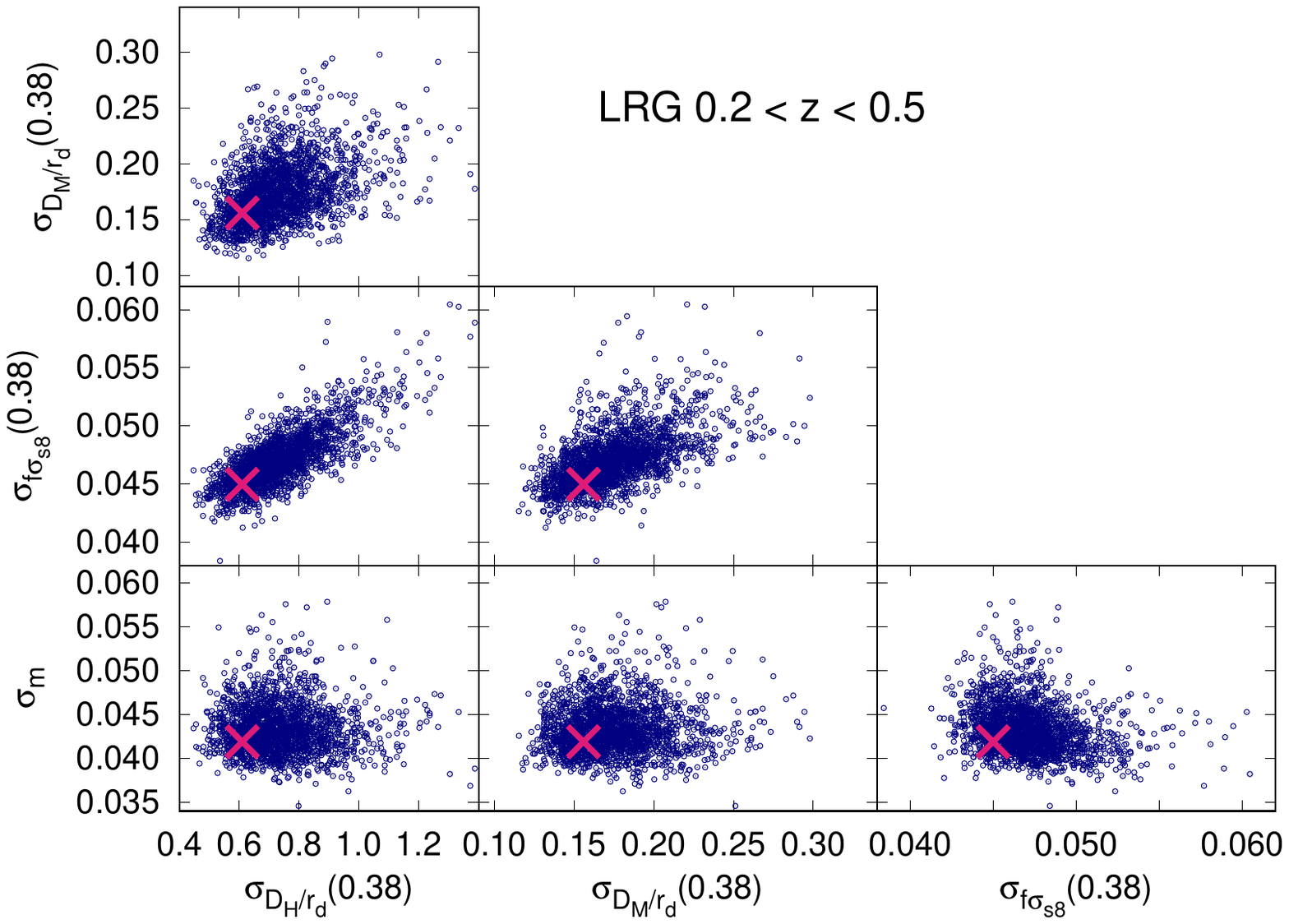}
        \includegraphics[scale=0.25]{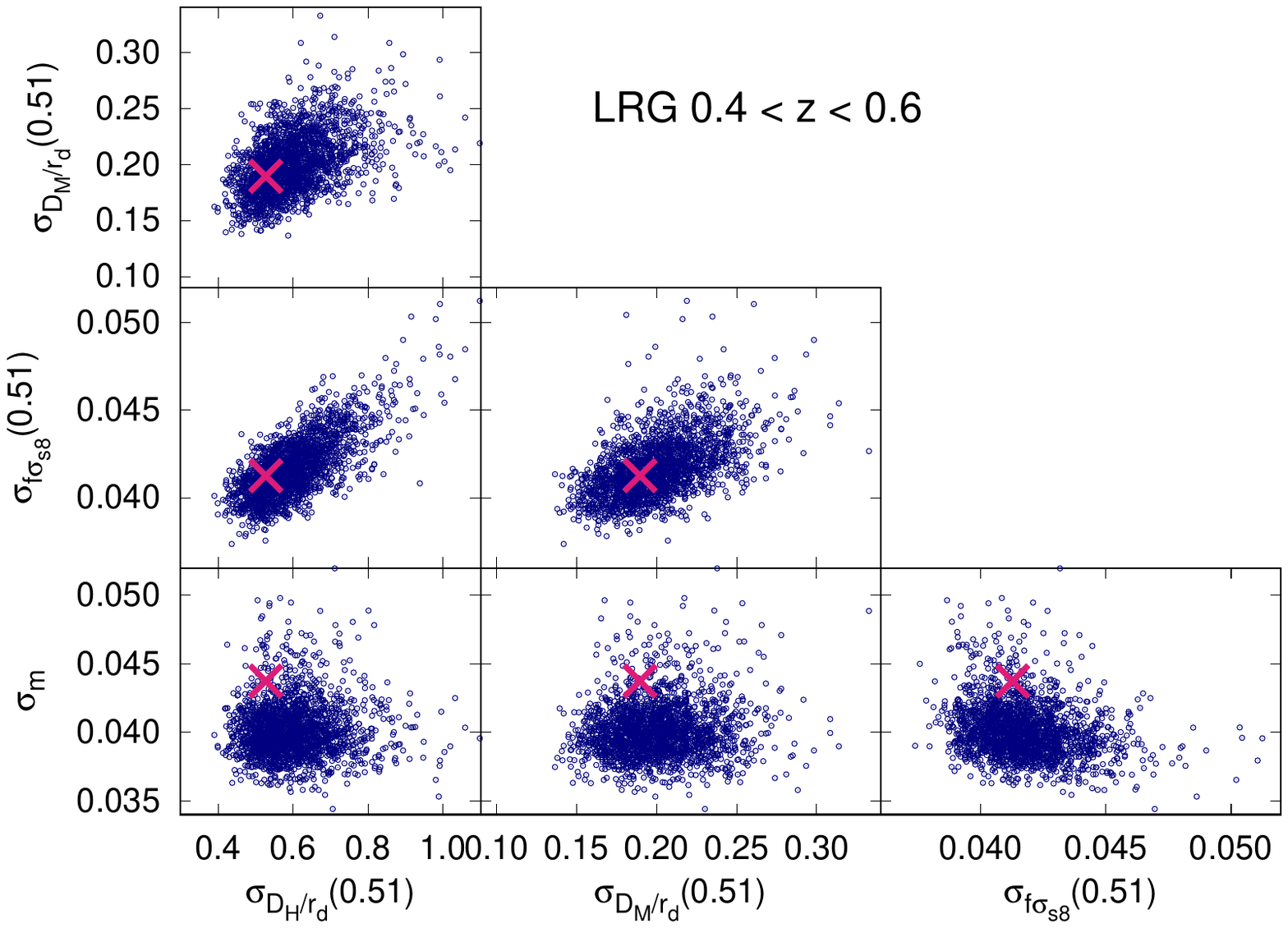}
    \includegraphics[scale=0.25]{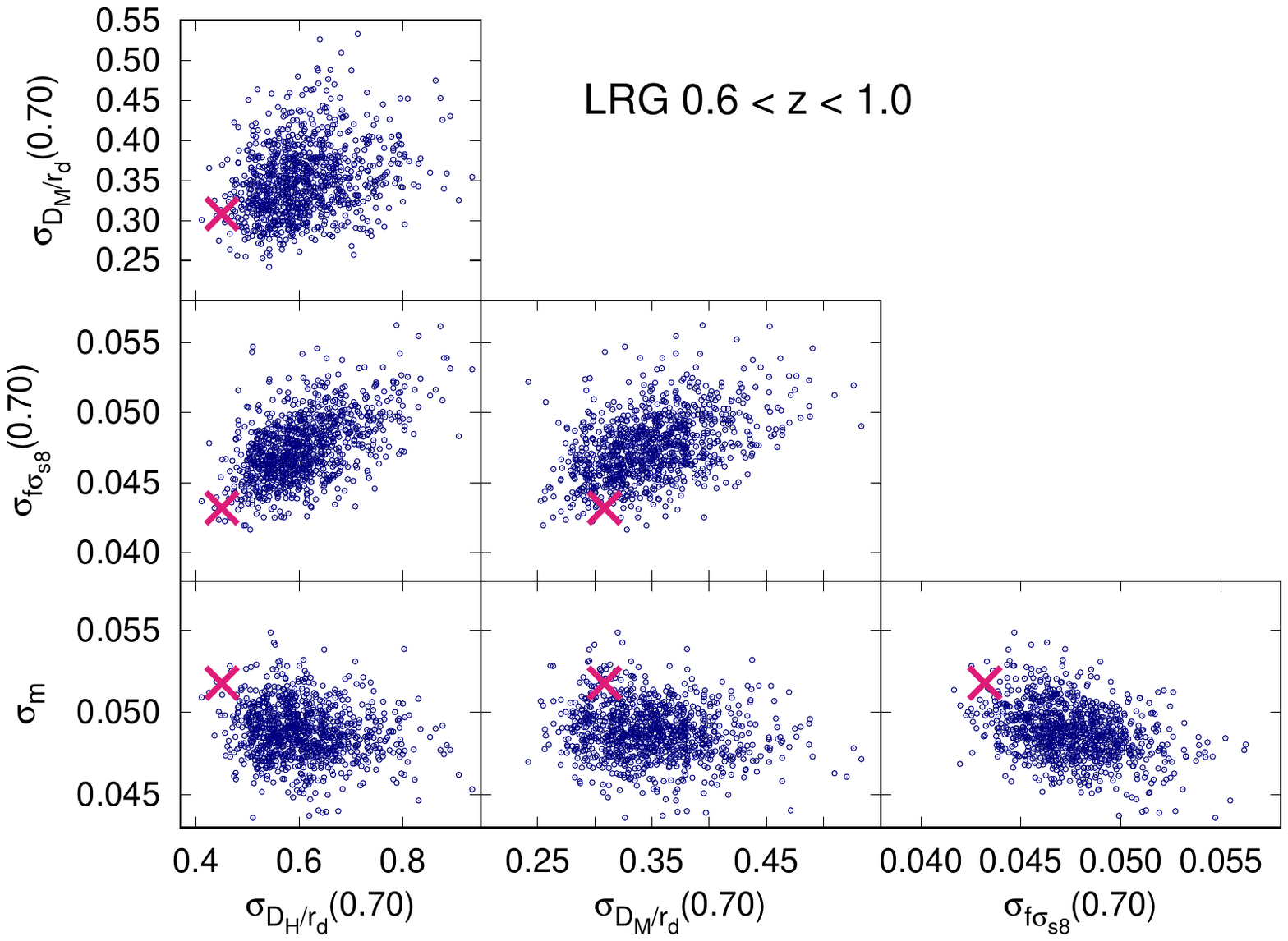}
        \includegraphics[scale=0.25]{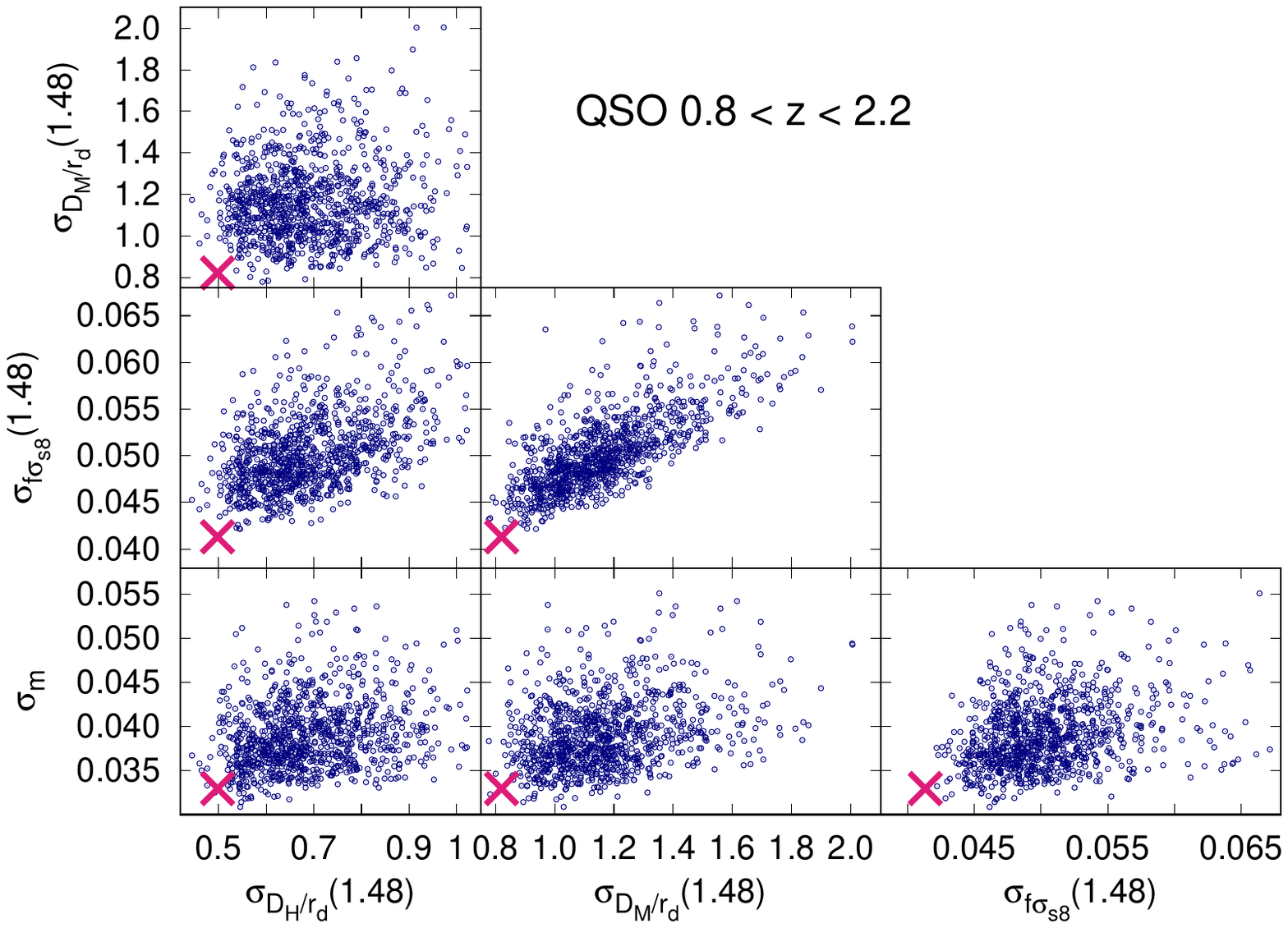}
    \caption{Same structure and notation as in figure~\ref{fig:signal}, but displaying the $1\sigma$ error-bar for each parameter of interest.}
    \label{fig:errors}
\end{figure}

The panels of figure~\ref{fig:errors} display the $1\sigma$ errors corresponding to the best-fit values presented in figure~\ref{fig:signal}.
For the BOSS LRG samples for $0.2<z<0.6$ (top-left and top-right panels) the errors for the data (red cross) are very typical compared to those of the mocks (blue dots). The bottom sub-panels, corresponding to eBOSS LRG and QSO samples, indicate that in some cases, the error of the data is significantly smaller than the one reported on the mocks. This happens for the longitudinal BAO distance, $D_H/r_{\rm d}$, and shape parameter, $m$, in the LRG $0.6<z<1.0$ sample, and up to some degree for all the parameters in the QSO $0.8<z<2.2$ sample. This behaviour was already reported by the eBOSS team (see for e.g., the discussion in section 5.1 of \cite{Gil-Marin:2020bct} for the LRG sample; and section 3.2.2 of \cite{neveuxetal2020} for the QSO sample) and is caused by an excess of significance in the BAO detection of both quasars and LRGs, likely originated by noise fluctuations in the data, which were different from those in the mocks. 

We aim to use the set of best-fit values (and their errors) of the mocks to perform quantitative tests: how their {\it rms} compares to the averaged errors, whether the distribution is Gaussian, as well as which are the typical deviations from their expected values. All this information is summarized in table~\ref{tab:mocks}, which displays the results for each of the four studied samples (each column), and for each of the four parameters of interest, $x=\alpha_\parallel,\,\alpha_\perp,\,f\sigma_{s8},\,m$. For each of these parameters we display in rows the systematic offset between the best-fit of the mean of the mocks and its expected value, $\Delta x^{\rm av.}$; the systematic offset between the mean of the best-fits of the individual mocks and its expected value, $\langle\Delta x\rangle$; the average of the error-bars of the best-fits, $\langle \sigma_x\rangle$, the {\it rms} among the best-fits, $S_x$; the mean of the $Z$-statistic, defined as $Z_{x_i} \equiv (x_i-\bar{x})/\sigma_{x_i}$, $\langle Z_x \rangle$; and the {\it rms} of $Z_{x_i}$, $S_{Z_x}$. All the $x$-variables of table~\ref{tab:mocks} are expressed in units of $10^3$ for better visualization.
The statistics displayed in table~\ref{tab:mocks} use all the available mock realizations, 2048 for the Patchy and 1000 for the EZmocks. The fits to the mean of the mocks offset-ed by the expected value, $\Delta x^{\rm av.}$, correspond to the pre-reconstructed signal only, and its covariance (and error) correspond to 100 times the covariance (10 times the errors) of one single realization (including both northern and southern caps), and not the error of the mean of all used mocks. This ensures that the statistical errors are sufficiently small to clearly identify systematics relevant for the actual data sample. Unlike $\Delta x^{\rm av.}$, the statistics related to the average of best-fits to individual mocks, $\langle \Delta x\rangle$, $\langle \sigma_x\rangle$, $S_x$, $Z_x$ and $S_{Z_x}$, contain both pre- and post-reconstruction information combined as described in section~\ref{sec:data-prepostcomb}. All analyses are performed using the same pipeline employed to analyse the data and described in section~\ref{sec:theory_Pell}.

\begin{table}[htb]
    \centering
    \begin{tabular}{|c|c|c|c|c|c}
    \hline
        Variable $\times10^3$ & LRG(0.38) & LRG(0.51) & LRG(0.70) & QSO(1.48)  \\
        \hline
        \hline
    $\Delta \alpha_\parallel^{\rm av.}$ & $13.7 \pm 4.0$ & $20.4\pm3.6$ & $18.5\pm3.8$ & $-15.7\pm5.3$ \\ 
   $\langle \Delta\alpha_\parallel\rangle$ & $7.4861$ & $12.3344$ & $14.6305$ & $15.0770$ \\ 
    $\langle \sigma_{\alpha_\parallel}\rangle$ & $29.9673$ & $26.5106$ & $30.2284$ & $53.8854$ \\ 
    $S_{\alpha_\parallel}$ & $33.4643$ & $27.4981$ & $31.4638$ & $51.8706$ \\ 
    $\langle Z_{\alpha_\parallel}\rangle$ & $-0.0546$ & $-0.0406$ & $-0.0501$ & $-0.0851$ \\ 
    $S_{Z_{\alpha_\parallel}}$ & $1.0723$ & $1.0313$ & $1.0055$ & $0.9794$ \\ 
    \hline
    $\Delta \alpha_\perp^{\rm av.}$ & $-2.4\pm2.4$ & $-6.2\pm2.2$ & $-6.8\pm2.6$ & $17.0\pm3.9$ \\ 
    $\langle \Delta\alpha_\perp\rangle$ & $-2.3970$ & $-4.5816$ & $-3.3037$ & $32.3815$ \\ 
    $\langle \sigma_{\alpha_\perp}\rangle$ & $16.9014$ & $14.9603$ & $19.9959$ & $38.9782$ \\ 
    $S_{\alpha_\perp}$ & $17.5587$ & $14.6142$ & $19.3600$ & $35.6125$ \\ 
    $\langle Z_{\alpha_\perp}\rangle$ & $0.0043$ & $0.0006$ & $-0.0050$ & $-0.0344$ \\ 
    $S_{Z_{\alpha_\perp}}$ & $1.0351$ & $0.9744$ & $0.9672$ & $0.9215$ \\ 
    \hline
    $\Delta f\sigma_{s8}^{\rm av.}$ & $3.9\pm5.7$ & $7.9\pm5.0$ & $12.7\pm6.0$ & $2.1\pm4.7$ \\ 
    $\langle \Delta f\sigma_{s8}\rangle$ & $5.8290$ & $11.6461$ & $14.5541$ & $7.8721$ \\ 
    $\langle \sigma_{f\sigma_{s8}}\rangle$ & $48.2098$ & $41.9628$ & $47.7991$ & $51.4069$ \\ 
    $S_{f\sigma_{s8}}$ & $48.4473$ & $40.6968$ & $45.6538$ & $49.9718$ \\ 
    $\langle Z_{f\sigma_{s8}}\rangle$ & $0.0032$ & $0.0018$ & $-0.0040$ & $0.0141$ \\ 
    $S_{Z_{f\sigma_{s8}}}$ & $1.0051$ & $0.9670$ & $0.9539$ & $0.9600$ \\ 
    \hline
    $\Delta m^{\rm av.}$ & $-25.0\pm5.2$ & $-26.0\pm5.0$ & $-27.2\pm5.5$ & $-19.0\pm3.9$ \\ 
    $\langle \Delta m\rangle$ & $-65.3894$ & $-56.1880$ & $-77.3976$ & $-58.9662$ \\ 
    $\langle \sigma_{m}\rangle$ & $43.3229$ & $40.1828$ & $48.8380$ & $39.0732$ \\ 
    $S_m$ & $34.1271$ & $31.2587$ & $40.9830$ & $31.6964$ \\ 
    $\langle Z_m \rangle$ & $-0.0199$ & $-0.0126$ & $-0.0111$ & $-0.0178$ \\ 
    $S_{Z_m}$ & $0.7819$ & $0.7709$ & $0.8451$ & $0.8216$ \\ 
    \hline
%    $\chi_{\rm av.}^2$ & $60.2/(78-12)$ & $83.3/(78-12)$ & $96.4/(78-12)$ & $132.1/(168-12)$ \\
 %   \hline
    \end{tabular}
    \caption{Statistics derived from the 2048 realizations of the Patchy mocks (for the LRG(0.38) and LRG(0.51) samples) and from the 1000 realizations of the EZmocks (LRG(0.70) and QSO(1.48) samples), for the four physical variables of interests, the BAO scales along and across the line of sight, $\alpha_\parallel$, $\alpha_\perp$, the growth of structure $f$ times $\sigma_{s8}$, and the shape parameter $m$. For each of these variables ($x=\alpha_\parallel,\,\alpha_\perp\,f\sigma_{s8},\,m$) we display the fit to the mean of the mocks minus its expected value, $\Delta x^{\rm av.}$; the average of individual fits minus its expected value, $\langle \Delta x\rangle$; the average of errors of the individual fits, $\langle \sigma_x\rangle$; the {\it rms} of all best-fits, $S_x$, the average of the $Z$-statistic (see text), $\langle Z_x \rangle$ and its {\it rms}, $S_{Z_x}$. The expected values for each of the samples can be found in table~\ref{tab:expected}.}
    \label{tab:mocks}
\end{table}

The results regarding the offset between the expected value and the best-fit to the mean of the mocks, $\Delta x^{\rm av.}$, show a $1 - 2\%$ deviation for $\alpha_\parallel$; a $0 - 1.7 \%$ deviation for $\alpha_\perp$; a $0 - 0.013$ deviation for $f\sigma_{s8}$; and a $-0.025$ deviation for $m$. Recall that the observed systematic offsets on the mocks should not be used to put constraints on the systematic error budget of the models, due to inaccuracies of the mocks when reproducing the actual clustering. However, we see a very good agreement with the expected value given the accuracy of the actual data, around $2\%$ for $\alpha_\parallel$, $1.5\%$ for $\alpha_\perp$, $0.040$ for $f\sigma_{s8}$ and $0.040$ for $m$. 

In ideal Gaussian conditions\footnote{When the distribution of best-fitting quantities is described by a Gaussian distribution, without any skewness or any anomalous kurtosis.} with sufficiently high signal-to-noise data (this is, large enough number of mock realizations) both $\langle \Delta x \rangle$ and $\Delta x^{\rm av.}$ should coincide. We indeed find a good agreement for $\alpha_\perp$ and $f\sigma_{s8}$, and reasonably good for $\alpha_\parallel$. On the other hand, $m$ shows significant differences, suggesting that the distribution of best-fitting values of $m$ is skewed towards negative values. 

We now focus on the error-related quantities, $\langle \sigma_x\rangle$ and the {\it rms} of the best-fits, $S_x$. When the distribution is Gaussian and the covariance correctly modelled, these two error estimates should be very close.
Indeed, we find a very good agreement among these two statistics for $\alpha_\parallel$, $\alpha_\perp$ and $f\sigma_{s8}$, indicating that not only the distribution for those variables is close to Gaussian, but also that there are no indications of covariance matrix-related issues. We do observe differences for $m$, but this is not a surprise, given the non-Gaussian signature presented when comparing the best-fit of the mean to the average of individual best-fits. Another way to quantify the agreement between $S_x$ and $\langle \sigma_x\rangle$ is through the variable $Z_x$, which for Gaussian distributions should return a mean value of 0 and a {\it rms} of 1. Indeed we find a very good agreement of $Z=0\pm1$ for $\alpha_\parallel$, $\alpha_\perp$ and $f\sigma_{s8}$. In particular, we find that $S_{Z_x}$ tend to be slightly above or below unity by just $5\%$ for $\alpha_\parallel$, $\alpha_\perp$ and $f\sigma_{s8}$, which gives an order of magnitude on the accuracy of the errors reported on the analysis of the actual data. For the variable $m$ we find an offset on $Z$ of around $-0.015$ and a $S_{Z_m}\sim0.8$ because of the non-Gaussian behaviour already described.

We conclude that the mocks describe well the expected clustering for the BAO variables, $\alpha_\parallel$ and $\alpha_\perp$, as well as for the anisotropic clustering, $f\sigma_{s8}$. They fail to reproduce accurately the signal of the shape parameter $m$, under-estimating its value by about $-0.025$, which corresponds to 1/2 to 2/3 of $1\sigma$ error for the data. We explain this effect as due to the inaccuracy of fast techniques implemented in the production of the mocks. Both Patchy and EZmocks are based on the Zeldovich approximation that does not fully account for the effect of pre-virialization, which slightly boosts the amplitude of the shape of the power spectrum at the scales where $m$ is sensitive, $0.05<k\,[h{\rm Mpc}^{-1}]<0.10$. 

The statistics of the errors for $\alpha_\parallel$, $\alpha_\perp$ and $f\sigma_{s8}$ show a very good agreement with the Gaussian statistics and confirm that the errors of the data are accurate at $5\%$ level. For the variable $m$ we report a non-Gaussian distribution which tend to skew the distribution towards negative values.

\subsection{Systematic error budget} \label{sec:sys_budget}

In the previous section we have reported how the pipeline used for the data performed on the EZ and Patchy mocks. However, these mocks are based on fast methods, such as the Zeldovich approximation, and they do not describe the power spectrum clustering at the precision level required for setting the systematic error budget. For this reason we aim to test the performance of the pipeline on full N-body mocks. We focus on two sets, the Nseries mocks and the PT challenge mocks, which we briefly describe below. 

The Nseries mocks\footnote{The Nseries mocks are publicly available in \cite{nseries:data}
.} consist of 84 pseudo-independent realizations of the BOSS CMASS northern geometry within redshifts $0.43<z<0.70;\,(z_{\rm eff}=0.56)$, which were used by the BOSS team to set the official systematic error budget of the BAO and RSD measurements in the final cosmology paper \cite{alam_clustering_2017}. The Nseries mocks have been generated out of 7 fully independent periodic boxes of $2.6\,{h}^{-1}{\rm Gpc}$ box side. For each of these cubic boxes 4 different orientations of the CMASS northern geometry are fitted, and each of these fitted orientations are applied on 3 pre-rotation positions of the box, where the 3 Cartesian positions and velocities are swapped, to extract all $7\times4\times 3=84$ pseudo-independent realizations with BOSS northern geometry. The mass resolution
is $1.5 \times 10^{11}\, {M_\odot h^{-1}}$, with $2048^3$ dark matter particles per box. The identified haloes are populated with galaxies following a HOD model tuned to match the
clustering of LRGs observed by BOSS. These mocks are analyzed using the covariance matrix extracted from the Patchy mocks with the same sky geometry as these Nseries mocks. The matrix elements of this covariance are rescaled by $10\%$ in order to account for the difference in particles between the Patchy mocks. This difference arises from veto mask effects which are not implemented in the Nseries mocks. The underlying cosmology of these mocks is consistent with the Wilkinson Microwave Anisotropy Probe (WMAP, \cite{2013ApJS..208...19H}) best-fitting cosmology, $h=0.7$, $\Omega_{\rm m}=0.286$, $n_s=0.96$, $\Omega_{\rm b}=0.047$, $\sigma_8=0.820$. In total, these 84 pseudo-independent mocks have an associated effective volume of $106\,{\rm Gpc}^{3}h^{-3}$. Many previous works \cite{gil-marin_clustering_2016a,griebetal17,Beutler:2016arn,Satpathyetal17,Gil-Marin:2020bct,Bautistaetal21,damicoetal22,philcoxetal22,deMattia:2019vdg,icaza-lizaolaetal20,Handetal17} make use of these 84 realizations, as if they were independent, to test their pipelines. This is only a good approximation when we exclusively focus on RSD- and BAO-related quantities. When we also want to test for the shape of the power spectrum, the additional realizations coming from the pre-rotations and the inherent real-space volume overlap among cuts, end up producing a severe under-estimation of the errors of parameters sensitive to the real space power spectrum on large scales, such as $m$. For this reason, we opt to carefully select a subset of these 84 skycuts to reduce the overlapping effect. In order to do so, we compute the cross power among the 12 different skycut orientations (including the pre-rotations) for each of the 7 boxes ($12\times11/2=66$ cross-power combinations for each box), and average each of these 66 cross spectra among the 7 independent realizations. We then select those 6 orientations per box whose $k$-mode-weighted added cross-power-spectrum-squared $k$-bins is closer to 0. This is motivated by the fact that partially overlapping samples should have cross-power signal different than 0.  We refer to this subset of $6\times7=42$ realizations as the Nseries-$\mathcal{Z}$\footnote{The indices for these realizations are $\{3,\,6,\,7,\,8,\,9,\,12\}+12i$, for $i=0,\,\ldots,6$.}, with a effective volume of $53\,{\rm Gpc}^{3}h^{-3}$. Additionally, we
consider the 7 fully independent Nseries cubic boxes, with a total effective volume of $80\,{\rm Gpc}^{3}h^{-3}$.

The PT challenge mocks
\cite{2020PhRvD.102l3541N,PTchallenge:data} consist of 10 independent realizations in periodic boxes whose comoving side length is $3840\,[h^{-1}{\rm Mpc}]$ with $3072^3$ particles, where the 3 input $\Lambda$CDM parameters, $\Omega_{\rm m}$, $A_s$ and $H_0$, were randomly selected from a Gaussian probability distribution centered at the Planck fiducial cosmology.\footnote{These randomly drawn values were kept secret (blind) and not publicly known. Other cosmological parameters such as the primordial tilt and the baryon-to-matter ratio were fixed to $\Omega_{\rm b}/\Omega_{\rm m}=0.1571$ and $n_s=0.9649$.} 
Halo catalogues are identified by using \textsc{Rockstar} halo finder \cite{rockstar}. Additionally, these haloes are populated using a HOD description roughly matching BOSS LRG galaxy data. Here we only focus on the snapshot produced at $z=0.61$. The mocks are analyzed using an analytically estimated covariance (also provided by the PT-Challenge team), where the correlation between different multipoles at the same wave-vector $k$ is non-zero, and the correlation between adjacent $k$-bins is ignored. In total, these mocks have an associated effective volume of $566\,{\rm Gpc}^{3}h^{-3}$.

\begin{table}[h]
    \centering
    \begin{tabular}{|c|c|c|c|c|}
    \hline
         $\Delta x\pm 2\sigma$ & Nseries-$\mathcal{Z}$ Sky &  Nseries Box & PT challenge & $1\sigma$ error of data\\
        \hline
        \hline
        $\alpha_\parallel$ & $0.0066 \pm 0.0088$ & $0.0082\pm0.0076 $ & $0.0077 \pm 0.0036$ & $[0.022 - 0.038]$\\
        $\alpha_\perp$ & $-0.0038\pm 0.0054$ & $-0.0021\pm0.0043$ & $-0.0003 \pm 0.0024$ & $[0.014 - 0.028]$ \\
        $ f\sigma_{s8}$ & $-0.0056 \pm  0.0115$ & $-0.007\pm0.010$ & $-0.0039 \pm 0.0049$ & $[0.041 - 0.045]$ \\
        $m$ & $-0.014 \pm 0.013$ & $-0.009\pm0.012$ & $-0.0012 \pm 0.0068$ & $[0.033 - 0.052]$ \\
        \hline
        $\Omega_{\rm m}$ & $-0.0048\pm0.0050$ & $-0.0026\pm0.0043$ & $0.0008\pm0.0022$ & $0.0057$ \\
        $H_0$ & $-0.10\pm0.75$ & $-0.16\pm0.70$ & $-0.24\pm0.36$ & $0.67$ \\
        $A_s\times 10^{9}$ & $0.033\pm0.103$ & $0.008\pm0.093$ & $-0.007\pm0.053$ & $0.20$\\
        %$\sigma_8$ & $-0.007\pm0.018$ & $-0.008\pm0.016$ &  & \\    
\hline
    \end{tabular}
    \caption{Systematic offsets found when fitting the mean of the 42 Nseries-$\mathcal{Z}$ skycut mocks ($z_{\rm eff}=0.56$; $V_{\rm eff}=53\,[{\rm Gpc}h^{-1}]^3$), the Nseries cubic mocks ($z_{\rm eff}=0.50$; $V_{\rm eff}=80\,[{\rm Gpc}h^{-1}]^3$), and the PT challenge mocks ($z_{\rm eff}=0.61$; $V_{\rm eff}=566\,[{\rm Gpc}h^{-1}]^3]$), for the four compressed physical variables of interest, $x=\alpha_\parallel$, $\alpha_\perp$, $f\sigma_{s8}$ and $m$; as well as for the three cosmology variables, $x=\Omega_{\rm m}$, $H_0$ and $A_s$, derived from the interpretation of the physical variables within a $\Lambda$CDM model. For both Nseries and PT challenge mocks, the error-bars represent the $95\%$ confidence level (2$\sigma$). For reference we also quote the $1\sigma$ statistical error for the full data sample, which spans along four LRG and QSO sub-samples within $0.2<z<2.2$, and therefore we just display the interval set by the largest and smallest error (see table~\ref{tab:boss_compare} for the statistical error for each of the samples). The errors of the cosmology variables have the $2.1>z$ Ly-$\alpha$ contribution and the BBN prior, as described in section~\ref{sec:data-early-time}. The analysis of data and mocks is done using the same pipeline assumptions as described in section~\ref{sec:data}. Unlike the LRG samples of the data, the Nseries or PT challenge mock data do not include the reconstruction signal.}
    \label{tab:final_sys}
\end{table}

Table~\ref{tab:final_sys} summarizes the observed offsets for these two sets of N-body mocks. The Nseries results are displayed for the power spectrum average of the 42 realizations of the `Nseries-$\mathcal{Z}$ Sky' sample, and for the power spectrum average of the 7 cubic boxes, `Nseries Box'. For reference the results from the average of the power spectra of the full 84 Nseries skycut mocks can be found in the panels of figure~12 of Ref. \cite{ShapeFit} in orange contours. On the other hand, the PT challenge results correspond to the DATA-like MIN in table 3 and red contours in fig. 2 of Ref. \cite{ShapeFitPT}. For each variable, we report the difference between the measured parameter and the expected quantity given the known cosmology of each set of mocks. The errors represent the $95\%$ confidence level (i.e., they represent the $2\sigma$ contours) and correspond to the error of the mean. In order to set the systematic error budget we follow a similar criterion to the eBOSS team analysis: for each variable, when the observed offset on the mean of the mocks is within the $2\sigma$ confidence interval, the precision of the mocks' effective volume is not sufficiently high to resolve potential systematics deviations. 
When a systematic offset detected on the mocks (at $>95\%$ confidence level, given the mocks statistics)
represents a significant fraction of the $1\sigma$ statistical error of the data, we add this contribution in quadrature to the $1\sigma$ statistical error of the data to account for the systematic error\footnote{The eBOSS team followed the conservative approach of adding the $2\sigma$ interval as a systematic error, when no-systematic error could be determined. We do not follow this approach in this paper and add no error budget when the deviation is found to be within $2\sigma$ of the mean of the mocks.}. For reference, we display the $1\sigma$ statistical error inferred from the data in the last column of table~\ref{tab:final_sys}. For the physical variables we display the interval between the smallest and largest error found for the four redshift bins, whereas for the cosmology variables we simply quote the errors when all the samples (including Ly-$\alpha$) are considered, along with the BBN prior, as described in section~\ref{sec:results} and \ref{sec:cosmo}.  Note that, unlike the PT- and Nseries mocks, the analysis of the data combines measurements at different redshifts, which allows to break degeneracies in the $\Lambda$CDM variables and greatly reduce the error bars. For this reason the data, with an effective volume of the galaxy sample corresponding to $2.82\,[{\rm Gpc}h^{-1}]^3$ has  errorbars on $\{\Omega_{\rm m},\,H_0,\,A_s\}$ comparable to the ones of the mock samples with much larger effective volumes.

Among the physical variables, we find that the Nseries-$\mathcal{Z}$ result for $m$ is off by $\sim2\sigma$, which represent a 1/2 to 1/4 contribution to the statistical $1\sigma$ error of the data. This discrepancy vanishes when $m$ is inferred from the Nseries boxes ($m$ is well within the $2\sigma$ expected deviation), and it also not present for the PT challenge mocks (well within $1\sigma$ expected deviation). Unlike the PT challenge mocks, the Nseries skycut realizations are not fully independent. When considering the full 84 realizations we find $\Delta m\pm2\sigma=-0.0172\pm0.0093$, a $3.8\sigma$ deviation from its expected value (see orange contours of fig. 12 of Ref. \cite{ShapeFit}), which is off by $1.5\sigma$ from the Nseries box results. Since both Nseries Sky and Nseries Box contain almost the same objects, this result is highly inconsistent. This discrepancy between skycuts and boxes motivated us to investigate effect of the overlap among the full 84 Nseries sky realizations. Indeed, because certain patches of the boxes are repeated among the 84 realizations, the parameter errors estimated from the mean of the 84 Nseries sky mocks is under-estimated, and in particular, this under-estimate is expected to be more severe at large scales, where the shape parameter $m$ is measured. By selecting those 42 realizations with less cross power among the realizations coming from the same box, we choose those realizations that are more independent (or less correlated), which helps to bring together the results from the Nseries Sky and boxes. We conclude that there may be still some residual overlap in the Nseries-$\mathcal{Z}$ sample, which slightly drives the deviation of $m$ away from the Nseries Box results, slightly outside the $2\sigma$ confidence interval. Since no relevant deviation is observed in the cases where the samples are fully independent, we do not consider necessary to add any systematic contribution to $m$. 

We find $\alpha_\parallel$ is biased by $4.2\sigma_{566}$ in the PT challenge (here the subscript 566 indicates that this is the statistical error for an effective volume of $566 [{\rm Gpc}h^{-1}]^3$), as well as $2.1\sigma_{80}$ on the Nseries Box mocks (again since the sigma refers to the statistical error of an effective volume of $80 [{\rm Gpc}h^{-1}]^3$), in both cases the systematic shift is of similar magnitude, $+0.0077$ and $+0.0082$, respectively. This represents a 1/3 to 1/5 of the statistical error of the data. As explained in Ref.~\cite{ShapeFitPT}, the origin of this systematic is related to the hexadecapole signal. Certainly, when repeating the fit on the mean using only the monopole and quadrupole signals we find an offset of $-0.003\pm0.011$ for the Nseries Boxes; and $0.0040\pm0.054$,\footnote{As in table~\ref{tab:final_sys}, the errors correspond to  the 95\% confidence level.} for the PT challenge mocks, both cases perfectly compatible with the expected result within the $2\sigma$ interval. However, the impact of this systematic error on the cosmology variables is non-existent. This may seem paradoxical at first sight, but is related to the effect of the model-prior. As discussed in  Ref.~\cite{ShapeFitPT}, both $\alpha_\parallel$ and $\alpha_\perp$ variables are related via geometrical arguments within the $\Lambda$CDM model (as well as many other cosmology models). This tight relation in the parameter space of the $\alpha$'s forbids the region where the systematic on $\alpha_\parallel$ extends (and also limits the constraining power of the hexadecapole when the geometrical relation between $\alpha_\parallel-\alpha_\perp$ is imposed by the model). This can be seen in the left panel of fig. 2 of Ref. \cite{ShapeFitPT} (see red contours corresponding to the pipeline choices used in this paper), where the dashed empty contours display the parameter space allowed by the $\Lambda$CDM model, and the solid filled contours are those directly inferred from the model-agnostic analysis. In any case, adding in quadrature the $4.2\sigma_{566}$ found for $\alpha_\parallel$ with its statistical error value for the four different samples, only yields an increase of  5\% and 2.6\% of the total error budget (from 0.022 to 0.023 and from 0.038 to 0.039 for the sample with the smallest and largest error, respectively) making this contribution irrelevant.

Focusing on the systematics of the $\Lambda$CDM parameters, we find that, from the PT challenge, Nseries-$\mathcal{Z}$ and Nseries Box mocks, all the three parameters are within the 95\% confidence level drawn from the set of mocks, showing no hint of systematics. Because of the large effective volume of $V_{\rm eff}=566\,[{\rm Gpc}h^{-1}]^3$ of the PT challenge mocks, the shift associated to the $95\%$ is also considerably smaller than the size of the $1\sigma$ statistical error-bar --3 times smaller for $\Omega_{\rm m}$, 2 times smaller for $H_0$ and 4 times smaller for $A_s$-- demonstrating that the total effective volume of these mocks is sufficiently high to limit the potential systematics on the data below a threshold which is significantly smaller than the statistical errors. Therefore we conclude that 1) we do not detect significant modeling systematics to be added to the cosmology variables, and 2) the precision in determining such systematics is small enough to make them negligible for the current BOSS+eBOSS combined catalogues. 

\section{Discussion and Conclusions}
\label{sec:conclusions}

We make use of the final BOSS and eBOSS catalogues, product of four generations of Sloan Digital Sky Survey, representing a two decades long effort. BOSS and eBOSS probe over 10 billion years of cosmic evolution through more than 2 million spectra and represents the state-of-the art three-dimensional map of LSS to date. It will remain unrivaled until the next generation surveys \cite{aghamousa_desi_2016,aghamousa_desi_2016-1, laureijs_euclid_2011} release their catalogues. 
The LSS clustering in BOSS and eBOSS catalogues has been analyzed by the SDSS collaboration using what we refer to as the `classic' approach, where the clustering information is compressed into three physical parameters as a function of redshift: the line-of-sight and plane-of-the-sky  apparent shift of the location of the standard ruler BAO feature in Fourier space which yields geometrical information on the expansion history and the amplitude of the redshift space distortions on linear scales which yields information on the linear growth rate of perturbations. But LSS clustering encloses  extra cosmologically-relevant signal, most of which can be extracted from the large-scales {\it Shape} of the power spectrum.  This is what the proposed ShapeFit \cite{ShapeFit} approach does. We extend the original ShapeFit formulation to include the post-recon BAO information, correctly accounting for the pre-recon and post-recon compressed variables covariance. 

Then we apply the ShapeFit analysis on the BOSS and eBOSS data and present the resulting constraints on the physical parameters, which are summarized in Fig. \ref{fig:results}. These can subsequently be interpreted in terms of a variety of cosmological models, to produce constraints on each model's parameters. The standard $\Lambda$CDM model, with parameters calibrated from CMB observations, provides a good fit to the recovered compressed parameters which were derived independently of the model, offering a powerful consistency check for the underlying model's assumptions. 

For the extensions to the $\Lambda$CDM model considered here (models with spatial curvature, dark energy equation of state parameter not equal to -1, both constant and varying in redshift according to the  $w_0-w_a$ parameterization, non-zero neutrino masses, and extra effective neutrino species), we find that the {\it Shape} improves the determination of the matter density parameter $\Omega_{\rm m}$, further helping to break parameter degeneracies. This additional information comes from the signature of physical processes around matter-radiation equality and is thus a late-time probe of early-time physics. This information is highly complementary to that captured by BAO and RSD signals.
When BOSS and eBOSS data are analyzed in combination with CMB data, the additional constraining power of the {\it Shape} compared to BAO+RSD is unimportant. 
It is worth mentioning, however, that ShapeFit -assuming a power law power spectrum with fixed spectral index and a BBN prior on the physical baryon density- provides from late-time observations alone a bound on the sum of neutrino masses of $\Sigma m_\nu<0.4$ eV (95\% confidence), consistent with $\Sigma m_\nu<0.24$ eV  provided by CMB alone including lensing.
ShapeFit in combination with Planck data yields $\Sigma m_\nu< 0.082$ eV, the tightest bound on the sum of neutrino masses from cosmology to date. The spatial curvature constraints from LSS alone improves by a factor 2.7 from the measurement of the {\it Shape}. 

For dark energy models with constant equation of state parameter and $w_0-w_a$ parameterization,  the  ShapeFit approach produces constraints on cosmological parameters comparable to those obtained by the combination of CMB+BAO+RSD.
In other words, ShapeFit provides a late-time probe that can, in some cases, substitute the use of CMB data and yet achieve comparable statistical power. 

It is important to stress that the compressed variables approach has several advantages compared to the alternative, full modeling (FM) approach where power spectra are directly compared to theory predictions for a given cosmological model to constrain the model's parameter. In the FM approach, the model needs to be chosen {\it ab inito}, hence the full analysis must be repeated to explore different models or families of models.  
As table~\ref{tab:other_results} summarizes, the compressed variables can be translated very quickly and at low computational cost into cosmological parameters for any cosmological model of choice. Section~\ref{sec:cosmo-comparison-FM} and Ref.~\cite{BriedenPRL21} further show that this compression is effectively lossless at least for $\Lambda$CDM. 

The compression aims at isolating the part of the signal whose information content is least affected by systematics. 
The compressed variables approach is model-agnostic; the model-dependence is introduced only at the interpretation stage, at the very end of the process.  
The physical parameters, capturing the effect on clustering of the physical processes at play, offer a simple way to disentangle late-time physics information from imprints of early-Universe physics. 
This is of value for going beyond simple parameter-fitting and for pursuing ways to test the model and its underlying assumptions. 
For a given cosmological model, constraints on the model's parameters come from a variety of signatures of disparate physical processing acting at different cosmic epochs; early-and late- time effects are intrinsically related. However, by measuring them separately  as in the compressed variable approach, the early- and late-time physical variables provide a powerful consistency test of the cosmological model.
The establishment of the $\Lambda$CDM model as the standard cosmological model, along with the avalanche of data in the first decade of the 2000', has transformed cosmology into a precision science. In precision cosmology the parameters of a given model (especially the $\Lambda$CDM model and simple, one-parameter extensions) are measured incredibly precisely. But precision is meaningless without accuracy: the standard cosmological model is likely an effective model, as a first-principle physical understanding of dark matter and dark energy is still lacking. To move beyond precision cosmology, it is important to devise analyses and approaches that are as much as possible model-independent or model agnostic, and that make possible to test (some of) the model's underlying assumptions.
ShapeFit represents an effort in this direction.

\appendix 
\section{Impact of the non-local Lagrangian bias in the shape parameter}\label{app:nonlocal_bias}

Here we explore the correlations arising between the shape parameter $m$ and the non-local bias parameters when we drop the local-Lagrangian bias assumption. 

We start by fitting the mean of the EZ mocks under different assumptions for $m$, $b_{\rm 3nl}$ and $b_{\rm s2}$. In order to avoid any interference with the survey geometry, or with the intrinsic redshift evolution within light-cone of the mock, we fit the averaged power spectrum of the 479 independent realizations of the EZmocks periodic cubic boxes. These boxes, whose length size is $L=5\,{\rm Gpc}h^{-1}$ and output redshift $z=0.675$, are one of the 5 snapshots used to construct the full EZ light-cone mocks, which are utilized to derive the covariance of the data. The covariance matrix for this specific analysis is derived from the 479 independent realizations of the boxes. As a data vector, we consider the mean of the 479 realizations, but use the scaled covariance corresponding to just one of the cubic boxes, which has around 100 times the number of galaxies of the full north and southern light-cones combined from $0.6<z<1.0$. This ensures that the statistical errors are sufficiently small for identifying systematics relevant for the actual data sample.

\begin{table}[htb]
    \centering
    \begin{tabular}{|c|c|c|c|c|}
    \hline
        Model & Baseline  & Expected value & Recovered value & $\chi^2/{\rm d.o.f.}$  \\
         & cosmology  & $m^{\rm exp}\times10^3$ & $m\times10^3$ &   \\
        \hline
        \hline
        Local, $m$ free & Fiducial  & $-11.5$ & $-36.7\pm 7.4$ & $46.0/(39-8)$ \\
        Local, $m$ free & EZ (true) & 0 & $-29.5 \pm 7.2$ & $46.0/(39-8)$ \\
        Local, $m\equiv0$ & EZ (true) & 0 & $0$ & $68.2/(39-7)$ \\
        \hline
         $b_{\rm 3nl}$ free, $m$ free & EZ (true) & 0 & $16 \pm 11$ & $38.1/(39-9)$ \\
         $b_{\rm 3nl}$ free, $m\equiv0$ & EZ (true) & 0 & $0$ & $38.5/(39-8)$ \\
         \hline   
         $b_{\rm s2}$ free, $m$ free & EZ (true) & $0$ & $-28 \pm 11$ & $44.5/(39-9)$ \\  
         $b_{\rm s2}$ free, $m\equiv0$ & EZ (true) & 0 & $0$ & $55.1/(39-8)$ \\
        \hline
    \end{tabular}
    \caption{Constraints derived from the mean of 479 independent  EZmock realizations of  periodic cubic boxes at $z=0.675$. We display the expected and measured values of $m$ and the minimum $\chi^2$ value for different setups and baseline cosmologies. `Local' stands for the fiducial setup used for the main analysis of the paper, in short this is $\ell=0,2,4$ and $0.02<k\,[h{\rm Mpc}^{-1}<0.15$ and assuming for $b_{s2}$ and $b_{3\rm nl}$ the local Lagrangian relations. Variations of this Local set up model are: $m\equiv0$ i.e., fixing the $m$ parameter to 0, and/or allowing $b_{\rm 3nl}$ and $b_{\rm s2}$ to float free. The associated error corresponds to just one of the boxes, which has around 100 times the number of galaxies of NGC+SGC $0.6<z<1.0$ LRG sample. When $b_{\rm 3nl}$ is set to be local, it takes the value of $\sim 0.11$ (given by the value of $b_1\sim 2.1$). When it is set to be free it is constrained to be $b_{3\rm nl}=-0.44\pm 0.12$ when $m$ is set to 0; and $b_{3\rm nl}=-0.54\pm 0.13$ when $m$ is free (in this case $b_{3\rm nl}$ and $m$ have a cross-correlation factor of $-0.6$). Similarly, when letting $b_{s2}$ free we find $b_{s2}=2.45\pm0.92$ for $m\equiv0$, and $b_{s2}=1.3\pm1.6$ when $m$ is 
    free to vary. For the local Lagrangian bias case we find $b_{s2}\sim -0.64$ given the value of $b_1\sim2.1$.
    }
    \label{tab:msys}
\end{table}

Table~\ref{tab:msys} reports the results of fitting the signal from the EZmocks cubic boxes. For conciseness we show the best-fit values of $m$ and the minimum $\chi^2$ obtained for different pipeline choices: `Local' is the fiducial pipeline used in the main analysis of this paper where the non-local biases are kept fixed to their local Lagrangian predictions; $m$ can be freely varied (along the rest of parameters as described in section~\ref{sec:theory_RSDcomp}) or can be fixed to $m\equiv0$ when doing the fit. In addition, we show the results for the baseline cosmology set to the fiducial case (used through the main text of this paper) and the own true cosmology of the EZmocks (see table~\ref{tab:cosmo} for the details on these two baseline cosmologies). Note that this baseline cosmology is  used  exclusively for computing the template of our model.\footnote{Since the signal is measured from the Cartesian positions of galaxies in the periodic boxes, we do not require any cosmology for transforming redshifts into comoving distances, and therefore no Alcock-Paczysnki effect is induced.}\footnote{In all the fits we include the grid correction for which $P(\langle {\bf k} \rangle)\neq \langle P({\bf k})\rangle$ as described in eq. 6.1-6.2 of \cite{ShapeFit}, although for such large box it turns out  to be a minor correction on $m$ for the $k$ range considered, $0.02<k\,[{\rm Mpc}^{-1}h]<0.15$, which shifts 4 units of $m\times 10^3$ towards positive values.} For reference, we add the expected value of $m$, which by definition is $0$ when the own true cosmology of the mocks is used as a baseline cosmology model. Under this simplistic approach we see how the recovered $m$ value is shifted away from its expected value when the `Local' set up with $m$ freely varied is used. For both fiducial and EZ cosmologies we find that $m$ moves towards negative values by about $0.025 - 0.030$. This result is in line with what was found in figure~\ref{fig:signal} when the individual light-cone Patchy and EZmocks were analyzed, and thus 
excludes that spurious effects of the modelling of the window, or the intrinsic redshift evolution for the light-cone mocks
can be the cause
for this mismatch when recovering $m$. 
When we force $m$ to be 0 we find that the value of best-fitting $\chi^2$ significantly increases, just confirming that $m=0$ is not a desired solution for the EZmocks. We then consider two follow up variations of this set up, allowing $b_{\rm 3nl}$ and $b_{\rm s2}$ to freely vary, with and without setting $m=0$. We find that allowing $b_{\rm s2}$ to be free, does not help significantly to recover the expected $m$. On the other hand, we find that when $b_{\rm 3nl}$ is freely varied, we recover a more consistent $m$, at the same time that the best-fitting $\chi^2$ reduces significantly. For this specific case we find that the best-fit value for $b_{\rm 3nl}$ is $b_{\rm 3nl}(m\equiv 0)=-0.44\pm0.12$ and $b_{\rm 3nl}(m\,{\rm free})=-0.54\pm0.13$; whereas for the local case $b_{\rm 3nl}({\rm local})\sim 0.11$. We discuss below the physical interpretation of these results.

\begin{figure}[htb]
    \centering
    \includegraphics[scale=0.55]{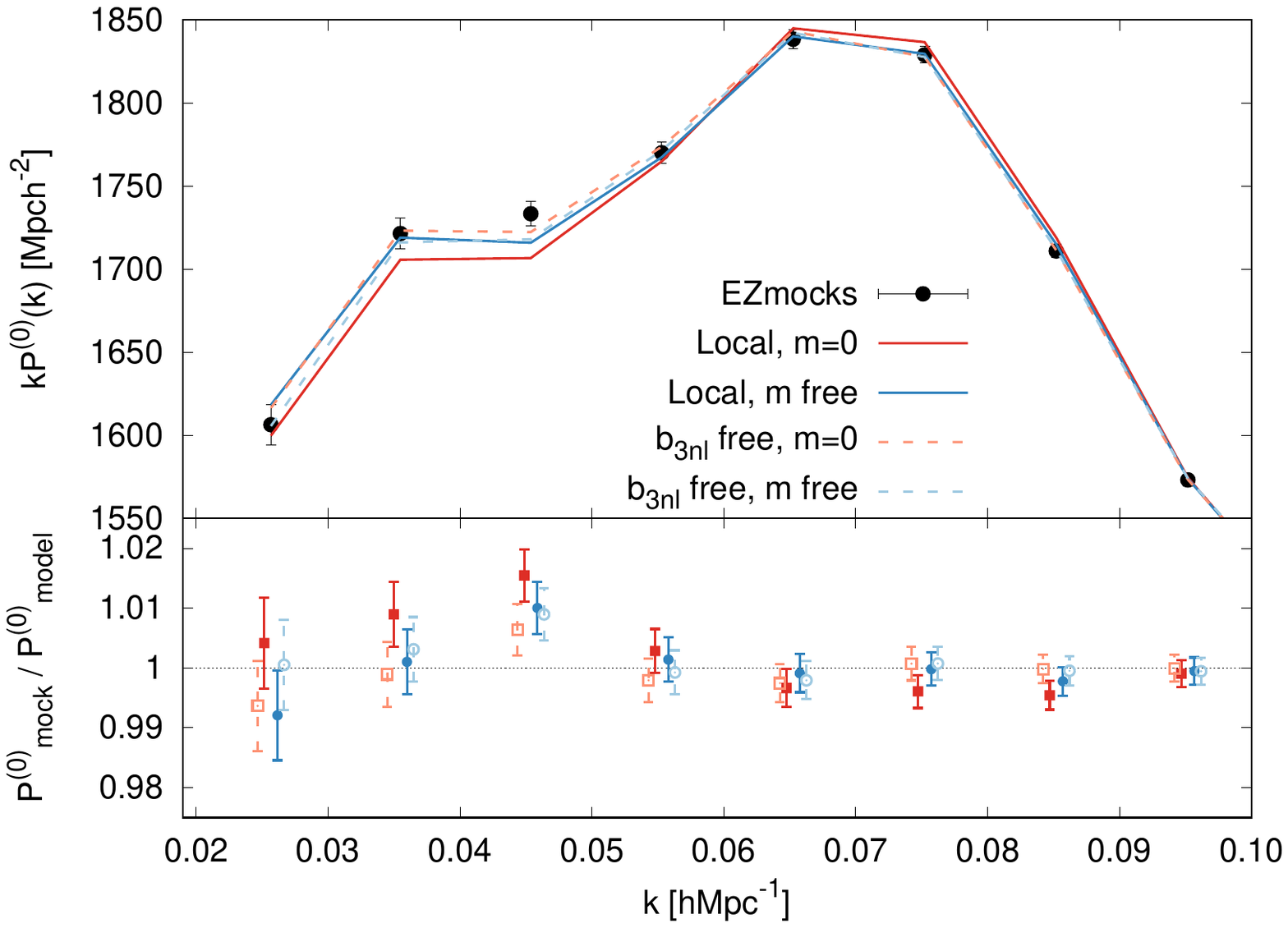}
    \caption{Performance of some of the models listed on table~\ref{tab:msys} (lines with colours) on the average of the 479 realizations of the periodic cubic EZmocks (black dots with error-bars). Red colour represents models where $m$ has been set to be 0 (its expected value), blue colour represents models with $m$ freely varied; solid lines represents models with Local Lagrangian conditions for $b_{3\rm nl}$, and dashed where $b_{3\rm nl}$ is freely varied. 
    In all cases the models employ the true base-line template from the EZmocks. Although only the monopole is shown, we simultaneously fit  the three relevant multipoles within the range $0.02<k\,[h{\rm Mpc}^{-1}]<0.15$, as for the data. The best-fitting values for $m$, $b_{\rm 3nl}$ and minimum $\chi^2$ are displayed in table \ref{tab:msys}.}
    \label{fig:m_test}
\end{figure}

Figure~\ref{fig:m_test} shows the performance of four representative models described in table~\ref{tab:msys}: (top panel) red lines correspond to the models with $m\equiv m^{\rm exp}=0$; blue lines to the models with $m$ as a free parameter; solid lines to models with $b_{\rm 3nl}$ set to its local Lagrangian prediction; dashed lines to the models where $b_{\rm 3nl}$ is free; for all  cases $b_{\rm s2}$ is set to is local Lagrangian prediction. Black dots with errorbars correspond to the measured signal from the cubic EZ mocks. As in table~\ref{tab:msys}, the error-bars correspond to $1\sigma$ of the volume of a single cubic realization. The bottom sub-panel display the ratio between the mocks measurement and the model best-fit, using the same color scheme. We see how floating $m$ free with local $b_{\rm 3nl}$ (solid blue line) produces a power spectrum  similar to that obtained by  setting $m=0$ and floating $b_{\rm 3nl}$ free (red dashed line), illustrating the degeneracy between $m$ and $b_{\rm 3nl}$. We also see very clearly how the signal (or feature in the power spectrum) responsible for obtaining a $m\neq m^{\rm exp}$ when $b_{\rm 3nl}$ is fixed to the local Lagrangian prediction, is localized mainly at  $k\simeq0.045\,h{\rm Mpc}^{-1}$ (solid red squares deviating from the horizontal black dotted line in the bottom sub-panel). As we have commented in Sec~\ref{sec:sys_budget} this behaviour is a spurious signal caused by the method adopted to produce the fast EZmocks. The EZmocks rely on the Zeldovic approximation and thus fail to include  key pre-virialization terms which slightly affect the amplitude of the power spectrum at large scales. These missing negative terms are responsible for an excess of power in the EZmocks at large scales,  mimicking the effect of a non-local bias, and thus pushing the recovered  $m$ to negative values under the local Lagrangian bias assumption for $b_{\rm 3nl}$ (or pushing $b_{\rm 3nl}$ away from its local prediction for $m=0$). This argument is supported by the results obtained  when fitting $m$ on N-body mocks, such as the PT challenge mocks (see \cite{ShapeFitPT} and section~\ref{sec:sys_budget}) where for a similar HOD we find results consistent with $m=0$, as well as for the local Lagrangian prediction on $b_{\rm 3nl}$ and $b_{\rm s2}$.

\begin{figure}[htb]
    \centering
    \includegraphics[scale=0.35]{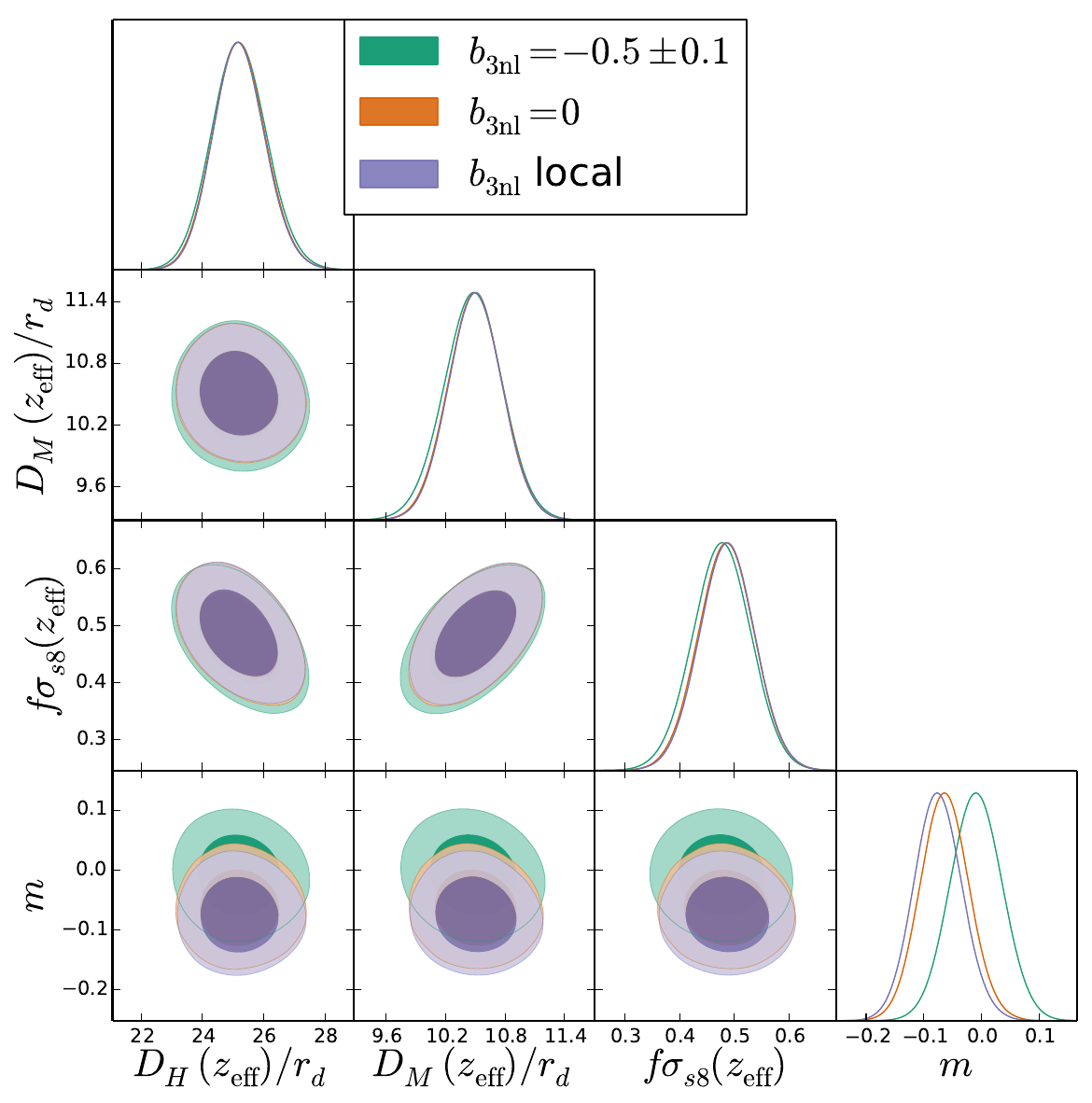}
        \includegraphics[scale=0.35]{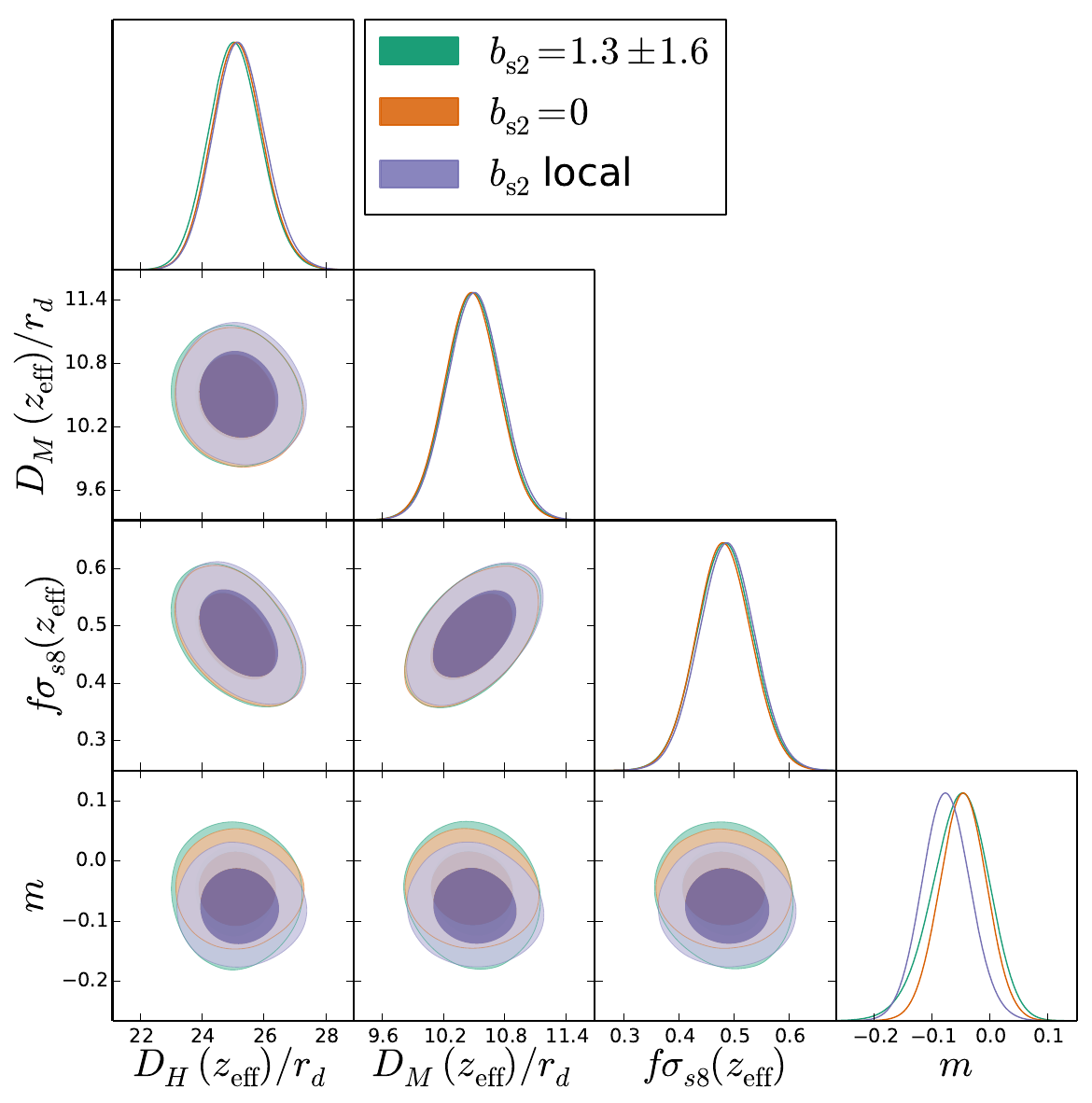}
    \caption{Impact of the assumptions for the non-local biases, $b_{\rm 3nl}$ (left panel) and $b_{s2}$ (right panel) on the cosmology parameters for the LRG sample $0.2<z<0.5$ data sample for the pre-reconstruction catalogue only. The colours represent different choices on these bias parameters, as indicated in the legend. The green contours display the results with the written Gaussian prior on these non-local bias parameters (motivated by the value preferred by the EZmocks), the orange contours the results when setting these biases to 0; the purple contours the results of setting them to the local Lagrangian prediction (fiducial pipeline choice for these work). See table~\ref{tab:b3nl_choice} for the numerical values of the left panel, along with the results when fitting the other data samples.}
    \label{fig:biases2l}
\end{figure}

The choice of assuming the local Lagrangian prediction on our main pipeline analysis applied for the data is hence well justified: when analyzing N-body mocks populated by galaxies following realistic HOD models, we find no sign of departure from locality. However, we want to test the impact of relaxing this assumption on the data,  to quantify  how much our results would change. This is displayed in figure~\ref{fig:biases2l} for the LRG $0.2<z<0.5$ sample:  the $b_{\rm 3nl}$($b_{\rm s2}$) bias parameter is allowed to depart from its local prediction in the left(right) panel. For each type of bias we show 3 cases: its local case (purple contours), the case where its value is set to 0 (orange contours) and the case where we float this parameter free with a tight Gaussian prior around the value preferred by the EZmocks. Although we know that the best-fit value of $b_{\rm 3nl}$ and $b_{\rm s2}$ for the EZmocks is an unphysical artifact of the fast techniques employed to generate these mocks, we assume those best-fitting non-local bias results as extreme cases if, mistakenly, the non-local parameters of the data were fixed by the signal of the EZmocks. In addition, table~\ref{tab:b3nl_choice} reports the best-fit values of $m$ for the 4 data samples when $b_{\rm 3nl}$ is kept to be local\footnote{Note that the values of $m$ for the $b_{\rm 3nl}$ local cases of the LRG samples are slightly different from those reported by table~\ref{tab:boss_compare}, as in this case we only use the pre-reconstruction catalogues for simplicity.}, when it is set to 0, and when is freely varied with a tight Gaussian prior around the preferred value when fitting the EZmocks.

From the right panel of figure~\ref{fig:biases2l} we see that the impact of the choosing $b_{\rm s2}$ to be local, 0, or to be set around the value preferred for the EZmocks, has a very minor impact on $m$, and no impact at all on the rest of physical parameters. From the left panel we conclude that, if $b_{\rm 3nl}$ departs from locality, fixing it to $b_{\rm 3nl}=0$ has no impact on the physical parameters, but setting it around the best-fit value preferred by the mocks, shift $m$ by 1.5$\sigma$. 

\begin{table}[htb]
    \centering
    \begin{tabular}{|c|c|c|c|}
    \hline
        Sample & $b_{\rm 3nl}$ local & $b_{\rm 3nl}=0$  & $b_{\rm 3nl}=-0.5\pm0.1$   \\
        \hline
        \hline
        LRG $0.2<z<0.5$ & $-0.075 \pm 0.042$ & $-0.063 \pm 0.042$ & $-0.009 \pm 0.045$  \\
        LRG $0.4<z<0.6$ & $-0.033 \pm 0.044$ & $-0.021 \pm 0.044$ & $0.041 \pm 0.047$  \\
        LRG $0.6<z<1.0$ & $-0.019 \pm 0.052$ & $-0.012 \pm 0.053$ & $0.017 \pm 0.055$  \\
        QSO $0.8<z<2.2$ & $-0.005 \pm 0.033$ & $0.002 \pm 0.033$ &  $0.028 \pm 0.035$  \\
        \hline
         
    \end{tabular}
    \caption{Best-fit $m$ values under different assumptions on $b_{\rm 3nl}$ for the four different data samples (pre-reconstruction only). $b_{s2}$ is assumed local Lagrangian. }
    \label{tab:b3nl_choice}
\end{table}

Table~\ref{tab:b3nl_choice} displays the effect of changing the assumption on the locality of $b_{\rm 3nl}$ for the four studied samples. As seen in the left panel of figure~\ref{fig:biases2l} for the LRG $0.2<z<0.5$ sample, only when $b_{\rm 3nl}$ departs from locality and it is set around the preferred value for the EZmocks, the locality assumption has a noticeable impact on the best-fit $m$, yet no impact on the rest of the physical parameters. This shift in $m$ oscillates between $0.5\sigma$ and $1.5\sigma$ with respect to the local case, always towards positive values. 

\begin{figure}[tb]
    \centering
    \includegraphics[scale=0.45]{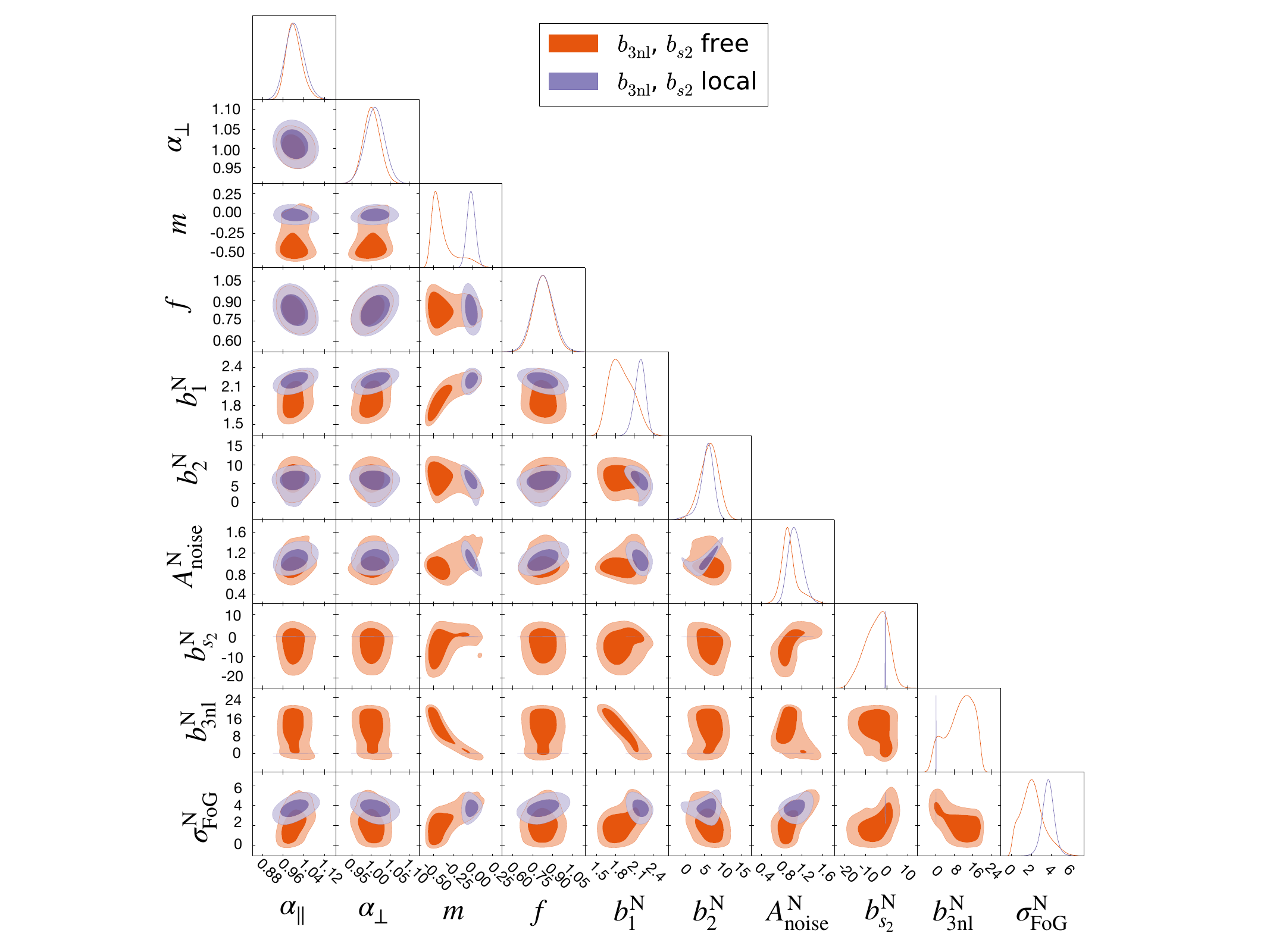}
    \caption{Posteriors for the pre-recon only BOSS+eBOSS LRG sample $0.6<z<1.0$. For clarity only the nuisance parameters of the NGC are shown (although the SGC is simultaneously constraint). The purple contours display the results when both $b_{\rm 3nl}$ and $b_{\rm s2}$ are set to their local Lagrangian predictions; the orange contours when they are freely varied with a wide uniform prior of $0\pm20$. }
    \label{fig:eboss_bfree}
\end{figure}

We conclude that the effect of $b_{s2}$ being local, or being set to the two explored non-local options has no impact at all on the physical parameters given the statistical precision of the data. The effect of $b_{\rm 3nl}$ is more important, as anticipated at the beginning of this appendix. Although the local and $b_{\rm 3nl}=0$ choices show no significant difference among them given the size of statistical errors of the data, setting it to $-0.5\pm0.1$ significantly shifts the $m$ best-fit values, by around $1.5\sigma$. Although there is no strong evidence that the observed galaxies have such strong non-local behaviour (such extreme $b_{\rm 3nl}$ values), the strong correlation found between $b_{\rm 3nl}$ and $m$ indicates that - for present and forthcoming surveys -  the interpretation of the shape $m$ as a cosmology observable, regardless whether is done via the ShapeFit compression or via direct fits to the full $P(k)$ shape, relies on certain assumptions about the bias properties of the studied sample of galaxies. The other compressed  parameters (and thus the cosmology extracted from them)  are insensitive to these assumptions.

Finally, in figure~\ref{fig:eboss_bfree} we display the full triangle plot of the pre-recon posteriors for the LRG eBOSS $0.6<z<1.0$ NGC sample, for the case where both $b_{\rm s2}$ and $b_{\rm 3nl}$ are set to their local Lagrangian prediction (purple contours) and where they are floated free with a wide uniform prior between $0\pm 20$. 
We see how in the case the two non-local parameters are allowed to float free the constrains on $m$ largely degrade, returning essentially no-information from the shape. In addition, we see that both $b_{s2}$ and $b_{\rm 3nl}$ hit the conservative prior limits, again indicating that the shape of the power spectrum is unable to simultaneously constrain $b_{\rm s2}$, $b_{\rm 3nl}$ and $m$ from power spectrum data alone. Because of this we do not report any preference of the data for non-local bias values. The inclusion of higher-order statistics in future data analysis may help to mitigate these large degeneration directions observed among $\{b_{\rm 3nl},\,b_{\rm s2},\,m\}$.

\section{Fiber collision effect in the quasar sample}

We aim to account for one of the main observational systematic effect of the eBOSS quasar samples, the fiber collisions. This effect is caused by the physical size of the optical fibers, which cannot be placed in the BOSS and eBOSS plates sufficiently close to each other to collect the spectra of imaging targets closer than a limiting angular size of 
62'', which for the quasar sample represents a physical distance of $\sim0.9\,{\rm Mpc}h^{-1}$. Imaging targets closer than this minimum angular separation will have spectroscopic redshifts  only  if they lie in regions of the sky that have observed by more than one plate.

Thus, the eBOSS quasar catalogues tends to miss the redshift information of objects in regions where the density of objects is high. Not correcting by the effect of fiber collisions induces a change in the 3D clustering of the quasars (especially along the LOS), which can potentially bias the physical parameters we measure. 

The fiber collision effect is imprinted in the EZmock quasar sample, and therefore by computing the physical parameters with and without the correction we can quantify the effect and validate the correction technique. We follow the approach proposed by \citep{Hahnetal17} which accounts for the effect of fibre collisions in the modelling part (i.e., it does not correct the data catalogues, but the fitted model), which is also followed by the eBOSS quasar team analysis in Fourier space \citep{neveuxetal2020}, as well as in configuration space \cite{houetal2021}.

In short, we modify the measured power spectrum by the following scale-dependent correction, $P^{\rm true}_\ell(k)=P^{\rm meas.}_\ell(k)-\Delta P_\ell(k)$, where $\Delta P_\ell(k)$ has two additive components, the uncorrelated and correlated,

\begin{eqnarray}
    \label{eq:cp1}\Delta P^{\rm uncorr}_\ell(k)&=&-f_{s}(2\ell+1)\mathcal{L}_\ell(0)\frac{(\pi D_{\rm fc})^2}{k} )W_{2\rm D}(kD_{\rm fc}),\\
    \nonumber\Delta P^{\rm corr}_\ell(k)&=&-\frac{(2\ell+1)f_s D_{\rm fc}^2}{4}\sum_{\ell'=0}^{\ell'_{\rm max}}\int_{k_{\rm min}}^{k_{\rm max}}qdq\,P_{\ell'}(q) \int^{\min(1,q/k)}_{\max(-1,-q/k)}d\mu \mathcal{L}_\ell(\mu) \\
    \label{eq:cp2}&\times& \mathcal{L}_{\ell'}(k\mu/q)W_{2\rm D}(qD_{\rm fc}),
\end{eqnarray}
where, $W_{\rm 2D}(x) \equiv 2J_1(x)/x$ is the top-hat function in 2D, and $J_1$ is a Bessel function of the first kind and of order 1. The total effect of these collisions is $\Delta P_\ell(k)=\Delta P_\ell^{\rm uncorr}(k)+\Delta P_\ell^{\rm corr}(k)$. As suggested in  \cite{neveuxetal2020} we take $f_c^{\rm NGC,SGC}=0.36,\,0.45$ (the fraction of objects affected by the fiber collisions in the northern and southern caps) and $D_{\rm fc}=0.9\,{\rm Mpc}h^{-1}$ (the collision radius at the effective redshift of the quasars, $z=1.48$). Although $\ell'_{\rm max}$ and $k_{\rm max}$ should extend to infinity, and $k_{\rm min}$ to 0, for practical reasons we choose $\ell'_{\rm max}=8$, $k_{\rm min}=10^{-3}\,h{\rm Mpc}^{-1}$ and $k_{\rm max}=0.4\,h{\rm Mpc}^{-1}$. As discussed in \cite{Hahnetal17} the effect of truncating $k_{\rm max}$ only affects the amplitude of the correction of the monopole, which is in any case re-absorbed by the free parameter which regulates the amplitude of shot noise. With these numbers, \citep{neveuxetal2020} determined that for the eBOSS quasars only $5\%$ of the collisions were correlated (being actually groups of quasars physically close to each other), which implies that the dominant part of the $\Delta P_\ell(k)$ shift is $P_\ell^{\rm uncorr}(k)$, which is independent of the underlying true clustering of the quasars. 

\begin{figure}[tb]
    \centering
    \includegraphics[scale=0.32]{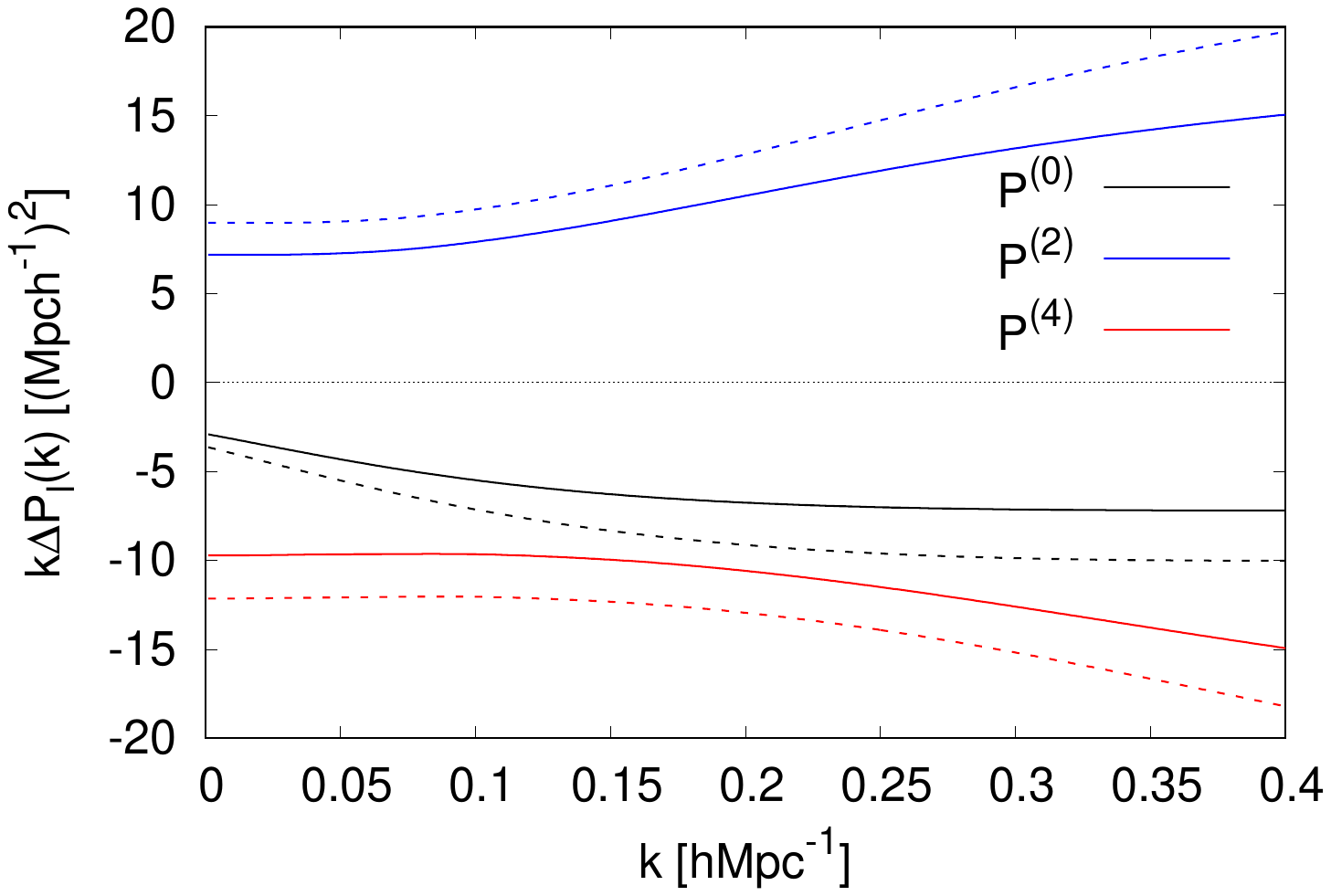}
        \includegraphics[scale=0.32]{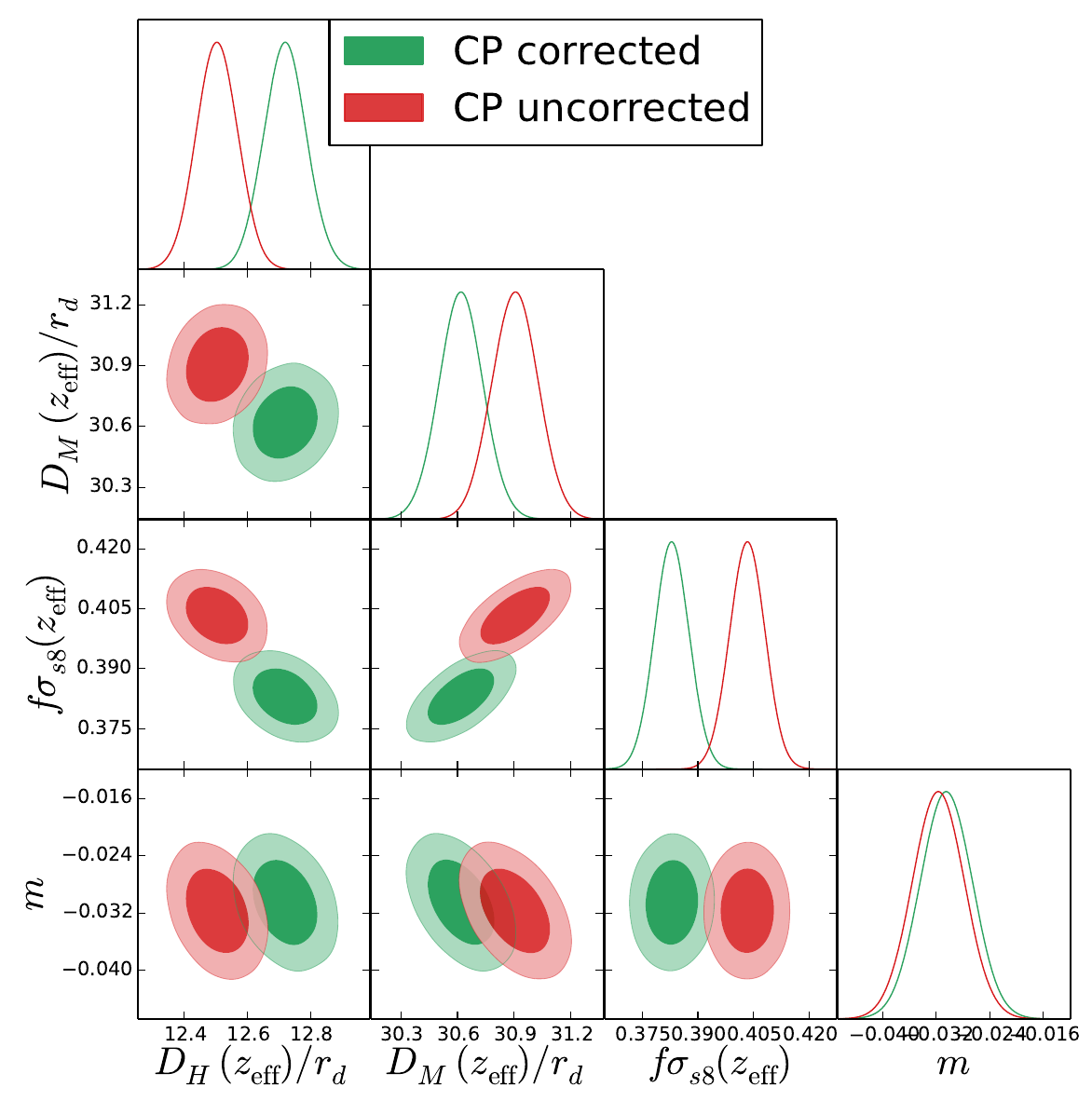}
    \caption{Left panel presents the close pair (CP in the figure legend)  correction proposed by \cite{Hahnetal17} implemented for the quasar sample according to eqs.~\eqref{eq:cp1}-\eqref{eq:cp2}. The solid/dashed lines display the total $\Delta P_\ell(k)$ correction for the northern/southern galactic cap, and the colors correspond to different $\ell$-multipoles, as displayed. The right panel displays the impact that ignoring the close pair correction has in physical parameters inferred from the mean power spectra of the 1000 realizations of quasar EZmocks: the fiber collision effect has been simulated in the EZmocks catalogues, and the (red) green contours show the analysis (not) correcting for them. The size of the contours correspond to a volume 100 times the volume of the data. Although not shown, the minimum $\chi^2$ improves when correcting for the close pairs from $202.7/(168-12)$ down to $132.1/(168-12)$, mainly due to the shift in the hexadecapole signal at all scales.}
    \label{fig:CP}
\end{figure}

The left panel of figure~\ref{fig:CP} displays the total shift $\Delta P_\ell(k)$, for the northern (solid) and southern (dashed) patches, where the different colors stands for the different $\ell$-multipole contributions as indicated. Our analysis is insensitive to an overall scale-independent shift in $P^{(0)}$ as we are marginalizing over the amplitude of the shot noise.

Relative to the statistical error-bar of the quasar sample, the largest effect turns out to be in the hexadecapole, as is the multipole less affected by the bias parameters, which can also in part absorb part of the fiber collision effects. We find results very consistent with those shown in fig. 8 of \cite{neveuxetal2020}. 

The right panel of figure~\ref{fig:CP} displays how the best-fit physical parameters shift when the mocks with the effect of fiber collisions imprinted are (not) corrected according to eqs.~\eqref{eq:cp1}-\eqref{eq:cp2} in (red) green contours. We fit the mean of the 1000 realizations of the EZmocks (northern and southern caps), although the plotted errors correspond to 10 times of the statistical error of one realization. We report a change on $D_M/r_{\rm d}$, $D_H/r_{\rm d}$ and $f\sigma_{s8}$, which shifts the uncorrected mocks best-fitting values towards the expected values according to table~\ref{tab:expected}. On the other hand, $m$ is barely affected by the fiber collision. This behaviour is  expected, as most of the signal of $m$ comes from the derivative of $P^{(0)}$ at scales of $k\sim0.05\,{\rm Mpc}^{-1}h$, which are barely affected by fiber collisions. Overall, the reported shift on the mocks represents a fairly small fraction of the error of the data, which is typically 10 times larger than the errors displayed by the plot. 

\begin{table}[htb]
    \centering
    \begin{tabular}{|c|c|c|c|c|c|}
    \hline
          & $D_H/r_{\rm d}$ & $D_M/r_{\rm d}$ & $f\sigma_{s8}$ & $m$ & $\chi_{\rm min}^2$  \\
         \hline\hline
    CP corr.  & $13.27 \pm 0.50$ & $31.01\pm 0.82$ & $0.468\pm 0.042$ & $-0.005\pm 0.033$ & $117.0$ \\
    CP uncorr. & $13.11 \pm 0.49$ & $31.23\pm 0.82$ & $0.482\pm 0.042$ & $-0.009\pm 0.034$ & $117.5$ \\
    \hline
    \end{tabular}
    \caption{Effect of the fiber collisions on the best-fitting physical parameters of actual quasar data sample. CP corr. stands for the results when the fitted model accounts for the close pairs corrections via eqs.~\eqref{eq:cp1}-\eqref{eq:cp2}; CP uncorr. display the best-fits when this correction is ignored. Along with the best-fitting values we also report the valaue of the minimum $\chi^2$ with 168-12 degrees of freedom.}
    \label{tab:cp}
    \end{table}
    
    Table~\ref{tab:cp} presents the results of performing the fit of the actual quasar data taking into account the close pair correction of eqs.~\eqref{eq:cp1}-\eqref{eq:cp2} (CP corr.), or ignoring it (CP uncorr.). We observe a fairly small shift, which cannot be distinguished from the intrinsic cosmic variance noise. However, as for the mocks, we see that $D_H/r_{\rm d}$ tends to increase by $1/3\sigma$, $D_M/r_{\rm d}$ and $f\sigma_{s8}$ tend to decrease by $1/4\sigma$ and $1/3\sigma$, respectively. On the other hand, $m$ is barely affected ($\sim1/10\sigma$). We also report a slight improvement of the minimum $\chi_{\rm min}^2$ value of the fit, mainly due to an increase of the hexadecapole signal. 
    
    In summary, we see that the close pairs correction has a measurable effect only when the effective volume of the sample is about $\times100$ the effective volume of the actual eBOSS data quasar sample, which shifts the posteriors of the BAO distances and $f\sigma_{s8}$ by a magnitude of $\lesssim 2\sigma$, whereas the shape parameter $m$ is unaffected. Nevertheless, for correctness, we do include this correction in our main analysis of the data, but we do not expect that inaccuracies on the exact free parameters of this model, such as the values of $f_c$ have any significant effect on the final results of this paper.

\section{Impact of the prior assumptions} \label{app:prior}
We discuss the motivations for and the effects of 
including certain priors 
in the analysis presented in the main text. We divide this appendix in two sections, the first one  about 
the priors on nuisance parameters for the model-agnostic fit to the measured power spectra, and the second  about 
the priors set during the interpretation of the compressed physical variables within a certain cosmology model. 

\subsection{Prior assumptions on the model-agnostic analysis}
\begin{figure}[tb]
    \centering
    \includegraphics[scale=0.45]{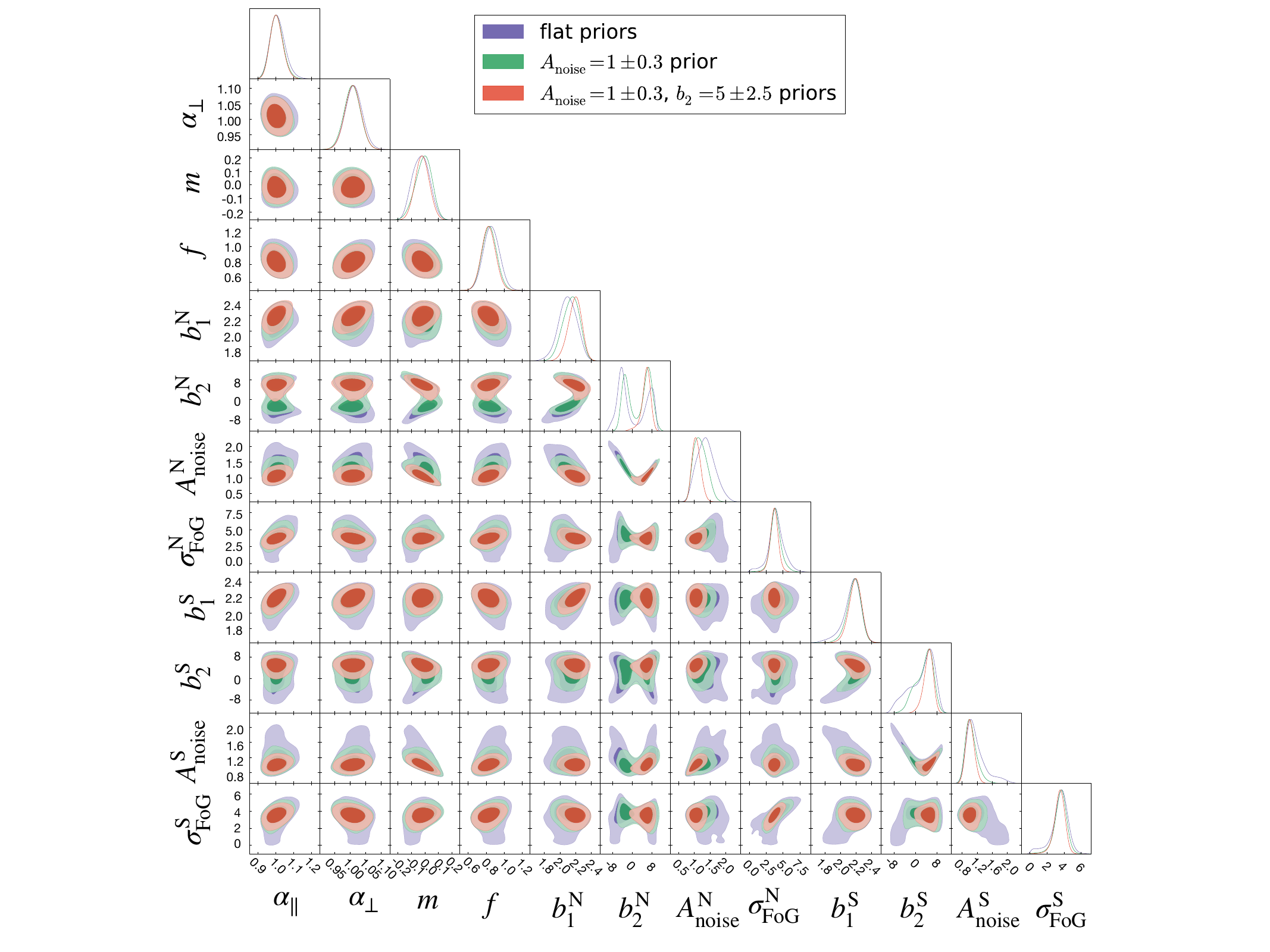}
    \caption{Posteriors for the pre-recon only BOSS+eBOSS LRG sample $0.6<z<1.0$. In purple the constraints using uniform wide priors for $A_{\rm noise}$ and $b_2$; in green when a Gaussian prior of $1\pm0.3$ is set for $A_{\rm noise}$; and in dark orange when additionally a Gaussian prior of $5\pm2.5$ is also set for $b_2$. For the rest of parameters uniform wide priors are used as presented in table~\ref{tab:results-priors}.}
    \label{fig:priors_LRGeboss}
\end{figure}

In section~\ref{sec:theory_RSDcomp} we have described the parametrization of the galaxy/quasar $P^{(\ell)}(k)$ model, with a set of physical and nuisance parameters. Among the nuisance parameters, the amplitude of shot noise $A_{\rm noise}$ and the second order bias $b_2$,  are not particularly well constrained by power spectrum data alone, especially when the signal-to-noise ratio is not very high. As shown in table~\ref{tab:results-priors} for all the samples we have set a Gaussian prior around the Poissonian noise prediction for the amplitude of shot noise, allowing a $30\%$ deviation at $1\sigma$. This is a Gaussian prior of $A_{\rm noise}=1\pm0.3$, for both northern and southern sample parameters. We know that halo-exclusion and the intrinsic clustering signal of the galaxy and quasar samples can induce some deviations from the pure Poissonian case, $A_{\rm noise}=1$, but having this noise deviating by more than $\sim0.6$ (2$\sigma$ of our Gaussian prior) is actually extremely unphysical. Nevertheless, we display in figure~\ref{fig:priors_LRGeboss} the effect of this Gaussian $30\%$ prior in green contours compared to the case with uniform wide priors on $A_{\rm noise}$ in purple contours, for the LRG $0.6<z<1.0$ sample. Although setting this Gaussian prior has some impact in some of the nuisance parameters, it barely has any impact on the physical variables, such as the BAO scaling distances, the growth, or the {\it Shape}.

In addition to the prior on $A_{\rm noise}$ we include a wide Gaussian prior around $b_2$ when analyzing the LRG $0.6<z<1.0$ sample. %The addition of this prior 
This choice is motivated by the multi-modal posterior this sample presents in the absence of this prior, and the strong degeneracy between the amplitude of shot noise and $b_2$ (see again purple and green in figure~\ref{fig:priors_LRGeboss} for $b_2$). In particular, we find that $b_2$ prefers either $b_2\sim5$ with $A_{\rm noise}\sim1$, or $b_2\sim-5$ with $A_{\rm noise}\sim1.5$. This long degeneracy could be also restricted by imposing a tighter prior on $A_{\rm noise}$, but we decide that setting an additional Gaussian prior on  $b_2$ around $5$, with a wide $\pm2.5$ deviation at $1\sigma$ amplitude is a cleaner approach to remove this spurious result. This feature only appears for the LRG sample at $0.6<z<1.0$, but not at the other two LRG samples due to the larger effective volume. 
We show in dark orange contours of figure~\ref{fig:priors_LRGeboss} the effect of the combination of both Gaussian priors on $A_{\rm noise}$ and $b_2$.

In summary, the inclusion of the Gaussian priors on $A_{\rm noise}$ (in all the samples) and of $b_2$ (in the LRG $0.6<z<1.0$ sample) help to increase the convergence of the Monte Carlo Markov Chains and barely affects the posteriors of the physical parameters. For both parameters, having this priors help to exclude the unphysical results where the amplitude of shot noise significantly departs from its Poissonian prediction. 

\subsection{Prior assumptions within a specific cosmology model}

To extract tight cosmological constraints from LSS data alone, in particular from the BOSS and eBOSS maps presented here, it is necessary to impose physical Gaussian priors on some parameters to break degeneracies.

One of these parameters is the baryon density $\omega_\mathrm{b}$, which is well measured by BBN and the CMB, so it is instructive to adopt this measurement within the cosmological analysis. The impact of this prior assumption is already discussed in depth in section \ref{sec:cosmo-baseline}, $\omega_\mathrm{b}$ is very degenerate with $h$ and $\omega_\mathrm{cdm}$, but it does not affect the constraints on $\Omega_\mathrm{m}$ and $\sigma_8$ (see figure \ref{fig:cosmo}). 

Another parameter that we do not vary freely in our baseline analysis is the spectral tilt $n_s$. Instead, we keep it fixed to the fiducial value $n_s^\mathrm{fid}=0.97$. The reasoning is that the limited range of scales we consider for galaxy clustering are not sufficient to determine the scale-independent tilt with high accuracy, while the Planck temperature and polarization power spectra
cover significantly larger, linear scales, that enable a precise determination of $n_s$. Due to its degeneracy with the shape $m$, allowing it to vary freely in the cosmological ShapeFit analysis would significantly degrade the constraining power of $m$. As shown in \cite{ShapeFit}, it is possible to measure both $m$ and $n_s$ from the galaxy power spectra independently, but in this work we rely on the well measured knowledge on the scale independent tilt from the CMB and infer cosmological constraints from $m$ only.

\begin{figure}[htb]
    \centering
    \includegraphics[width=0.49\textwidth]{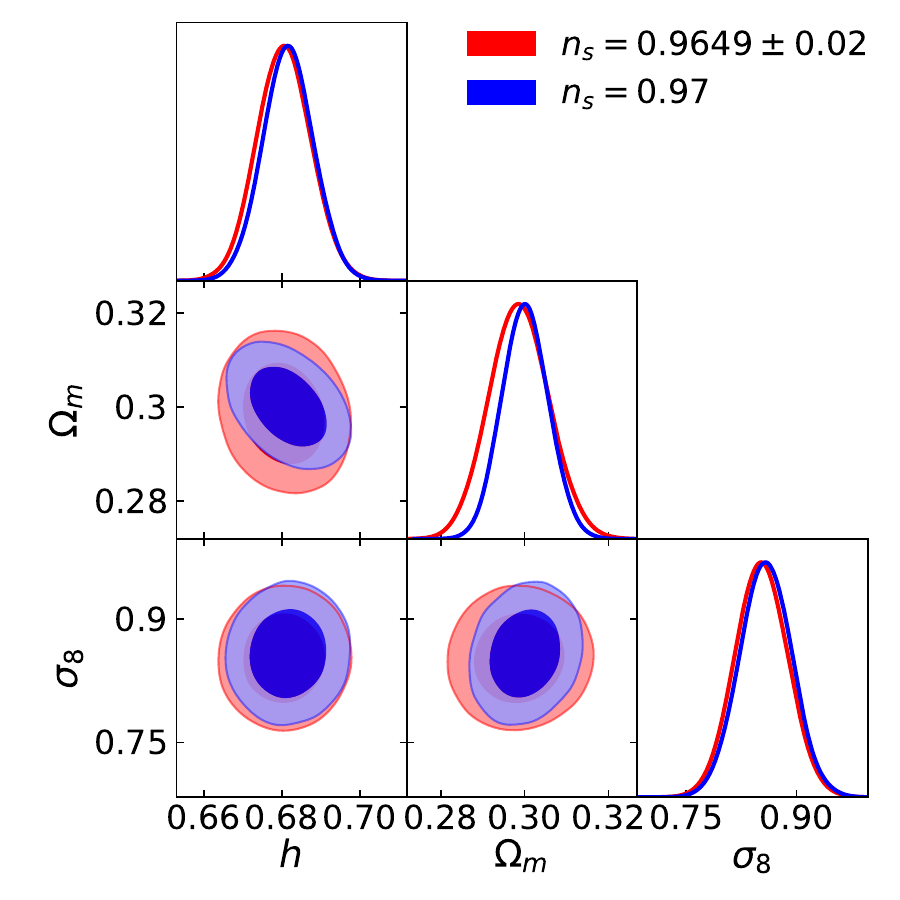}
        \includegraphics[width=0.49\textwidth]{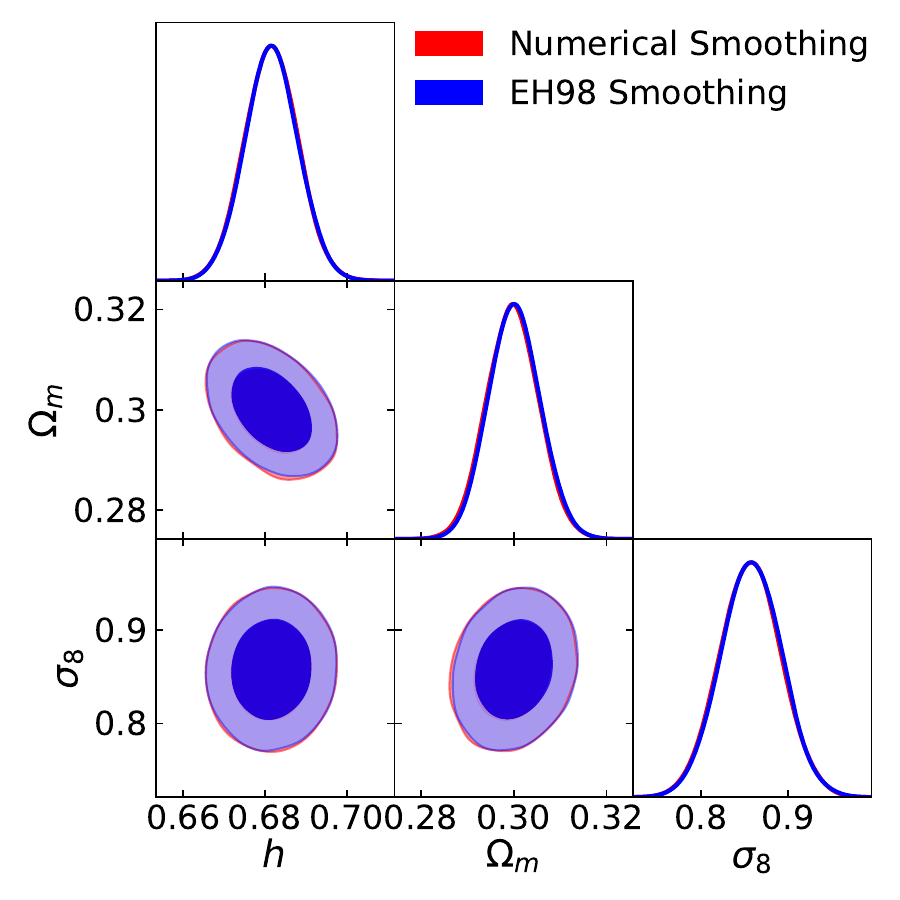}
    \caption{Left panel: Comparison between cosmological $\Lambda$CDM parameter constraints when varying the spectral index $n_s$ under a Gaussian prior (red contours) with respect to our baseline of fixing it to the fiducial value (blue contours). Right panel: Comparison of constraints for different BAO smoothing methods to determine the shape parameter $m$. We use either a numerical method to remove the BAO feature from the linear power spectrum (red contours) or the EH98 fitting function (blue contours).}
    \label{fig:prior-ns-and-smoothing}
\end{figure}

In the left panel of figure \ref{fig:prior-ns-and-smoothing} we show the impact of relaxing the prior $n_s=n_s^\mathrm{fid}$ (blue contours) by allowing to vary $n_s$ within a Gaussian prior probability distribution centered at the value $n_s^\mathrm{Planck}=0.9649$ measured by Planck, with a width of $\Delta n_s = 0.02$, which is $5\times$ larger than the Planck measured standard deviation (red contours). We can see that this has no impact on $h$ and $\sigma_8$, but slightly degrades our $\Omega_{ \mathrm m}$ constraints. While our baseline analysis returns $\Omega_{\mathrm m} = 0.3001 \pm 0.0057$, here we find $\Omega_{\mathrm m} = 0.2987 \pm 0.0073$.

We hence conclude that our baseline $\Lambda$CDM results are robust even under systematic deviations of $n_s$ from the Planck measured value. 

\section{Smoothing the BAO wiggles}\label{app:smoothBAO}

To obtain the shape $m$ prediction for any combination of cosmological parameters we need to infer the matter transfer function `no-wiggle' shape $T_\mathrm{nw}$ (or power spectrum shape $P_\mathrm{nw}$),  entering eq. \eqref{eq:theory_m}. However, this quantity with removed BAO signal is not a direct output of Boltzmann codes, and must hence be inferred in an additional step.

Several numerical methods exist to extract the BAO signal from a fiducial power spectrum template, for example, in our BAO analysis we use the method from \cite{Kirkby_BAOsmoothing}. However, this method involves a polynomial fitting routine, which does not return stable results for all cosmological parameter combinations. Even for methods that rely on simple numerical functions such as fast Fourier transforms or interpolation (see for instance \cite{Hamann_shape}) it is a non-trivial task to obtain stable results across cosmological parameter space. As argued extensively in appendix A of \cite{Hinton:2016jfq}, this is because the BAO are a relatively minor signal compared to the broadband. Finding a universal numerical BAO extraction method is further complicated by the fact that other features in the linear power spectrum -determined by the matter-radiation equality turnover and baryon suppression scales- show a rich cosmology dependence. Any numerical method relying on a set of hyperparameters, such as the one presented in \cite{Hinton:2019nky}, must therefore be validated to deliver standardizable results against that cosmology dependence. Since such a validated method does not yet exist in the literature, in this work we use the analytic EH98 formula \cite{EH_TransferFunction} to obtain our {\it Shape} predictions.

The drawbacks of the EH98 formula are its inaccuracy of up to 5\% and its lack of validity beyond $\Lambda$CDM and simple parameter extensions of varying neutrino mass $\Sigma m_\nu$, effective number of neutrino species $N_\mathrm{eff}$, curvature $\Omega_\mathrm{k}$ or dark energy equation of state $w$. While the EH98 formula is still sufficient for the range of models explored in this paper, an alternative numerical method would be beneficial for future applications. 

Here, we present a new numerical method as such an alternative and compare its performance with respect to the baseline EH98 formalism. The building block of this numerical smoothing is given by the following \texttt{scipy} (and \texttt{numpy}) functions:
\begin{itemize}
    \item[\it i)] We use the \texttt{find\_peaks} function from \texttt{scipy.interpolate} to find the minima and maxima of the gradient (using \texttt{numpy.gradient}) of the input spectrum. The scales at which these minima and maxima are found represent the nodes, that we require the output smooth function to cross. We start looking for peaks at a certain scale $k_\mathrm{start}$, which is our first hyperparameter. By default, we set $k_\mathrm{start}=0.02 \hoverMpc$
    \item[\it ii)] We use the \texttt{interp1d} function from \texttt{scipy.signal} to interpolate two smooth functions crossing all minima (maxima) identified with the help of the peak finder in {\it i)} and combine both curves into the final output by computing their average. The order $j$ of the spline interpolation is our second hyperparameter. By default we perform quadratic interpolation, i.e. we set $j=2$.
\end{itemize}
As argued before, we aim to establish a ``standardizable'' method, which delivers results independent of the hyperparameters.\footnote{While we leave the demonstration of its independence from the hyperparameters for future work, here we argue why our method is well suited to be "standardizable'': because we cancel out the standard ruler $r_\mathrm{d}$ and turnover scale $k_\mathrm{eq}$ power spectrum dependence.} To achieve this goal we proceed as follows:
\begin{enumerate}
    \item We start with our fiducial linear matter power spectrum $P_\mathrm{lin}^\mathrm{fid}$ obtained from the Boltzmann code CLASS \cite{2011JCAP...07..034B} using the fiducial values of table \ref{tab:cosmo} and calculate the corresponding smooth EH98 power spectrum $P_\mathrm{EH98}^\mathrm{fid}$. 
    \item We calculate the ratio $P_\mathrm{lin}^\mathrm{fid}/P_\mathrm{EH98}^\mathrm{fid}$ to divide out the power spectrum dependence on the turnover scale $k_\mathrm{eq}$, which is excellently modeled via the EH98 formula.
    \item We apply the numerical smoothing (steps {\it i)} and {\it ii)}) with that ratio as input.
    \item We multiply the de-wiggled output with $P_\mathrm{EH98}^\mathrm{fid}$ to obtain $P_\mathrm{num}^\mathrm{fid}$. 
    \item Given a cosmology $\mathbf{\Omega}$ different from the fiducial cosmology $\mathbf{\Omega}^\mathrm{fid}$, we compute the ratio  $P_\mathrm{lin}/P_\mathrm{EH98}$ for that cosmology after rescaling them by the sound horizon ratio $s=r_\mathrm{d}/r_\mathrm{d}^\mathrm{fid}$ (as for the nominator of eq. \eqref{eq:theory_m}).
    \item We ``standardize'' that ratio by dividing it by the pre-computed ratio for the fiducial cosmology, i.e. we compute $(P_\mathrm{lin}/P_\mathrm{EH98})/(P_\mathrm{lin}^\mathrm{fid}/P_\mathrm{EH98}^\mathrm{fid})$. 
    \item We use this ``standardized'' ratio as input for the numerical smoothing, steps {\it i)} and {\it ii)}.
    \item We multiply the de-wiggled output with $P_\mathrm{EH98} \cdot (P_\mathrm{lin}^\mathrm{fid}/P_\mathrm{EH98}^\mathrm{fid})$ to obtain the final numerically smoothed power spectrum $P_\mathrm{num}$.
\end{enumerate}
Note that steps 1.-4. are carried out only once to generate the fiducial templates, while steps 5.-8. need to be performed at each step in cosmlogical parameter space. From our tests using MCMC's we found that the ``double standardization'' procedure described here is sufficient to deliver stable results. In this way we can get rid of the dependence on $k_\mathrm{eq}$ (through the division by the EH98 formula) and the baryon suppression and onset of oscillation (through the division by the fiducial template ratio). If not taken into account, both of these effects can lead to artificial discontinuities in the determination of the shape $m$.

\begin{figure}[t]
    \centering
    \includegraphics[width=\textwidth]{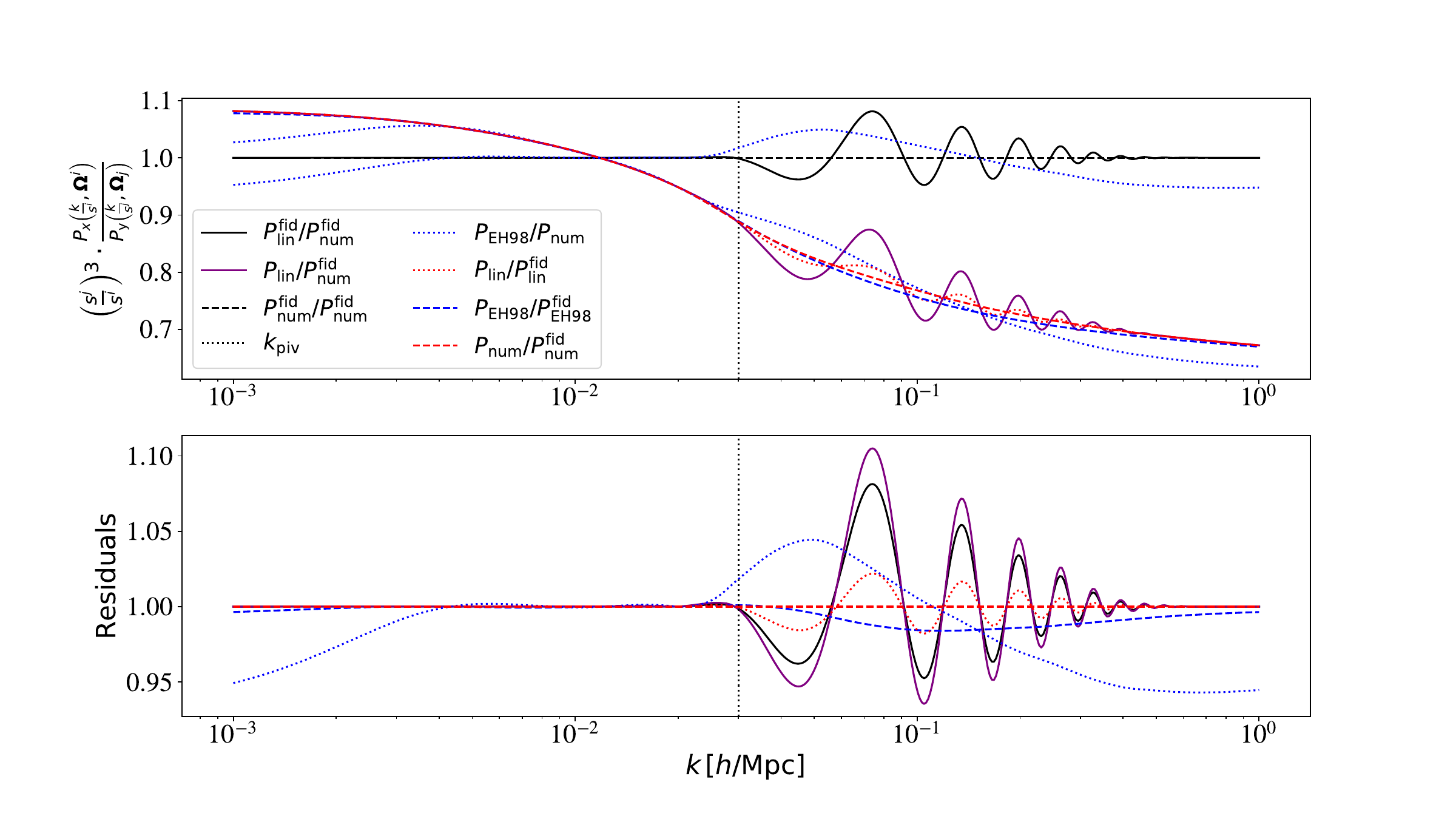}
    \caption{Comparison between the different BAO smoothing methods, the baseline EH98 method (blue, labeled `EH98') and the numerical method (red, labeled `num'), where the final results relevant for ShapeFit are represented by the dashed lines and the dotted lines display intermediate steps. In the top panel we show the fiducial power spectrum in black and the $s$-rescaled spectrum of a different cosmology with reduced matter density $\Delta \omega_\mathrm{m}=-0.02$ in purple. Both are divided by the numerically smoothed fiducial power spectrum $P_\mathrm{num}^\mathrm{fid}$ to visualize their difference in {\it Shape}. The ratio of both spectra, which serves as input for our numerical smoothing method (step 6 in text, where here we exclude the EH98 factors in the top panel, but include them in the bottom panel) is given by the red dotted line. The blue dotted lines show the ratio of the EH98 method with respect to the numerical method, which deviate from each other by up to 5\%. Note, however, that ShapeFit relies on the relative shift in {\it Shape} with respect to the fiducial, which does not show any difference between methods at the pivot scale $k_p$ (black, dotted lines). 
    }
    \label{fig:smoothing-visualization}
\end{figure}

We explain this procedure visually and compare it to the pure EH98 smoothing method in figure \ref{fig:smoothing-visualization}. The top panel shows the ratio of $s^{i/j}$-rescaled power spectra $P_x(\mathbf{\Omega}^i)/P_y(\mathbf{\Omega}^j)$, where the subscripts $x,y$ either stand for the linear power spectrum (`lin') or for the no-wiggle linear power spectrum obtained via the analytic (`EH98') formula or the numerical method (`num') described here. The superscripts $i,j$ either stand for the fiducial cosmology (`fid') of table \ref{tab:cosmo} or a cosmology (empty superscript) with displaced matter density $\Delta \omega_\mathrm{m}=-0.02$. In most cases the denominator is given by $P_\mathrm{num}^\mathrm{fid} = P_\mathrm{num}(k,\mathbf{\Omega}^\mathrm{fid})$ as reference, but also other cases are displayed for comparison. The exact values of the sub- and superscripts are specified in the legend, where the dependencies $P(\dots)$ are not written out for simplicity. 

From the top panel of figure \ref{fig:smoothing-visualization} we can appreciate the difference in {\it Shape} between the $s$-rescaled power spectrum of the displaced cosmology (purple line) and the fiducial power spectrum (black line). Their ratio (red, dotted line) still shows some residual wiggles due to the difference in BAO amplitude between the two cosmologies. Its residual with respect to the EH98 function (red, dotted line of bottom panel) is used as input for our numerical smoothing function (step 6). The advantage of using the residual power spectrum ratio as input is that the numerical method is more precise (and in particular more independent of the hyperparameters) the smaller the BAO amplitude. This is simply because with smaller amplitude there is less room left for possible numerical smoothing solutions. In fact, if the residual BAO amplitude is zero, all numerical smoothing operations would deliver identical results. The final numerically smoothed spectrum $P_\mathrm{num}/P_\mathrm{num}^\mathrm{fid}$ (red, dashed lines) can be compared to the baseline EH98 result $P_\mathrm{EH98}/P_\mathrm{EH98}^\mathrm{fid}$ (blue, dashed lines) in both panels. We see that they are in nearly perfect agreement for $k<0.05 \hoverMpc$ and reach at most a $2\%$ discrepancy at  $k=0.1 \hoverMpc$. Since we measure the slope at pivot scale $k_p=0.03 \hoverMpc$ (black, dotted lines), both methods deliver the exact same cosmological constraints, as demonstrated for our baseline $\Lambda$CDM model in the right panel of figure \ref{fig:prior-ns-and-smoothing}. However, to achieve this result it is very important to use for each cosmology the same BAO wiggle removal procedure as for the fiducial power spectrum template. This is because, in absolute terms, the EH98 prediction (blue, dotted lines) is quite different from the numerical calculation, in this case up to 5\% and even up to 2\% at $k_p$.

\section{Full data-vectors and covariances}\label{app:datavectors}

We list the full covariances and data-vectors corresponding to the compressed set of variables of the samples used in this paper. These results correspond to those listed in table~\ref{tab:boss_compare}, but explicitly displaying the cross correlation among parameters. Unlike in table~\ref{tab:boss_compare} we display the $f\sigma_{s8}$ parameter (see section~\ref{sec:theory_shapefit} for its definition), which is slightly different from the usual $f\sigma_8$ parameter which has been reported before. 

The two BOSS LRG overlaping samples at $0.2<z<0.5;\, z_{\rm eff}=0.38$ and $0.4<z<0.6;\, z_{\rm eff}=0.51$ correspond to, 
 \begin{equation}
   D^{\rm LRG}(0.38,0.51)=
 \begin{pmatrix}
D_H/r_{\rm d}(0.38) \\
D_M/r_{\rm d}(0.38) \\
f\sigma_{s8}(0.38) \\
m(0.38) \\
D_H/r_{\rm d}(0.51) \\
D_M/r_{\rm d}(0.51) \\
f\sigma_{s8}(0.51) \\
m(0.51) \\
 \end{pmatrix}=
  \begin{pmatrix}
24.979 \\
10.239 \\
0.47323 \\
-0.066479 \\
22.256 \\
13.302 \\
0.48109 \\
-0.022856 
 \end{pmatrix},
 \end{equation}
with  covariance
   \begin{eqnarray} 
&C^{\rm LRG}& (0.38;0.51) = 10^{-3} \cdot \\
\nonumber& & \begin{pmatrix}
370.75 & -29.539 & -11.436 & -1.3729 & 108.63 & -8.6873  &  -3.2297  &     -0.87764\\
 & 24.259 & 2.9207 & -0.52414 & -8.0277  &  -4.2352 & 1.0342  & -0.25930 \\
 & & 2.1106 & -0.073861 & -3.1726  &  -0.29013   &    0.70911 & 0.071011 \\
 & & & 1.7474 & -0.94588  &  -0.58482 &   0.097818 & 0.74692 \\
  &  &  &  & 277.21 & -34.533 & -7.9761 & -1.5555 \\
  &  &  &  & & 35.805 & 2.8011 & -0.36639 \\
   &  &  &  & & & 1.6569 & -0.18942 \\
    &  &  &  &  &  &  & 1.9244
 \end{pmatrix}.
\end{eqnarray}
The BOSS and eBOSS LRG sample, $0.6<z<1.0;\,z_{\rm eff}=0.698$ has,
 \begin{equation}
   D^{\rm LRG}(0.698)=
 \begin{pmatrix}
D_H/r_{\rm d} \\
D_M/r_{\rm d} \\
f\sigma_{s8} \\
m \\
 \end{pmatrix}=
  \begin{pmatrix}
19.542 \\
17.700 \\
0.47947 \\
-0.008019
 \end{pmatrix},
 \end{equation}
 and the following covariance,
   \begin{equation} 
C^{\rm LRG} (0.698) = 10^{-3}
 \begin{pmatrix}
201.75 & -46.049 & -5.4088 & -2.7122 \\
 & 93.889 & 5.0808 & -0.51303 \\
 & & 1.8683 & -0.70222 \\
 & & & 2.6965 
 \end{pmatrix}.
\end{equation}
Finally, the quasar sample at $0.8<z<2.2;\,z_{\rm eff}=1.48$ has, 
 \begin{equation}
   D^{\rm QSO}(1.48)=
 \begin{pmatrix}
D_H/r_{\rm d} \\
D_M/r_{\rm d} \\
f\sigma_{s8} \\
m \\
 \end{pmatrix}=
  \begin{pmatrix}
13.274 \\ 
31.007 \\
0.46773 \\
-0.005327
 \end{pmatrix},
 \end{equation}
 and covariance, 
  \begin{equation} 
C^{\rm QSO} (1.48) = 10^{-3}
 \begin{pmatrix}
245.99 & -16.412 & -7.8840 & -4.4936 \\
 & 669.76 & 20.710 & -5.9721 \\
 & & 1.7861 & 0.086278 \\
 & & & 1.0934 
 \end{pmatrix}.
\end{equation}
We provide these compressed parameter covariances, the power spectra measurements, power spectrum covariances and window matrices employed in this analysis in \cite{shapefit:data}.

\acknowledgments
We thank the anonymous referee for useful comments and suggestions that helped to improve the manuscript.
We thank Davide Gualdi for providing the building block of our numerical BAO smoothing function. This work makes use of the \textsc{fftw} and \textsc{gsl} libraries, \texttt{scipy} and \texttt{numpy} \texttt{python} packages and plots were generated using \texttt{getdist} and \texttt{gnuplot}. We also thank the developers of \textsc{CLASS} and \textsc{MontePython}.
H.G-M. and S.B. acknowledges the support from ‘la Caixa’ Foundation (ID100010434) with code LCF/BQ/PI18/11630024. L.V., H.G-M. and S.B. acknowledge support of European Union's Horizon 2020 research and innovation programme ERC (BePreSySe, grant agreement 725327). Funding for this work was partially provided by the Spanish MINECO under project PGC2018-098866-B-I00 MCIN/AEI/10.13039/501100011033 y FEDER “Una manera de hacer Europa”, and the “Center of Excellence Maria de Maeztu 2020-2023” award to the ICCUB (CEX2019-000918-M funded by MCIN/AEI/10.13039/501100011033).

The MultiDark-Patchy mocks was an effort led from the IFT UAM-CSIC by F. Prada’s group (C.-H. Chuang, S. Rodriguez-Torres and C. Scoccola) in collaboration with C. Zhao (Tsinghua U.), F.-S. Kitaura (AIP), A. Klypin (NMSU), G. Yepes (UAM), and the BOSS galaxy clustering working group.

We thank Cheng Zhao for providing the EZmock periodic cubic boxes for the quasar sample, as well as for the effort of providing the whole set of EZmocks light-cones for the eBOSS samples. We also thank Jeremy Tinker for leading the Nseries mocks production within the BOSS team effort.

We thank the whole BOSS and eBOSS teams for their valuable effort in creating the BOSS and eBOSS catalogues and making it publicly available for the community.

We acknowledge the IT team at ICCUB for the
help with the \textsc{Aganice} and \textsc{Hipatia} clusters where all the calculations presented in this paper where done.

%
%  These Macros are taken from the AAS TeX macro package version 4.0.
%  Include this file in your LaTeX source only if you are not using
%  the AAS TeX macro package and need to resolve the macro definitions
%  in the BibTeX entries returned by the ADS abstract service.
%
%  For more information on the AASTeX macro package, please see the URL
%	http://www.aas.org/publications/aastex.html
%  For more information about ADS abstract server, please see the URL
%	http://adswww.harvard.edu/ads_abstracts.html
%

% Abbreviations for journals.  The object here is to provide authors
% with convenient shorthands for the most "popular" (often-cited)
% journals; the author can use these markup tags without being concerned
% about the exact form of the journal abbreviation, or its formatting.
% It is up to the keeper of the macros to make sure the macros expand
% to the proper text.  If macro package writers agree to all use the
% same TeX command name, authors only have to remember one thing, and
% the style file will take care of editorial preferences.  This also
% applies when a single journal decides to revamp its abbreviating
% scheme, as happened with the ApJ (Abt 1991).

\def\jnl@style{\it}
%commente par Seb
\def\aaref@jnl#1{{\jnl@style#1}}
%ref remplace par aaref pour eviter conflit...

\def\aaref@jnl#1{{\jnl@style#1}}

\def\aj{\aaref@jnl{AJ}}                   % Astronomical Journal
\def\araa{\aaref@jnl{ARA\&A}}             % Annual Review of Astron and Astrophys
\def\apj{\aaref@jnl{ApJ}}                 % Astrophysical Journal
\def\apjl{\aaref@jnl{ApJ}}                % Astrophysical Journal, Letters
\def\apjs{\aaref@jnl{ApJS}}               % Astrophysical Journal, Supplement
\def\ao{\aaref@jnl{Appl.~Opt.}}           % Applied Optics
\def\apss{\aaref@jnl{Ap\&SS}}             % Astrophysics and Space Science
\def\aap{\aaref@jnl{A\&A}}                % Astronomy and Astrophysics
\def\aapr{\aaref@jnl{A\&A~Rev.}}          % Astronomy and Astrophysics Reviews
\def\aaps{\aaref@jnl{A\&AS}}              % Astronomy and Astrophysics, Supplement
\def\azh{\aaref@jnl{AZh}}                 % Astronomicheskii Zhurnal
\def\baas{\aaref@jnl{BAAS}}               % Bulletin of the AAS
\def\jrasc{\aaref@jnl{JRASC}}             % Journal of the RAS of Canada
\def\memras{\aaref@jnl{MmRAS}}            % Memoirs of the RAS
\def\mnras{\aaref@jnl{MNRAS}}             % Monthly Notices of the RAS
\def\pra{\aaref@jnl{Phys.~Rev.~A}}        % Physical Review A: General Physics
\def\prb{\aaref@jnl{Phys.~Rev.~B}}        % Physical Review B: Solid State
\def\prc{\aaref@jnl{Phys.~Rev.~C}}        % Physical Review C
\def\prd{\aaref@jnl{Phys.~Rev.~D}}        % Physical Review D
\def\pre{\aaref@jnl{Phys.~Rev.~E}}        % Physical Review E
\def\prl{\aaref@jnl{Phys.~Rev.~Lett.}}    % Physical Review Letters
\def\pasp{\aaref@jnl{PASP}}               % Publications of the ASP
\def\pasj{\aaref@jnl{PASJ}}               % Publications of the ASJ
\def\qjras{\aaref@jnl{QJRAS}}             % Quarterly Journal of the RAS
\def\skytel{\aaref@jnl{S\&T}}             % Sky and Telescope
\def\solphys{\aaref@jnl{Sol.~Phys.}}      % Solar Physics
\def\sovast{\aaref@jnl{Soviet~Ast.}}      % Soviet Astronomy
\def\ssr{\aaref@jnl{Space~Sci.~Rev.}}     % Space Science Reviews
\def\zap{\aaref@jnl{ZAp}}                 % Zeitschrift fuer Astrophysik
\def\nat{\aaref@jnl{Nature}}              % Nature
\def\iaucirc{\aaref@jnl{IAU~Circ.}}       % IAU Cirulars
\def\aplett{\aaref@jnl{Astrophys.~Lett.}} % Astrophysics Letters
\def\apspr{\aaref@jnl{Astrophys.~Space~Phys.~Res.}}
                % Astrophysics Space Physics Research
\def\bain{\aaref@jnl{Bull.~Astron.~Inst.~Netherlands}} 
                % Bulletin Astronomical Institute of the Netherlands
\def\fcp{\aaref@jnl{Fund.~Cosmic~Phys.}}  % Fundamental Cosmic Physics
\def\gca{\aaref@jnl{Geochim.~Cosmochim.~Acta}}   % Geochimica Cosmochimica Acta
\def\grl{\aaref@jnl{Geophys.~Res.~Lett.}} % Geophysics Research Letters
\def\jcp{\aaref@jnl{J.~Chem.~Phys.}}      % Journal of Chemical Physics
\def\jgr{\aaref@jnl{J.~Geophys.~Res.}}    % Journal of Geophysics Research
\def\jqsrt{\aaref@jnl{J.~Quant.~Spec.~Radiat.~Transf.}}
                % Journal of Quantitiative Spectroscopy and Radiative Transfer
\def\memsai{\aaref@jnl{Mem.~Soc.~Astron.~Italiana}}
                % Mem. Societa Astronomica Italiana
\def\nphysa{\aaref@jnl{Nucl.~Phys.~A}}   % Nuclear Physics A
\def\physrep{\aaref@jnl{Phys.~Rep.}}   % Physics Reports
\def\physscr{\aaref@jnl{Phys.~Scr}}   % Physica Scripta
\def\planss{\aaref@jnl{Planet.~Space~Sci.}}   % Planetary Space Science
\def\procspie{\aaref@jnl{Proc.~SPIE}}   % Proceedings of the SPIE
\def\jcap{\aaref@jnl{J. Cosmology Astropart. Phys.}}
                % Journal of Cosmology and Astroparticle Physics

\let\astap=\aap
\let\apjlett=\apjl
\let\apjsupp=\apjs
\let\applopt=\ao

\newcommand{\etal}{et al.\ }

\newcommand{\mpc}{\, {\rm Mpc}}
\newcommand{\kpc}{\, {\rm kpc}}
\newcommand{\hmpc}{\, h^{-1} \mpc}
\newcommand{\ihmpc}{\, h\, {\rm Mpc}^{-1}}
\newcommand{\ikms}{\, {\rm s\, km}^{-1}}
\newcommand{\kms}{\, {\rm km\, s}^{-1}}
\newcommand{\hkpc}{\, h^{-1} \kpc}
\newcommand{\lya}{Ly$\alpha$\ }
\newcommand{\lyb}{Lyman-$\beta$\ }
\newcommand{\lyaf}{Ly$\alpha$ forest}
\newcommand{\lr}{\lambda_{{\rm rest}}}
\newcommand{\bF}{\bar{F}}
\newcommand{\bS}{\bar{S}}
\newcommand{\bC}{\bar{C}}
\newcommand{\bB}{\bar{B}}
\newcommand{\vdF}{{\mathbf \delta_F}}
\newcommand{\vdS}{{\mathbf \delta_S}}
\newcommand{\vdf}{{\mathbf \delta_f}}
\newcommand{\vdn}{{\mathbf \delta_n}}
\newcommand{\vdC}{{\mathbf \delta_C}}
\newcommand{\vdX}{{\mathbf \delta_X}}
\newcommand{\xrei}{x_{rei}}
\newcommand{\lrmin}{\lambda_{{\rm rest, min}}}
\newcommand{\lrmax}{\lambda_{{\rm rest, max}}}
\newcommand{\lmin}{\lambda_{{\rm min}}}
\newcommand{\lmax}{\lambda_{{\rm max}}}
\newcommand{\hi}{\mbox{H\,{\scriptsize I}\ }}
\newcommand{\heii}{\mbox{He\,{\scriptsize II}\ }}
\newcommand{\vp}{\mathbf{p}}
\newcommand{\vq}{\mathbf{q}}
\newcommand{\vxperp}{\mathbf{x_\perp}}
\newcommand{\vkperp}{\mathbf{k_\perp}}
\newcommand{\vrperp}{\mathbf{r_\perp}}
\newcommand{\vx}{\mathbf{x}}
\newcommand{\vy}{\mathbf{y}}
\newcommand{\vk}{\mathbf{k}}
\newcommand{\vR}{\mathbf{r}}
\newcommand{\tdtwo}{\tilde{b}_{\delta^2}}
\newcommand{\tstwo}{\tilde{b}_{s^2}}
\newcommand{\tbthree}{\tilde{b}_3}
\newcommand{\tadtwo}{\tilde{a}_{\delta^2}}
\newcommand{\tastwo}{\tilde{a}_{s^2}}
\newcommand{\tabthree}{\tilde{a}_3}
\newcommand{\tpsi}{\tilde{\psi}}
\newcommand{\vv}{\mathbf{v}}
\newcommand{\fnl}{{f_{\rm NL}}}
\newcommand{\tfnl}{{\tilde{f}_{\rm NL}}}
\newcommand{\gnl}{g_{\rm NL}}
\newcommand{\orderfour}{\mathcal{O}\left(\delta_1^4\right)}
\newcommand{\SDSSPF}{\cite{2006ApJS..163...80M}}
\newcommand{\PF}{$P_F^{\rm 1D}(k_\parallel,z)$}
\newcommand\ionalt[2]{#1$\;${\scriptsize \uppercase\expandafter{\romannumeral #2}}}%  
\newcommand{\vxone}{\mathbf{x_1}}
\newcommand{\vxtwo}{\mathbf{x_2}}
\newcommand{\vRot}{\mathbf{r_{12}}}
\newcommand{\cm}{\, {\rm cm}}

%\bibliographystyle{JHEP}
%\bibliography{bibliography}

\providecommand{\href}[2]{#2}\begingroup\raggedright\endgroup

\end{document}